\documentclass[%
 aip,
 amsmath,amssymb,
 reprint,%
floatfix]{revtex4-1}

\usepackage{graphicx}
\usepackage{dcolumn}
\usepackage{bm}
\usepackage[utf8]{inputenc}
\usepackage[T1]{fontenc}
\usepackage{mathptmx}
\usepackage{pbox}
\usepackage{array}
\usepackage{makecell}
\usepackage{caption}
\usepackage{graphicx}
\usepackage{dcolumn}
\usepackage{bm}
\usepackage{amsmath,empheq}
\usepackage{bigints}
\usepackage{float}
\usepackage{epsfig}
\usepackage{epstopdf}
\usepackage{eqnarray}
\usepackage{mathrsfs}
\usepackage{natbib}
\usepackage{mathtools}
\usepackage{subcaption}
\begin{document}

\preprint{AIP/123-QED}

\title[]{Chimera states for a globally coupled sine circle map lattice: spatiotemporal intermittency and hyperchaos}

\author{Joydeep Singha}
\email{joydeep@physics.iitm.ac.in}
\affiliation{Department of Physics, Indian Institute of Technology Madras, Chennai, 600036, India}
\author{Neelima Gupte}%
\email{gupte@physics.iitm.ac.in}
\affiliation{Department of Physics, Indian Institute of Technology Madras, Chennai, 600036, India}

\date{\today}

\begin{abstract}
We study the existence of chimera states, i.e. mixed states, in a globally coupled sine circle map lattice, with different strengths of inter-group and intra-group coupling. We find that at specific values of the parameters of the CML, a completely random initial condition evolves to chimera states, having a phase synchronised and a phase desynchronised group, where the space time variation of the phases of the maps in the desynchronised group shows structures similar to spatiotemporally intermittent regions. Using the complex order parameter we obtain a phase diagram that identifies the region in the parameter space which supports chimera states of this type, as well as other types of phase configurations such as globally phase synchronised states, two phase clustered states and fully phase desynchronised states. We estimate the volume of the basin of attraction of each kind of solution. 
The STI chimera region is studied in further detail via numerical and analytic stability analysis, and the Lyapunov spectrum is calculated. This state is identified to be hyperchaotic as the two largest Lyapunov exponents are found to be positive. The distributions of laminar and burst lengths in the incoherent region of the chimera show exponential behaviour. The average fraction of laminar/burst sites is identified to be the important quantity which governs the dynamics of the chimera. After an initial transient, these settle to steady values which can be used to reproduce the phase diagram in the chimera regime.
\end{abstract}
\maketitle

\begin{quotation}
The study of chimera states, i.e. mixed states where synchronised and desynchronised dynamics coexist, has been at the forefront of studies in nonlinear dynamics involving both theoretical and experimental systems. A variety of classes of chimera states, i.e. states which contain co-existing domains of distinct kinds of spatiotemporal behaviour can be seen. These include multi-headed chimera states, travelling chimera states, amplitude chimera states, twisted chimera states etc, and have been seen in coupled oscillator models such as the Kuramoto model,coupled Ginzburg-Landau oscillators and other systems. Here, we investigate chimera and other states in a coupled sine circle map lattice which is a discrete version of coupled oscillator systems. The CML consists of two populations of globally coupled identical sine circle maps with distinct values for the intergroup and intragroup coupling. We observe spatiotemporally intermittent chimeras, i.e. states which consist of a synchronised subgroup, and a state where coherent (phase synchronised) and incoherent (phase incoherent) domains co-exist, at low values of the nonlinearity map parameter. Such STI chimeras have been observed earlier in coupled oscillators models such as Stuart-Landau oscillators, Ginzburg-Landau oscillators, coupled optical resonators, chemical reactions etc. We analyse the STI chimera seen in the CML system by plotting the phase diagram of the system using the global order parameter, and identifying  the region where STI chimeras can be seen. The basin stability of the STI chimera state, as opposed to other states e.g. fully synchronised states, fully desynchronised states and two cluster states,  which can be seen in the phase diagram is established. The linear stability analysis of the chimera region is carried out, using analytic and numerical methods. The Lyapunov exponents obtained via this analysis establish that the STI chimera is hyperchaotic. Further, the pairwise order parameter is used to distinguish laminar and burst sites, and the time evolution of the laminar and burst sites show that the fraction of laminar and burst sites in the system reaches a steady state. The phase diagram obtained from these stationary states matches the phase diagram obtained from the complex order parameter exactly. We also study the distributions of laminar and burst sites in the system, and find that they fall off exponentially due to the globally coupled nature of the system.       
\end{quotation}

\section{\label{sec: introduction} Introduction}
\par The chimera phase pattern is a remarkable spatiotemporal property found in spatially extended dynamical systems. This phase pattern has been seen in systems of coupled phase oscillators \cite{kuramoto2002, strogatz2004, strogatz2005, strogatz2008, strogatz2010, sethia2008, sheeba2009, sheeba2010, omelchenko2008, laing2009, wang2011, showalter2012, showalter2013, showalter2018, shashi2013, omelchenko2018, bountis2014, panaggio2016, tareda2016, dai2018, wu2018, maistrenko2014, jaros2015, xie2014, dudkowski2014, hovel2015, yao2015, xie2015} and was recently discovered in coupled map lattice models  \cite{nayak2011, neelima2016, hagerstrom2012, li2018}. In the context of dynamical systems, the `chimera'  state is defined to be a state with the characteristic stable coexistence of a synchronous group of oscillators together with a desynchronised group of oscillators. Similar dynamical behaviour was found in early studies of unihemispheric sleep \cite{rattenborg2000} and the asynchronous eye closure \cite{mathews2006} of sea mammals, birds and reptiles. In addition to the phase coupled oscillator systems mentioned above, this kind of  spatio-temporal behaviour has also been seen to exist in other oscillator systems. These include non-locally coupled complex Ginzburg-Landau oscillators \cite{kuramoto2002}, delay-coupled rings of phase oscillators \cite{sethia2008}, bipartite oscillator populations \cite{sheeba2009, sheeba2010}, Stuart-Landau oscillators \cite{omelchenko2008}, networks of Kuramoto oscillators \cite{laing2009, wang2011}, coupled chemical oscillators \cite{showalter2012, showalter2013, showalter2018}, and mechanical oscillator networks \cite{shashi2013}. The detailed analysis of oscillator systems with different kinds of coupling has been reviewed recently by Omel'chenko \cite{omelchenko2018}.

\par Here, we study the existence of chimera states in a coupled map lattice which is a discrete analog of coupled phase oscillator system where both space and time are considered to be discrete. The chimera phase state as well as other other mixed states were reported in specific systems of coupled map lattices in both theoretical \cite{nayak2011, neelima2016} models and  experimental systems \cite{hagerstrom2012, li2018}. The CML, used here, is of the form used in Refs. \cite{nayak2011, neelima2016} and consists of two populations of globally coupled identical sine circle maps where the strength of the coupling within each population and that between the maps belonging to distinct populations take different values. Oscillator models with two species of identical dynamical units, leading to chimera states have been explored earlier in refs. \cite{strogatz2008, bountis2014, panaggio2016, tareda2016, dai2018} for phase oscillators and in Ref.  \cite{wu2018} for Fitzhugh-Nagumo oscillators.  The existence of chimera states in globally coupled systems has also been reported for systems of Stuart-Landau oscillators and for the complex Ginzburg-Landau equation \cite{sethia2014, schmidt2015, konrad2014}.

\par We note that different types of chimera states with interesting spatio-temporal behaviours have been studied in various contexts. These include multiheaded chimera states \cite{maistrenko2014, jaros2015}, travelling chimera states\cite{xie2014, dudkowski2014}, multi-chimera states \cite{hovel2015, yao2015}, twisted chimera states \cite{xie2015}, and amplitude chimera states \cite{sathi2018}. It was also shown earlier that the specific CML which we study here can support another kind of mixed state, namely the splay-chimera state where the  coexistence of a phase synchronised group of maps and a phase desynchronised group of maps consisting of splay phase configurations was reported \cite{neelima2016}. In this paper, we report the existence of yet another kind of chimera state for this system, where the evolution of random initial conditions in certain regions of the parameter space results in a new class of chimera solutions where the space time variation of the desynchronised group shows spatiotemporally intermittent behaviour. In addition to the chimera states described here, this system supports various other kinds of phase configurations viz. globally synchronised states, two phase clustered states, fully phase desynchronised states, etc. We define a complex order parameter for the entire system as well as for each group. We show that the transition between these phase configurations upon the change of the parameters can be identified from these order parameters which take unique values for each of these states. We thus obtain the phase diagram of the coupled map lattice and identify the regimes which support chimera states of this type, and regimes which support other phase configurations. Subsequent analysis focusses on the chimera region of the phase diagram  and its neighbourhood. We note that chimeras with co-existing coherent and incoherent regions with spatiotemporally intermittent structures have also been seen in systems of coupled oscillators with global \cite{azamat2014, konrad2014, gregory2010, krischer2015, lennart2014} and local coupling \cite{lennart2015, clerc2016, clerc2017}.

\par We carry out the stability analysis of each solution thus identified with special focus on the analysis of the chimera states having spatiotemporally intermittent structures. We note that the phase space is high dimensional, leading to the existence of multiattractor solutions at identical parameter values. We find the relative volume of the basin of attraction of all these solutions including the STI chimera and its mirrored version by estimating the fraction of initial condition which evolve to each state.

\par The linear stability analysis of the STI chimera can be carried out analytically due to the low values of the nonlinearity parameter. The values of the Lyapunov spectrum obtained analytically in this regime, match the numerically obtained values. Two of the Lyapunov exponents of the system turn out to be positive, implying that the temporal evolution of the STI chimera is hyperchaotic. Thus, this is one of the very few hyperchaotic chimera solutions seen so far \cite{wolfrum2011}. The laminar (coherent) and burst (incoherent) sites are identified using a pairwise version of the global order parameter. The distribution of the length of laminar and turbulent segments shows exponential behaviour with a higher probability of longer turbulent segments. Due to the global nature of the coupling, the spatiotemporal evolution of the STI chimera depends only on the fraction of laminar and turbulent sites in each subgroup. The average fraction of laminar and turbulent sites in each subgroup saturates to steady state values after an initial transient. These steady state values are used to recreate the phase diagram in this regime. This phase diagram matches exactly the phase diagram obtained via the global and subgroup order parameters, confirming that the average fraction of laminar and turbulent sites in each subgroup is the crucial factor which governs the dynamics of our system. We discuss the implications of our results in practical contexts.      

\par Our paper is organised in the following manner: Section \ref{sec: model} discusses the coupled sine circle map lattice model under study.  In section \ref{sec: PD}, we introduce the complex order parameters and obtain a phase diagram using their calculated values. We also discuss here the variety of phase configurations that can be found when the system is evolved using random initial conditions. Section \ref{sec: BS} discusses the basin stability of each of attractors including the chimera states. In section \ref{sec: chim} we discuss the behavior of the chimera consisting of a  phase synchronised group and desynchronised group with spatio-temporally intermittent regions and obtain the Lyapunov exponents in section \ref{sec: LS and LE}. A method of identifying and labelling the laminar and burst sites is outlined in section \ref{sec: laminar_burst} and the distribution of laminar and burst segments is discussed in section \ref{sec: LB}. The evolution of the fraction of laminar and turbulent sites is discussed in section \ref{sec: transition} and the phase diagram is obtained in terms of their steady-state values. Section \ref{sec: conclusion} summarises our conclusions.

\section{\label{sec: model}The model}
\begin{figure}[t]
\includegraphics[scale = 0.35]{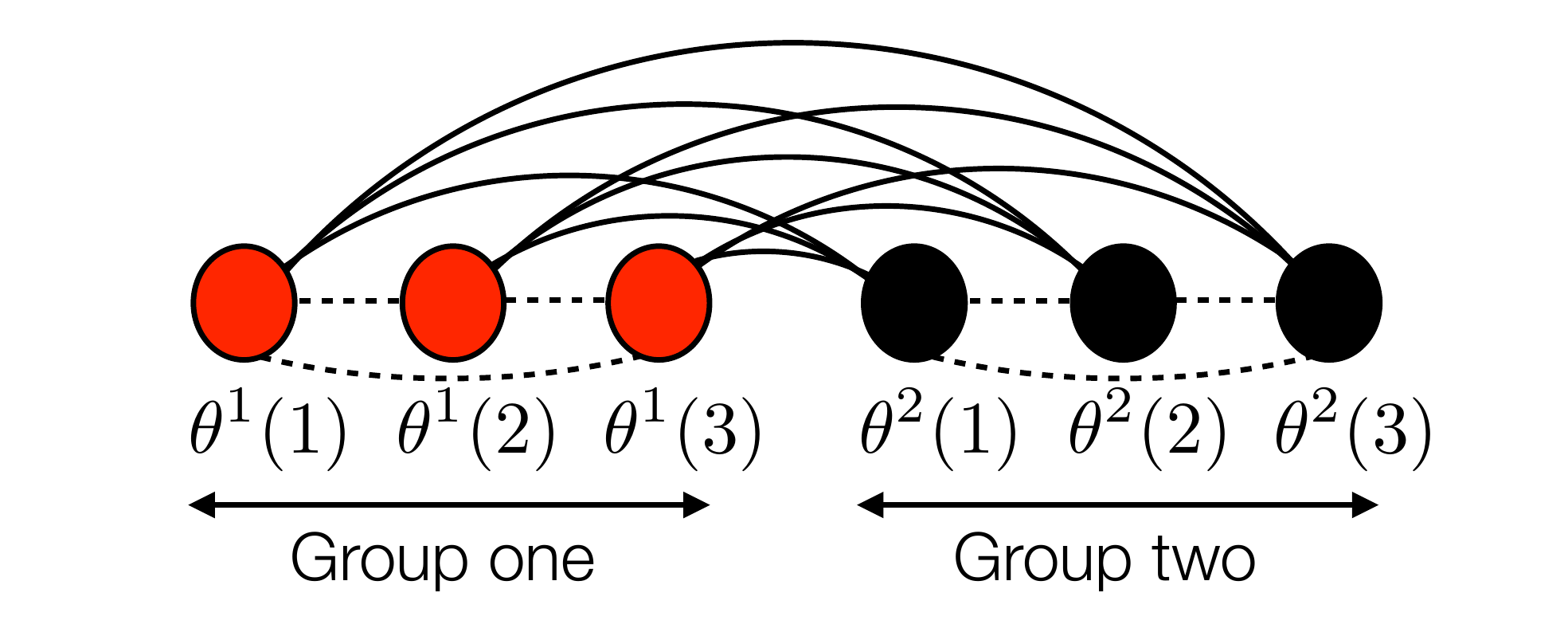}
\caption{\label{fig: topology}\footnotesize (color online) The schematic of the system of Eq. \ref{sinecml} with 3 maps in each group. The intergroup coupling is shown by solid lines and the intra-group coupling is denoted by dotted lines.}
\end{figure}

Here, we study a lattice of coupled sine circle maps, where the maps are distributed into two groups, which are globally coupled, but with two distinct values for the intragroup and intergroup coupling. The evolution equation for a single sine circle map is given by,
\begin{equation}
\theta_{n + 1} = \theta_{n} + \Omega -\frac{K}{2\pi}\sin(2\pi\theta_{n}) \mod{1}
\label{sinecir}
\end{equation}
where $\theta$ is the phase of the map, $0<\theta<1$ and $n$ is the time step. The parameter $\Omega$ denotes the frequency ratio in the absence of nonlinearity and $K$ determines the strength of nonlinearity. A single sine circle map shows Arnold tongues organised by frequency locking and quasi-periodic behaviours \cite{ott1993}. It shows universality in the mode locking structure prior to both the period doubling route to chaos and quasi-periodic route to chaos depending on the value of $\Omega$ \cite{jensen1983, ott1993}. The evolution equation for the coupled sine circle map lattice considered here is given by,
\begin{widetext}
\begin{equation}
\theta_{n+1}^{\sigma}(i) = \theta_{n}^{\sigma}(i)  + \Omega - \frac{K}{2\pi} \sin(2\pi \theta_{n}^{\sigma}(i)) + \sum\limits_{\sigma' = 1}^{2} \frac{\epsilon_{\sigma \sigma'}}{N} \left[ \sum\limits_{j = 1}^{N}(\theta_{n}^{\sigma'}(j) + \Omega - \frac{K}{2\pi} \sin(2\pi\theta_{n}^{\sigma'}(j)))\right] 
\mod 1
\label{sinecml}
\end{equation}
\end{widetext}
The equation above defines the evolution of the $i$th map in the group $\sigma$, where $\sigma$ takes values $1,2$, and $N$ is the number of maps in each of the groups. We also define the coupling parameters to be $\epsilon_{11} = \epsilon_{22} = \epsilon_{1}$ and  $\epsilon_{12} = \epsilon_{21} = \epsilon_{2}$ with the constraint $\epsilon_1 + \epsilon_2 = 1$. Therefore,  our model consists of two groups of identical sine circle maps where $N$ is the number of maps in each group. Each map in a given group is coupled to all the maps in its own group by the parameter $\epsilon_1$ whereas it is coupled to the maps in the other group by the parameter  $\epsilon_2$. We note that the evolution equation is completely symmetric under interchange of the group labels, $\sigma = 1, 2$. Thus the system in equation \ref{sinecml} is controlled by three independent parameters, $K, \Omega, \epsilon_1$. A schematic of the CML of Eq. \ref{sinecml} with three lattice sites in each group is shown in figure \ref{fig: topology}. 

\par This CML is a discrete version of globally coupled oscillator models with two populations, which have been motivated by biological examples of chimera states, such as the unihemispherical sleep patterns of sea mammals \cite{rattenborg2000, mathews2006}. In the oscillator context a model consisting of two groups of identical Kuramoto oscillators representing each hemisphere of brain was proposed by Abrams et al. \cite{strogatz2008} and showed chimera states. The CML which we discuss has a similar coupling topology, and couples identical sine circle maps, which represent discrete versions of phase oscillator systems.

\par The system under consideration has many degrees of freedom with maps that are coupled globally with two groups which differ in their intergroup and intragroup coupling. As a consequence of this, different initial conditions generally evolve to distinct attractors with different spatiotemporal properties; e.g. an initial condition where an identical phase is assigned to each site will always evolve to a globally synchronised state. In \cite{nayak2011} it was shown that an initial condition, where all the phases of the maps in one group are identical while the maps in the other group are set to random phases between zero and one, evolves to chimera states, clustered chimera states, clustered states etc. at different region in the parameter space.Another initial condition with a system wide splay phase configuration was shown to evolve to a splay phase state, and to splay chimera states depending on the parameters \cite{neelima2016}. Initial conditions such as these break the symmetry between the groups. Here, we explore this CML using a very general initial condition where the phases of each of the maps in both of the groups are randomly distributed between zero and one. 

\par
\begin{figure*}
\centering \begin{tabular}{ccc}
\includegraphics[scale = 0.45]{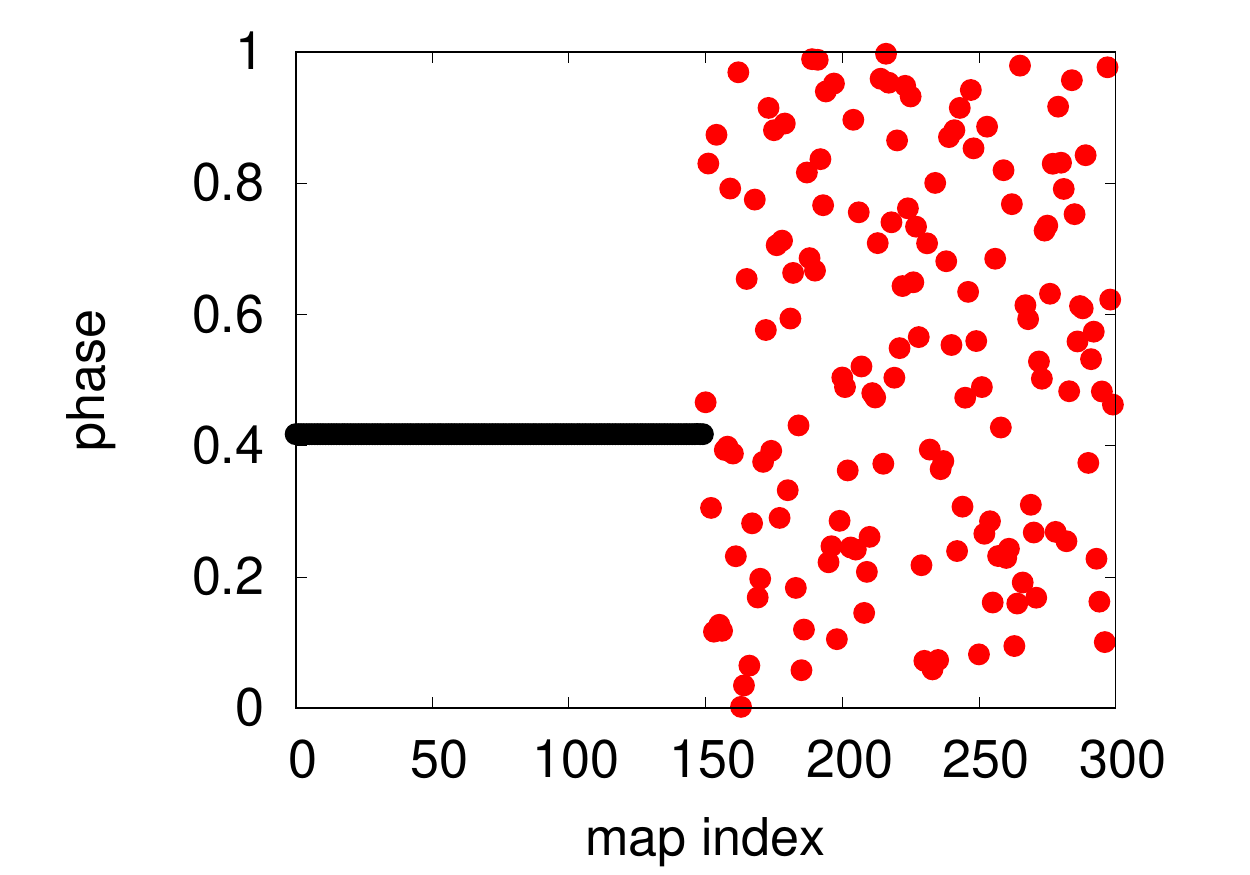}&
\hspace{-0.8cm}
\includegraphics[scale = 0.5]{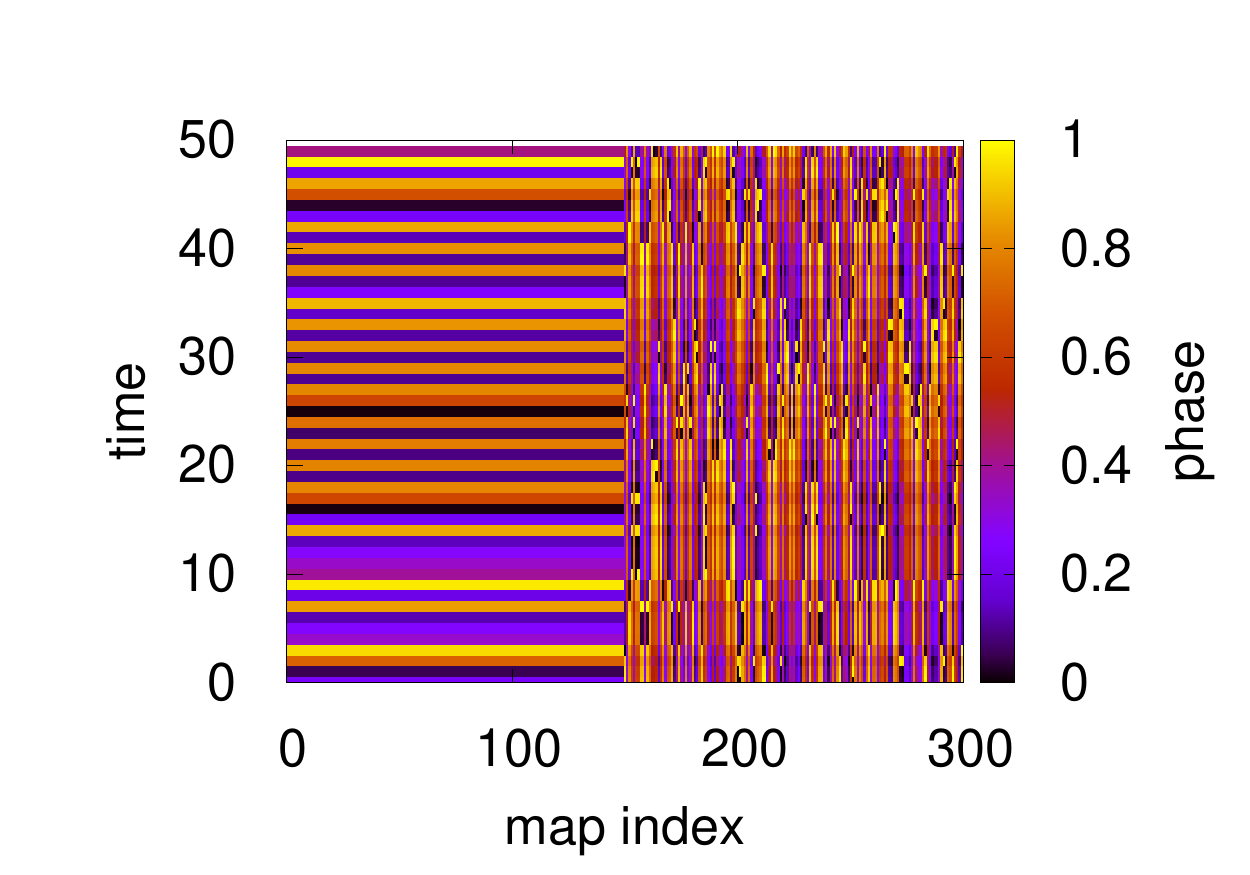}&
\hspace{-0.5cm}
\includegraphics[scale = 0.45]{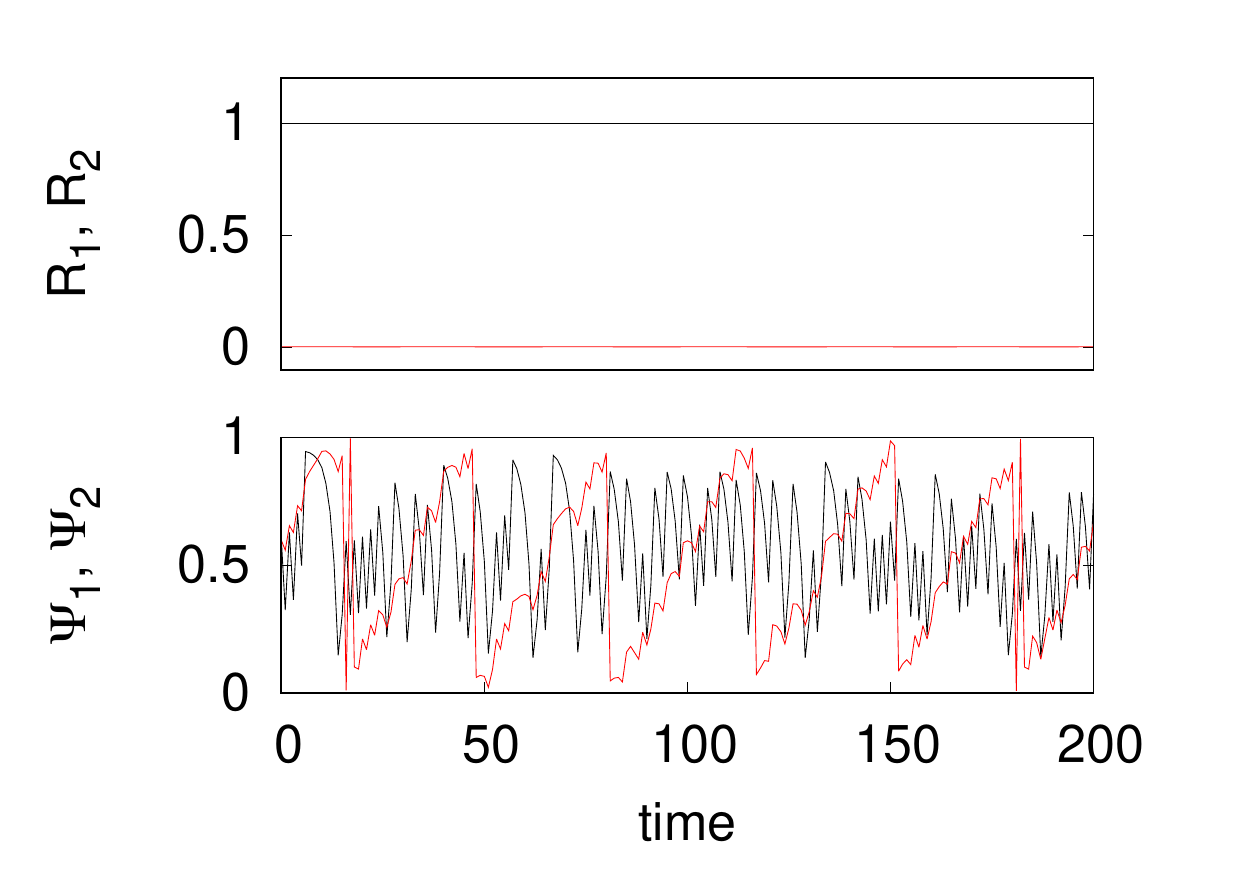}\\
(a) & (b) & (c)\\
\includegraphics[scale = 0.45]{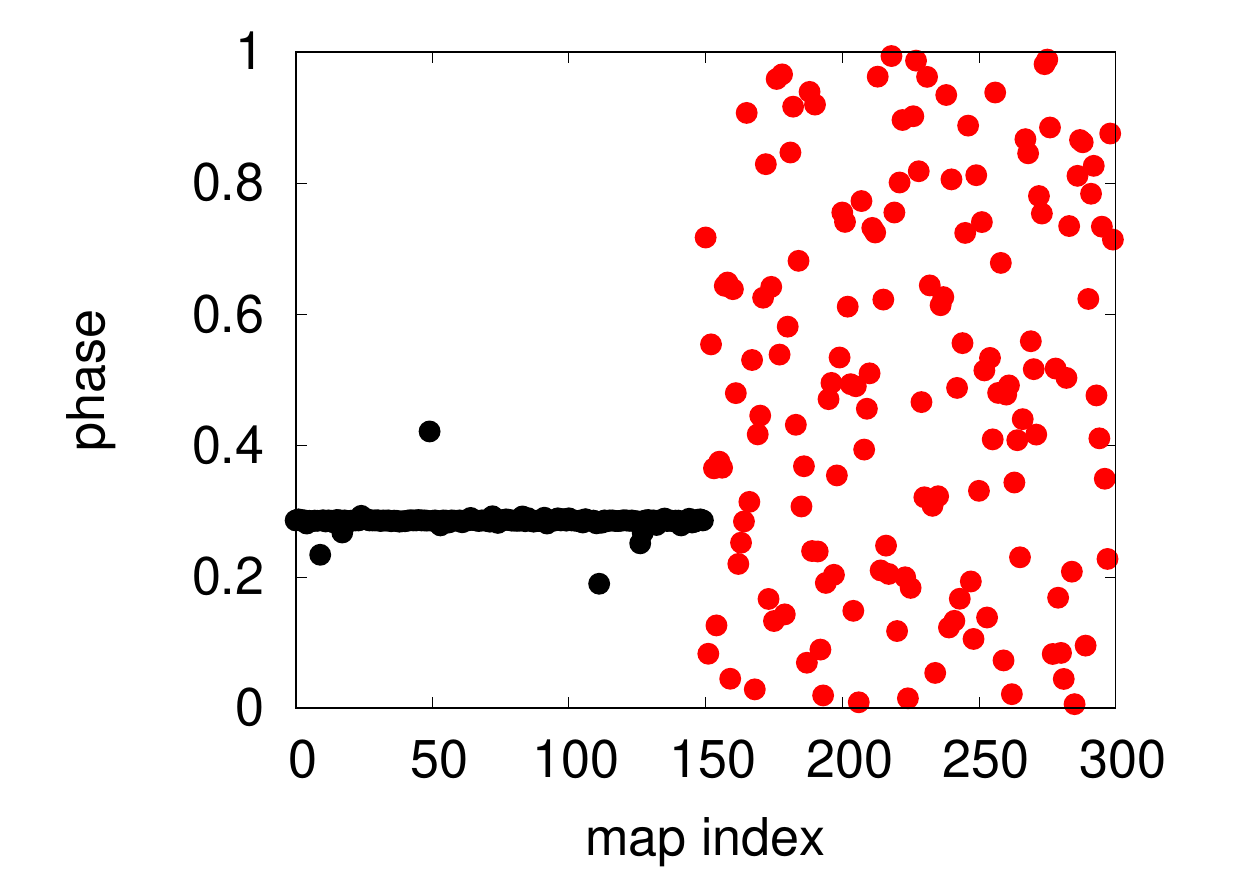}&
\hspace{-0.8cm}
\includegraphics[scale = 0.5]{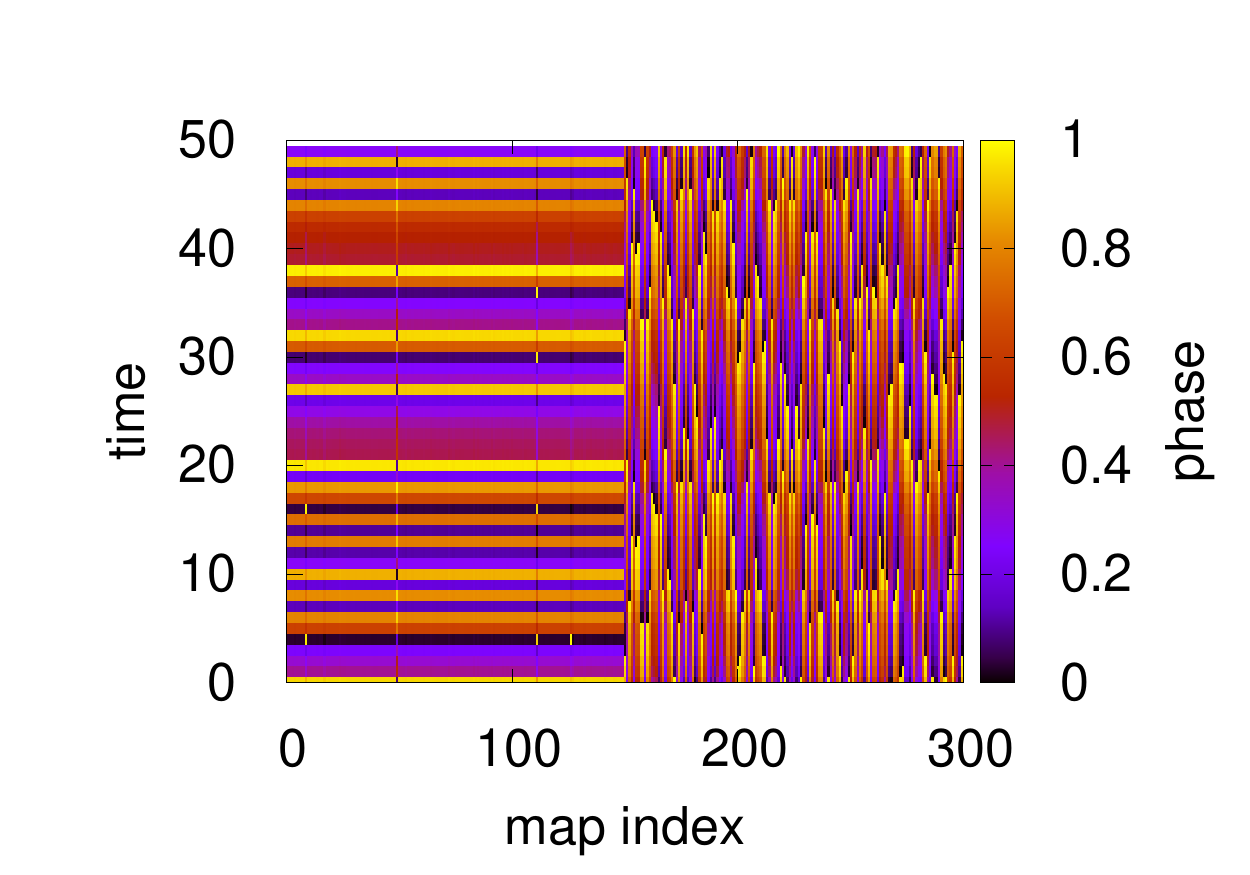}&
\hspace{-0.5cm}
\includegraphics[scale = 0.45]{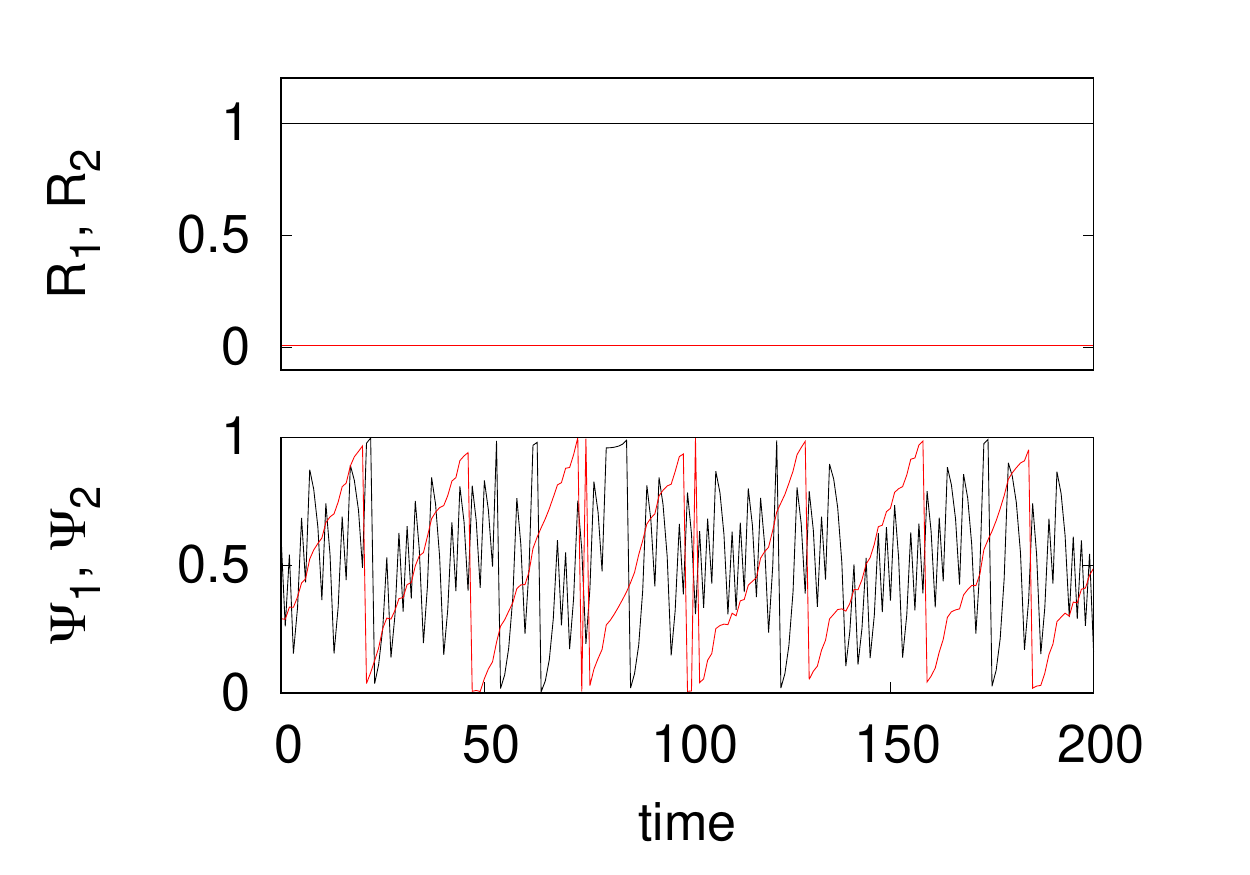}\\ 
(d) & (e) & (f)\\
\end{tabular}
\caption{\label{fig: initial}\footnotesize (color online) (a) The snapshot of the chimera state with a purely synchronised group is shown. The parameters are $K = 10^{-5}$, $\Omega = 0.27$, $\epsilon_1 = 0.82$, $N = 150$ (b) The space time plot of the chimera state without the defects in the synchronised group. (c) The temporal variation of $R^{1}$, $R^{2}$, $\Psi^{1}$, $\Psi^{2}$ for the chimera states with complete phase synchronisation in group one. (d) A snapshot of the chimera state with defects in the synchronised group. The parameters are $K = 10^{-5}, \Omega = 0.27, \epsilon_1 = 0.93, N = 150$. The sites between 1 to 150 belong to group one and the sites between 151 to 300 belong to group two. (e) The space time plot of the chimera state shown in (d). (f) The temporal variation of order parameter $R^{1}$, $R^{2}$ and average phases, $\Psi^{1}$, $\Psi^{2}$ (defined in Eq. \ref{ord_ang_1} and \ref{ord_ang_2} for group one and two respectively) for the chimera states with defects in the synchronised group shown in (d). In both (c) and (f), $R^1$ and $\Psi^1$ are shown in black whereas $R^{2}$ and $\Psi^{2}$ are denoted in red.}
\end{figure*}

 We report that at certain parameter values, the fully random initial condition evolves to a chimera state which consists of a spatially phase synchronised group and a spatially and temporally phase desynchronised group (figure \ref{fig: initial}). At particular values of $K, \Omega, \epsilon_1$ and $N$ we find a chimera phase state with a purely synchronised subgroup where all maps in group one belong to a phase synchronised cluster (see figure \ref{fig: initial}(a)) whereas at other parameters we observe chimera states, where the spatially phase synchronised subgroup has defects, as the phases of a small fraction of circle maps do not belong to the synchronised cluster (figure \ref{fig: initial}.(d)).  We also see in figure \ref{fig: initial}.(b) and (e) that the space time variation of the desynchronised group in both type of chimera states shows spatiotemporally intermittent structures, as synchronised islands in the shape of cones can be observed within the desynchronised phases. Other states can be seen at other parameter values which are discussed in the next section.  

\section{Phase diagram}\label{sec: PD}
We note that the system is controlled by the parameters $K, \Omega, \epsilon_1$. Apart from this set of parameters, the system dynamics also depends on the size, $2N$ of the system and the initial condition. We fix the size of the system at $N = 150$ and vary the parameters to look for the chimera phase configuration. To identify the chimera states as seen in figure \ref{fig: initial} we use the order parameters, $R_{n}^{1}, R_{n}^{2}, R_{n}$ and the average phase, $\Psi^{1}_{n}, \Psi^{2}_{n}, \Psi_{n}$ defined respectively for each of the groups at time step $n$ as, 
\begin{equation}
 R_{n}^{1}\exp{\left( i2\pi\Psi^{1}_{n} \right)} = \frac{1}{N}\sum\limits_{j = 1}^{N} \exp{\left( i2\pi\theta^{1}_{n}(j) \right)}
 \label{ord_ang_1}
\end{equation}
\begin{equation}
R_{n}^{2}\exp{\left( i2\pi\Psi^{2}_{n} \right)} = \frac{1}{N}\sum\limits_{j = 1}^{N} \exp{\left( i2\pi\theta^{2}_{n}(j) \right)}
\label{ord_ang_2}
\end{equation}
\begin{equation}
R_n \exp{\left( i2\pi\Psi_{n} \right)} = \frac{1}{2N}\sum\limits_{\sigma = 1}^{2}\sum\limits_{j = 1}^{N}\exp{\left( i2\pi\theta_{n}^{\sigma}(j) \right)}
\label{global}
\end{equation}

It is clear that $R_{n}^{1}$, $R_{n}^{2}$ becomes one when the phases of the maps in the corresponding group are synchronised at time step $n$. In that case, the phases at which the groups synchronise are given by $\Psi^{1}_{n}, \Psi^{2}_{n}$ respectively. Similarly their values become approximately zero when the phases are uniformly distributed between zero and one. Similar conclusions can be drawn for $R_n, \Psi_n$ if the whole system is phase synchronised or desynchronised. If all the maps are fully phase synchronised at a time step, then $\Psi^{1}_{n}$ and $\Psi^{2}_{n}$ become equal at that time step, while $R_{n}^{1}, R_{n}^{2}, R_{n}$ become one. These properties of these quantities enable us to look for the chimera states of the types shown in figure \ref{fig: initial}.(c) and (d), as we vary the parameters $K, \Omega, \epsilon_1$.

\par
\begin{figure*}
\centering
\begin{subfigure}{0.30\textwidth}
\centering\includegraphics[scale = 0.4]{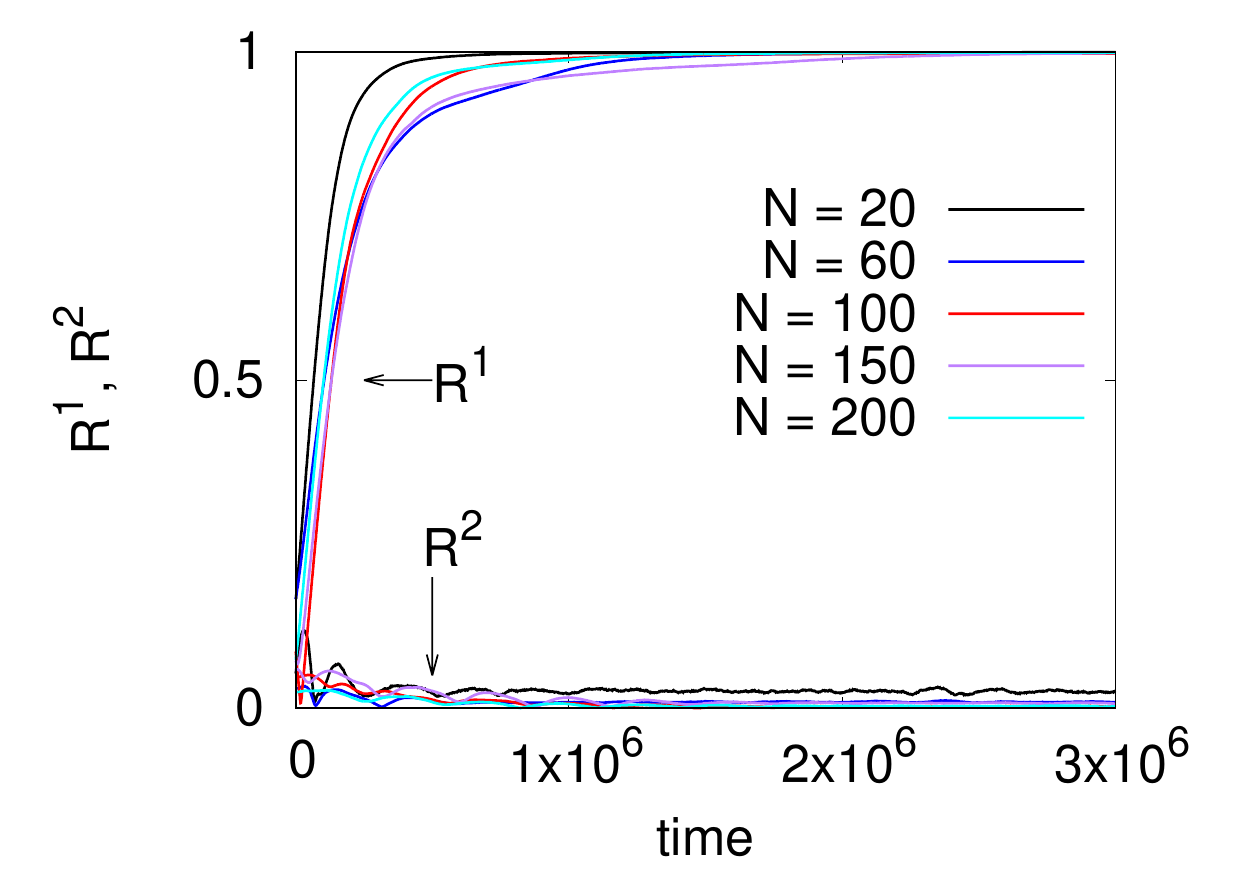}
\caption{}
\end{subfigure}%
\begin{subfigure}{0.30\textwidth}
\centering\includegraphics[scale = 0.4]{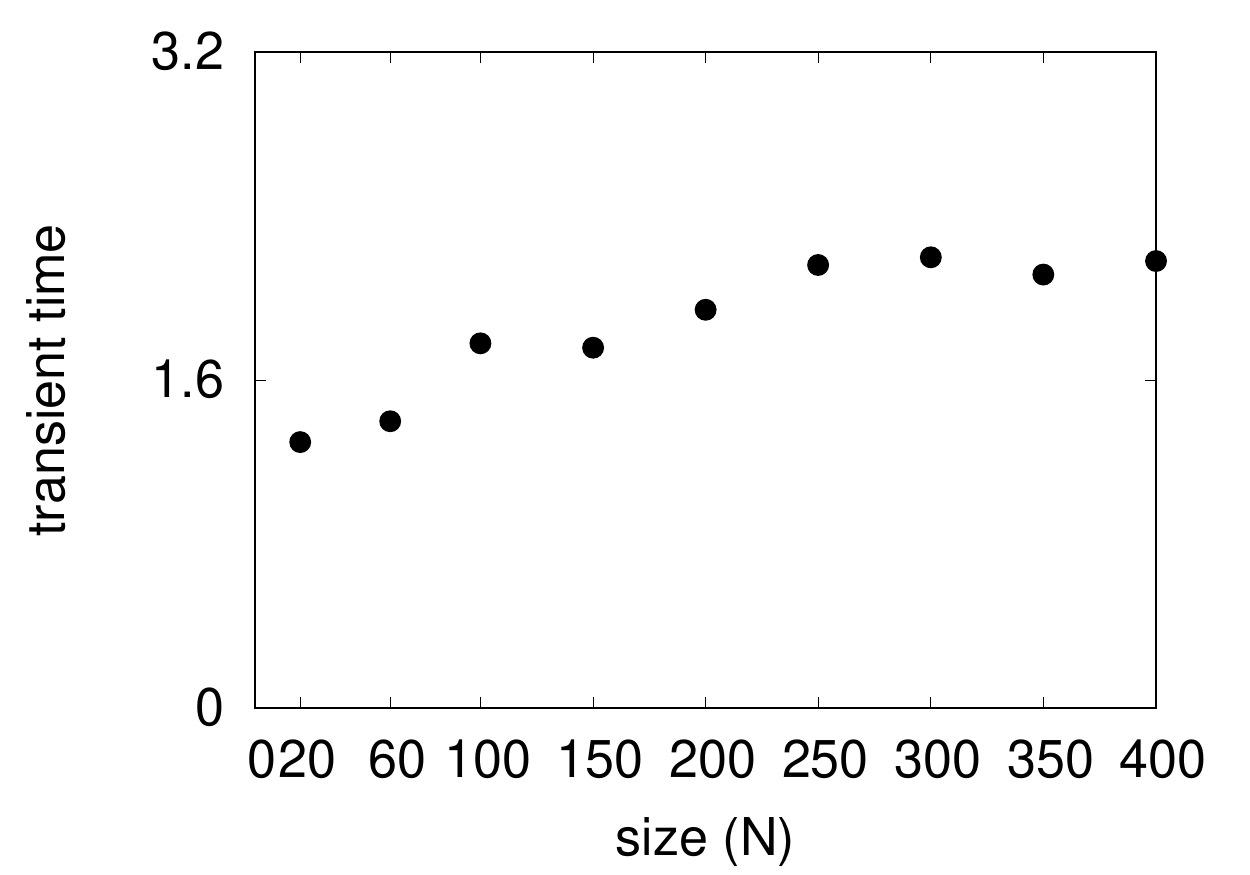}
\caption{}
\end{subfigure}
\caption{\label{fig: trans} \footnotesize (color online) (a) The variation of the order parameters, $R, R_{n}^{1}, R_{n}^{2}$ with time for $N = 20, 60, 100, 150$ and $200$. (b) Average transient time is plotted for different sizes. An overall trend of increasing transient time can be observed.}
\end{figure*}

\begin{figure}
\centering
\begin{subfigure}{0.35\textwidth}
\centering\includegraphics[scale = 0.5]{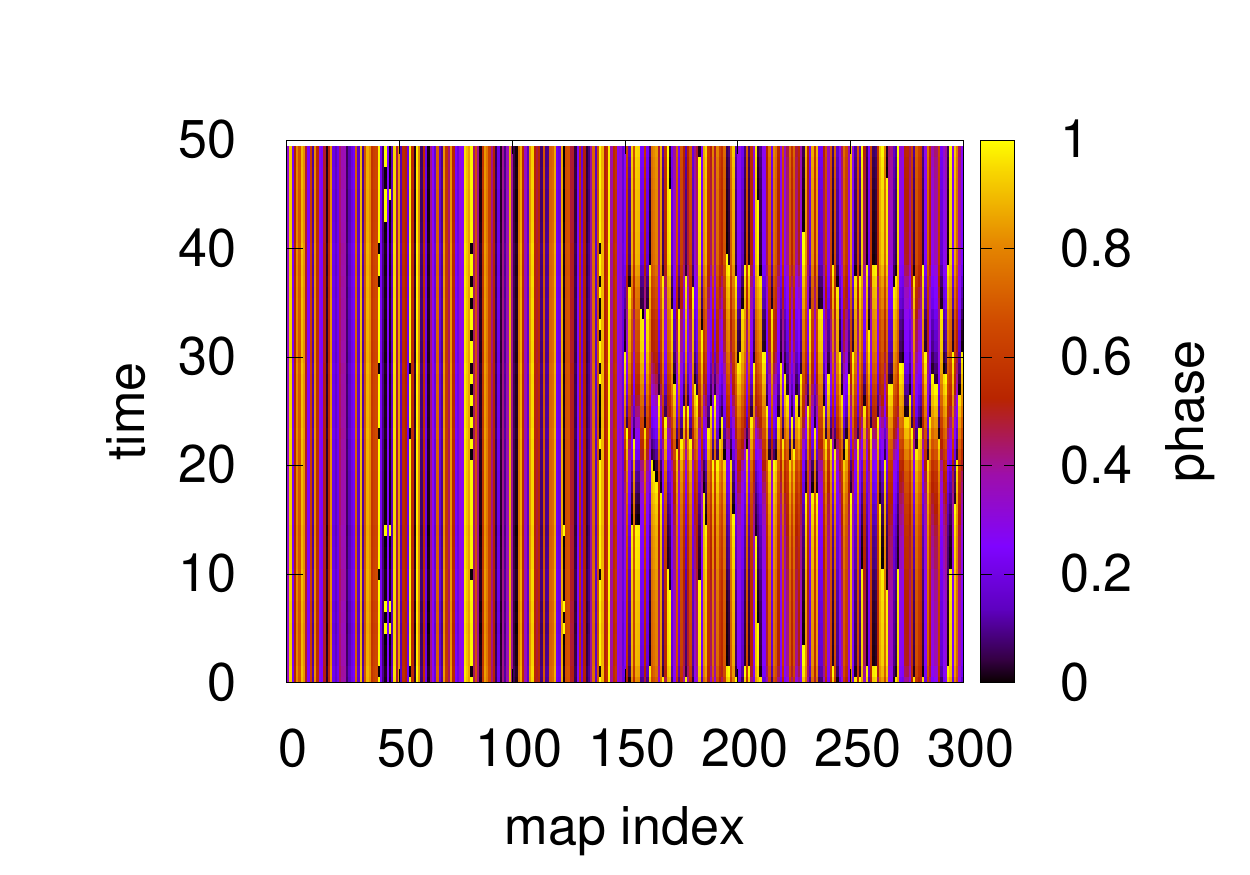}
\caption{}
\end{subfigure}\\%
\begin{subfigure}{0.35\textwidth}
\centering\includegraphics[scale = 0.5]{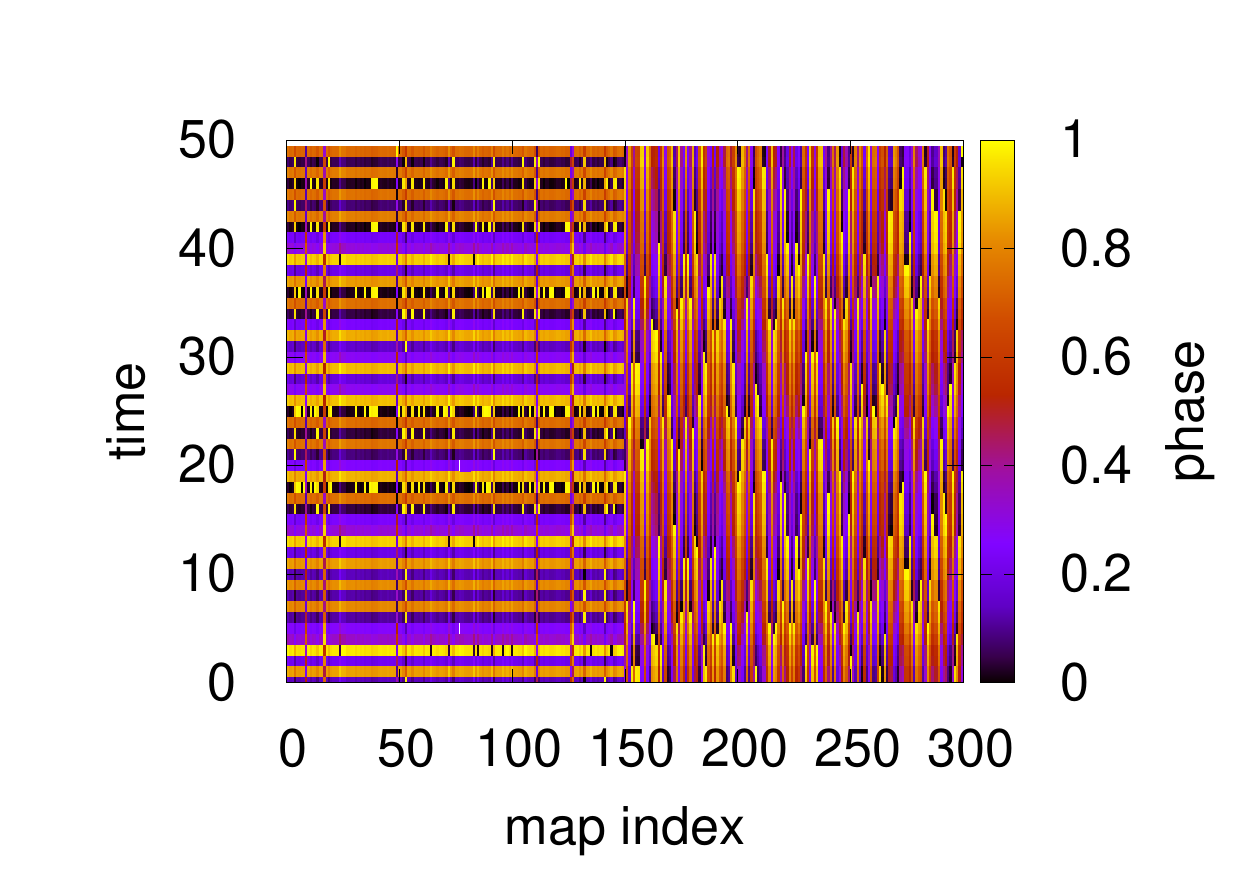}
\caption{}
\end{subfigure}
\caption{\label{fig: trans_state} \footnotesize (color online) The space time plot of the system after (a) 20000 time steps and (b) 500000 time steps for $N = 150$. For the above plots we use the parameters $K = 10^{-5}, \Omega = 0.27, \epsilon_1 = 0.93$.}
\end{figure}

 It is clear that the minimum number of time steps required for the system to settle into chimera states of interest here is a function of the system size. Figure \ref{fig: trans}.(a) shows the variation of the order parameters $R_{n}^{1}, R_{n}^{2}, R_{n}$ with time for different system sizes, $N = 20, 60, 100, 150, 200$. It is clear that the Eq. \ref{sinecml} settles from a completely random initial condition to the chimera state shown in Fig. \ref{fig: initial}. Fig. \ref{fig: initial}.(b) show that the average transient time for systems of smaller sizes is shorter than that required by larger system. Overall we see that the subgroup order parameter $R^{1}_{n}$ rises to values above 0.8 after three hundred thousand time steps, and slowly tends to one approximately after three million time steps, while the subgroup order parameter $R_{n}^{2}$ becomes zero. Such values of the group wise order parameters imply the existence of chimera phase configurations. 
The space time variation of the phases of the maps at intermediate time steps show that the CML is in mixed configurations which are different (see Fig. \ref{fig: trans_state}.(a), (b)) from the chimera states under consideration. Here we always evolve the system for  $3\times 10^6$ iterations or more, in all our subsequent numerical calculations. 

\begin{figure}
\centering
\begin{subfigure}{0.5\textwidth}
\centering\includegraphics[scale = 0.65]{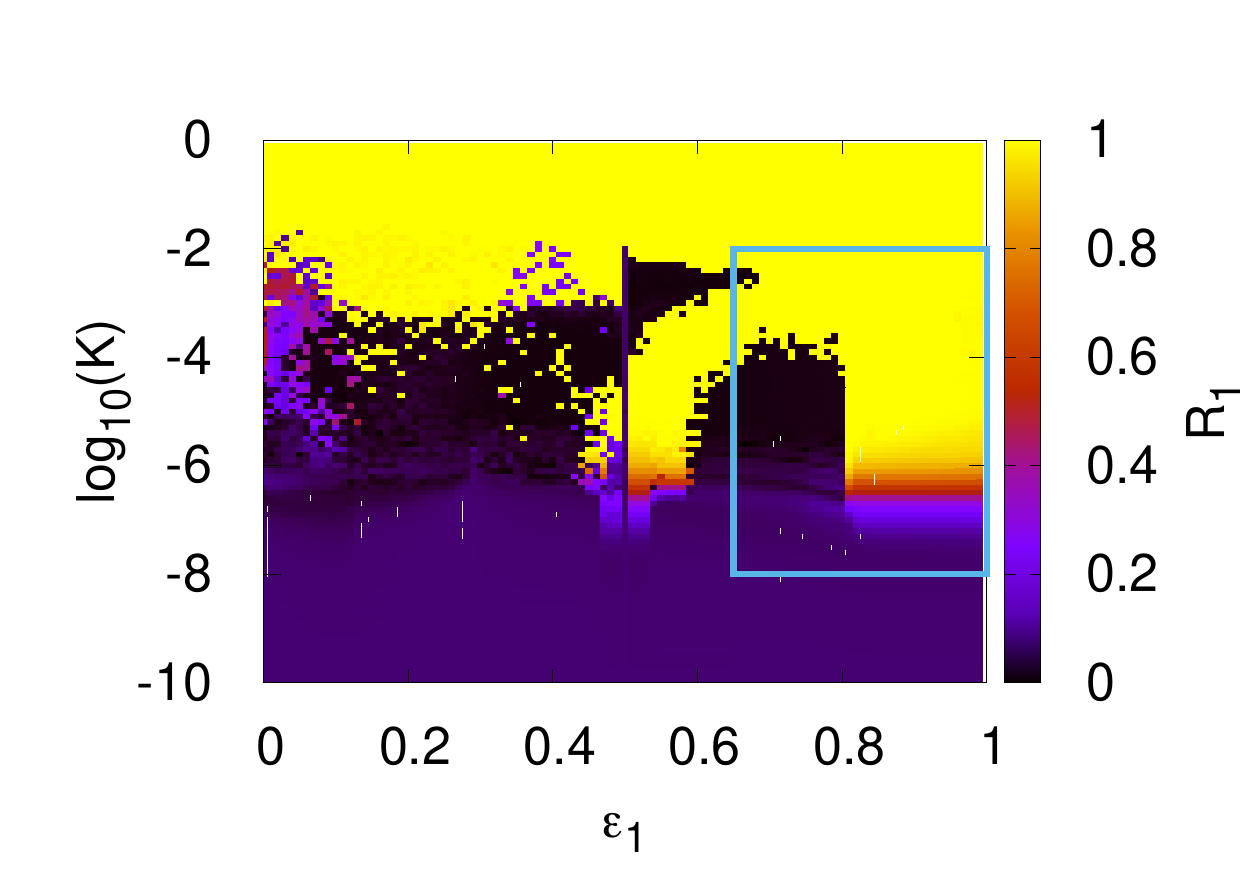}
\caption{}
\end{subfigure}\\%
\begin{subfigure}{0.5\textwidth}
\centering\includegraphics[scale = 0.65]{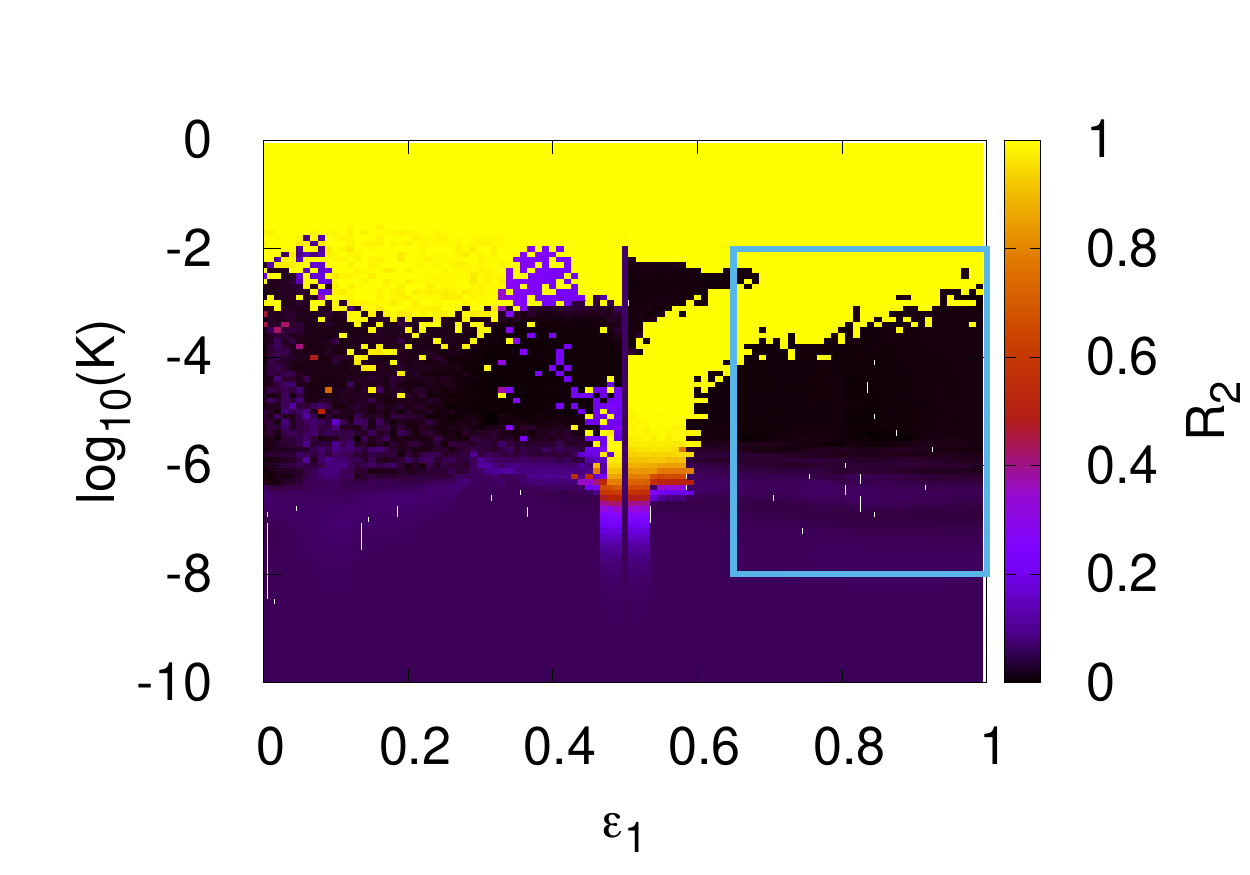}
\caption{}
\end{subfigure}\\%
\begin{subfigure}{0.5\textwidth}
\centering\includegraphics[scale = 0.65]{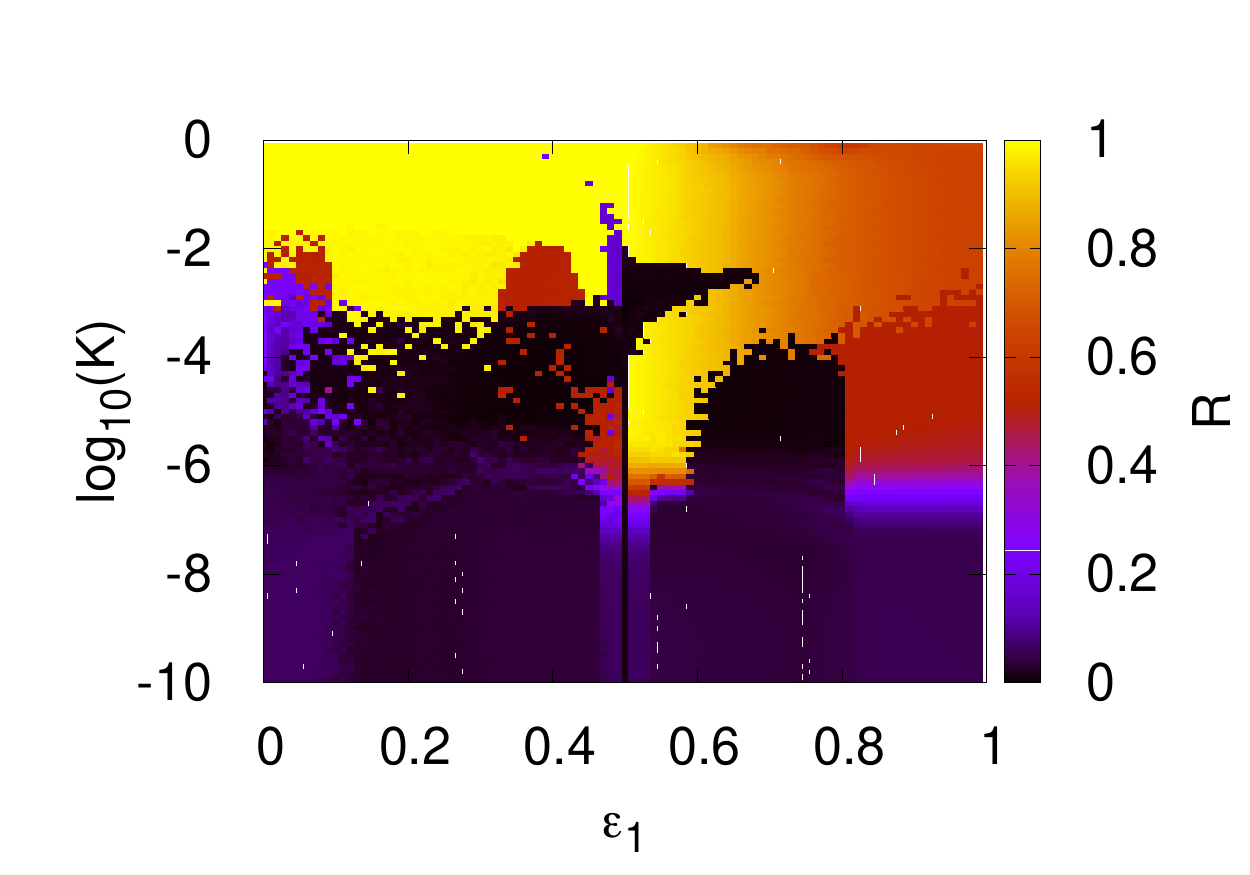}
\caption{}
\end{subfigure}\\
\caption{\label{fig: order_1} \footnotesize (color online) The order parameters (a)$R^{1}$, (b)$R^{2}$ and (c)$R$ are plotted for the values  $N = 150, \Omega = 0.27$. The color code for the values of the order parameter is indicated in the vertical bar in each plot. At each parameter value we use a random initial condition and iterate the system initially for $4 \times 10^6$ time steps, after which the order parameters $(R^1, R^2, R)$ are calculated and averaged over $10^5$ time steps. The region where chimera states are seen is identified  by the order parameter values $R^1 \approx 1$ and $R^2 \approx 0$. A magnified version of this region (enclosed by the rectangle) along with the pure states in the boundary is given in Fig. \ref{fig: order_zoom}.}
\end{figure}

We obtain a phase diagram for the parameter value $\Omega = 0.27$ and vary the parameters $K, \epsilon_1$ in the range $10^{-8} < K < 1$ and $0 < \epsilon_1 < 1$. At each values of these parameters we use a fixed set of initial phase values which are randomly distributed between zero and one. We calculate $R_{n}^{1}$, $R_{n}^{2}$, $R_n$ for $10^5$ time steps and calculate the average after the system of Eq.\ref{sinecml} is iterated for three million time steps. Figure \ref{fig: order_1} show the values of $R^{1}$, $R^{2}$ and $R$ respectively with the variation of $K, \epsilon_1$ at $\Omega = 0.27$. 
  
\begin{figure*}
  \centering \begin{tabular}{ccc}
 \hspace{-.5cm}
\includegraphics[scale = 0.43]{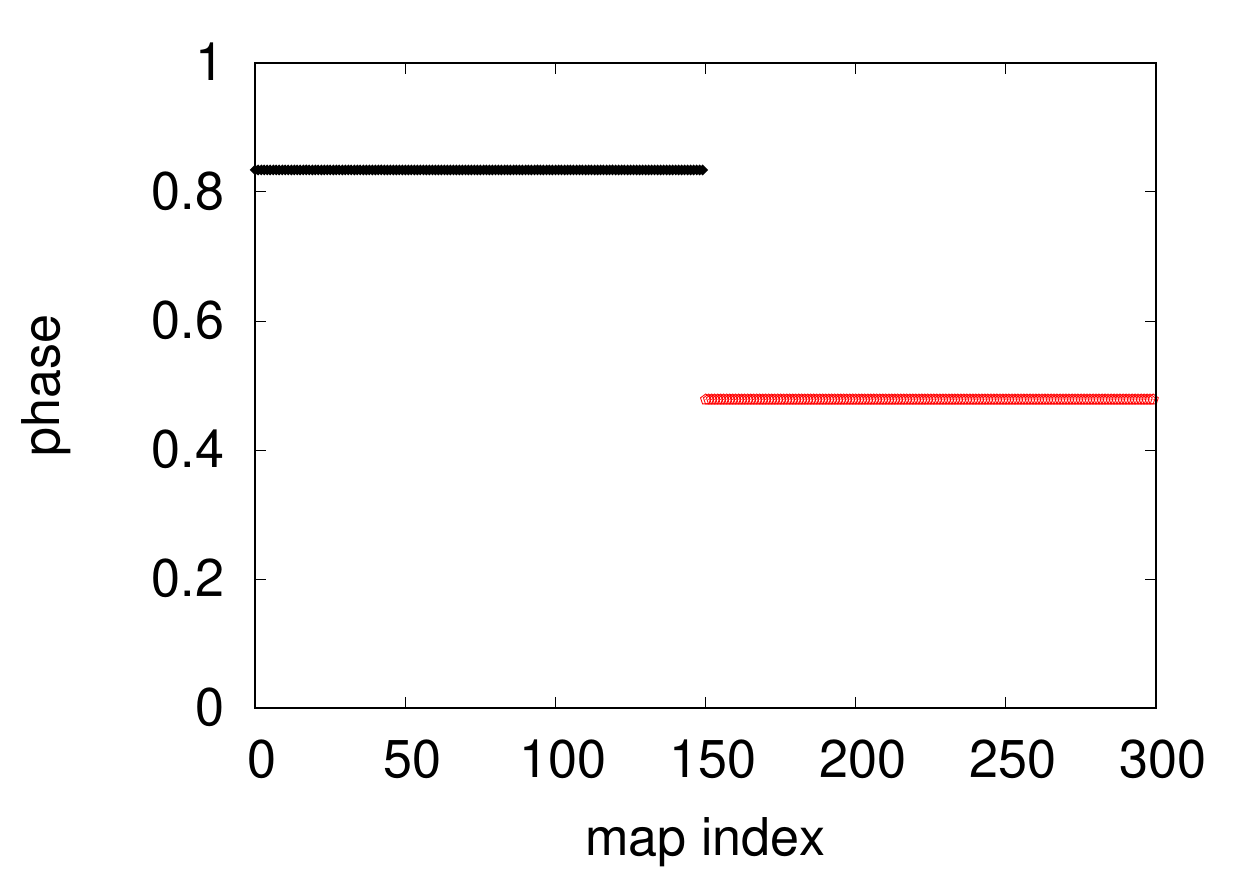}&
\hspace{-0.5cm}\includegraphics[scale = 0.43]{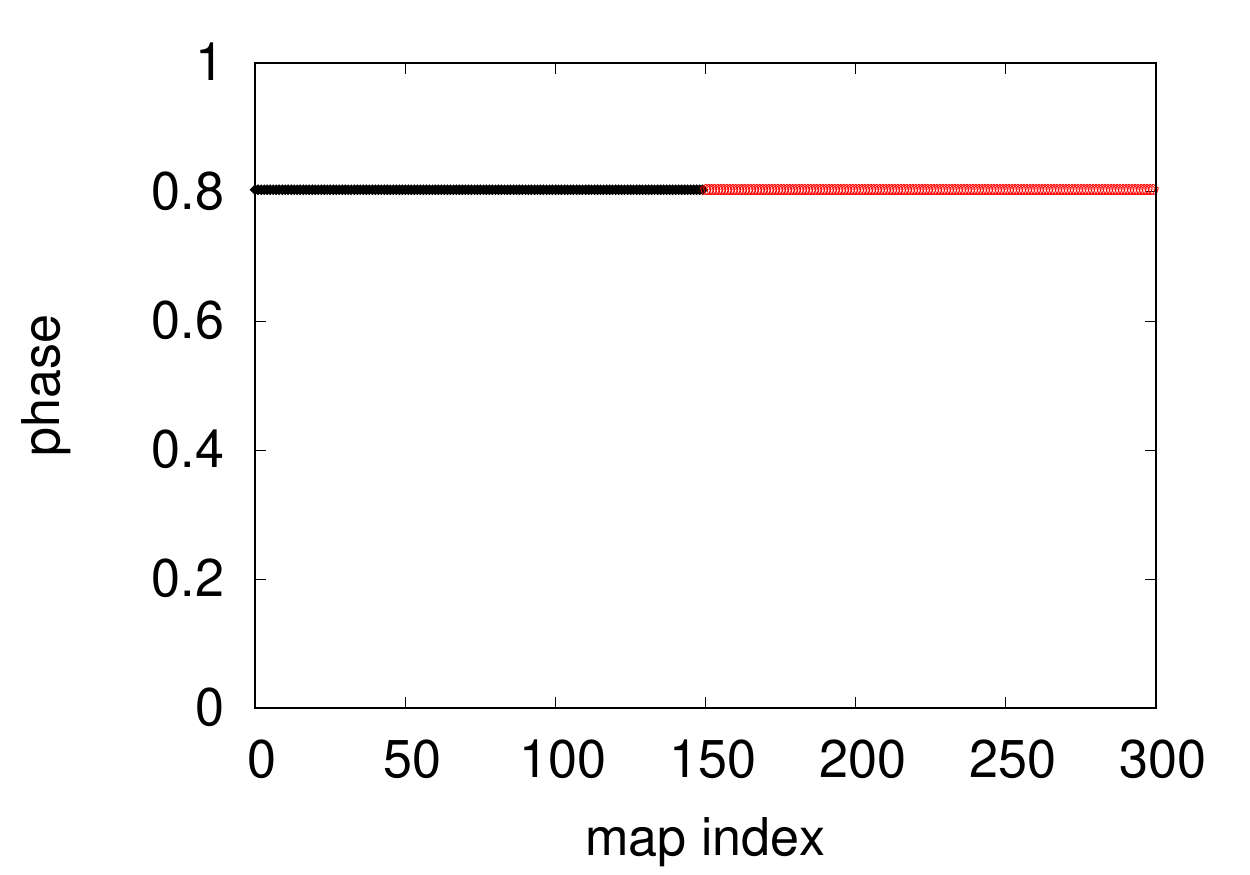}&
\hspace{-0.5cm}\includegraphics[scale = 0.43]{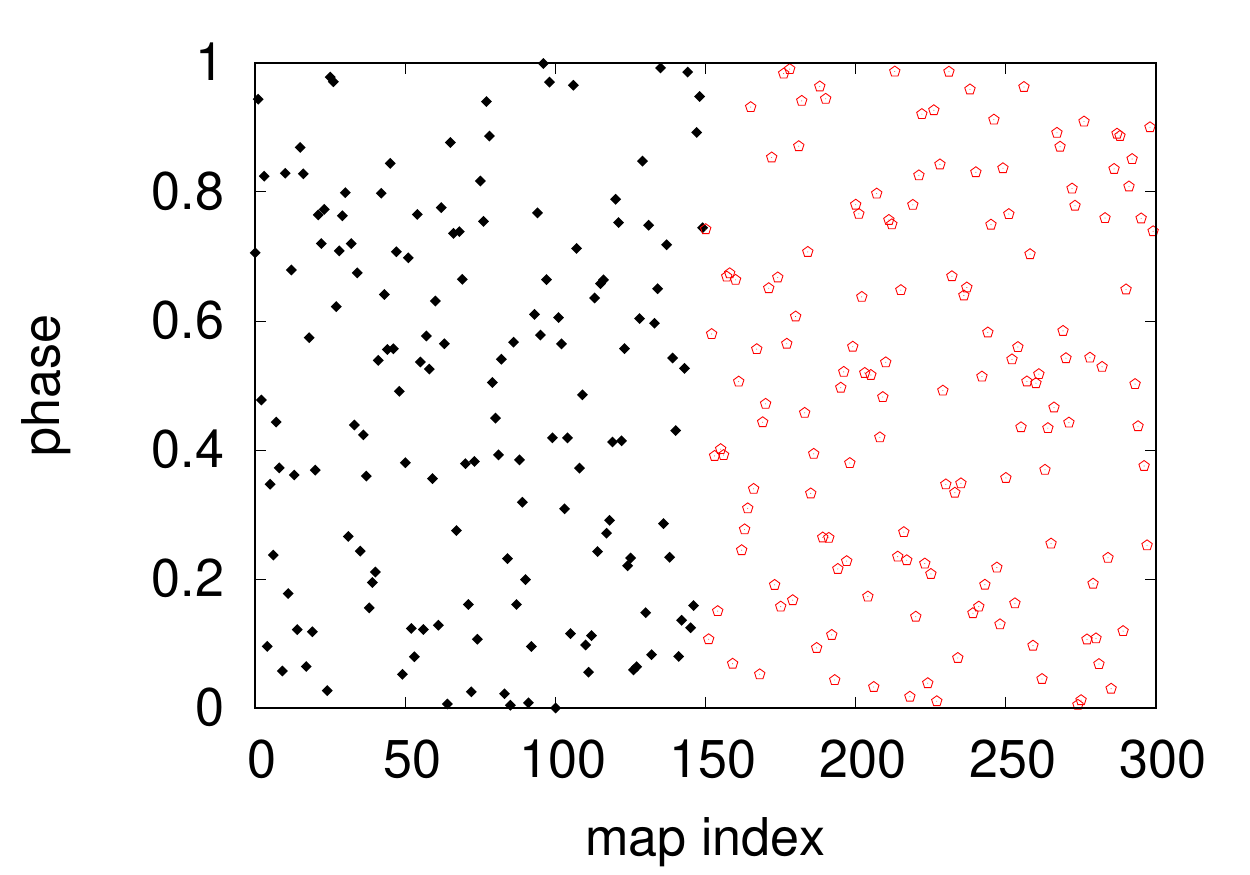}\\
(a) & (b) & (c)\\
\hspace{-0.5cm}\includegraphics[scale = 0.44]{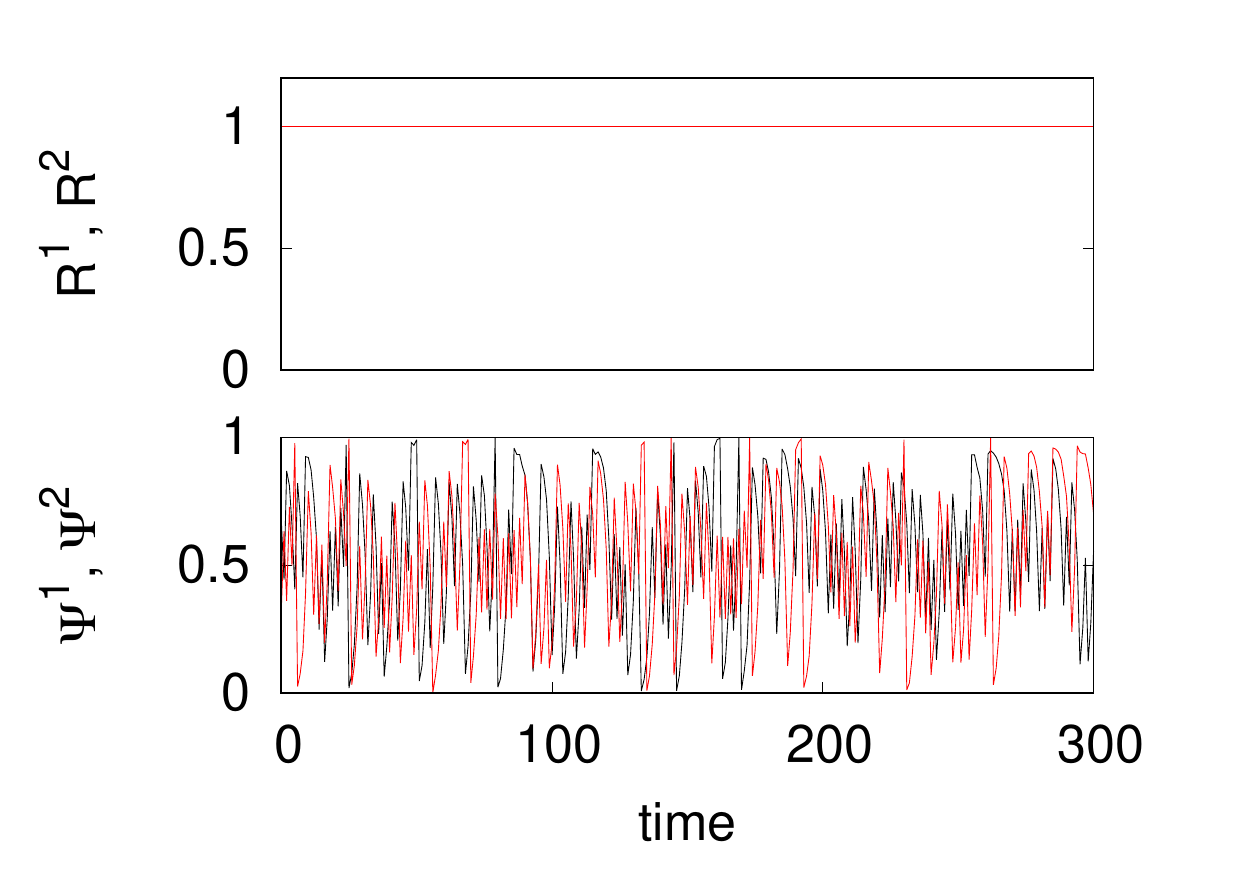}&
\hspace{-.5cm}\includegraphics[scale = 0.44]{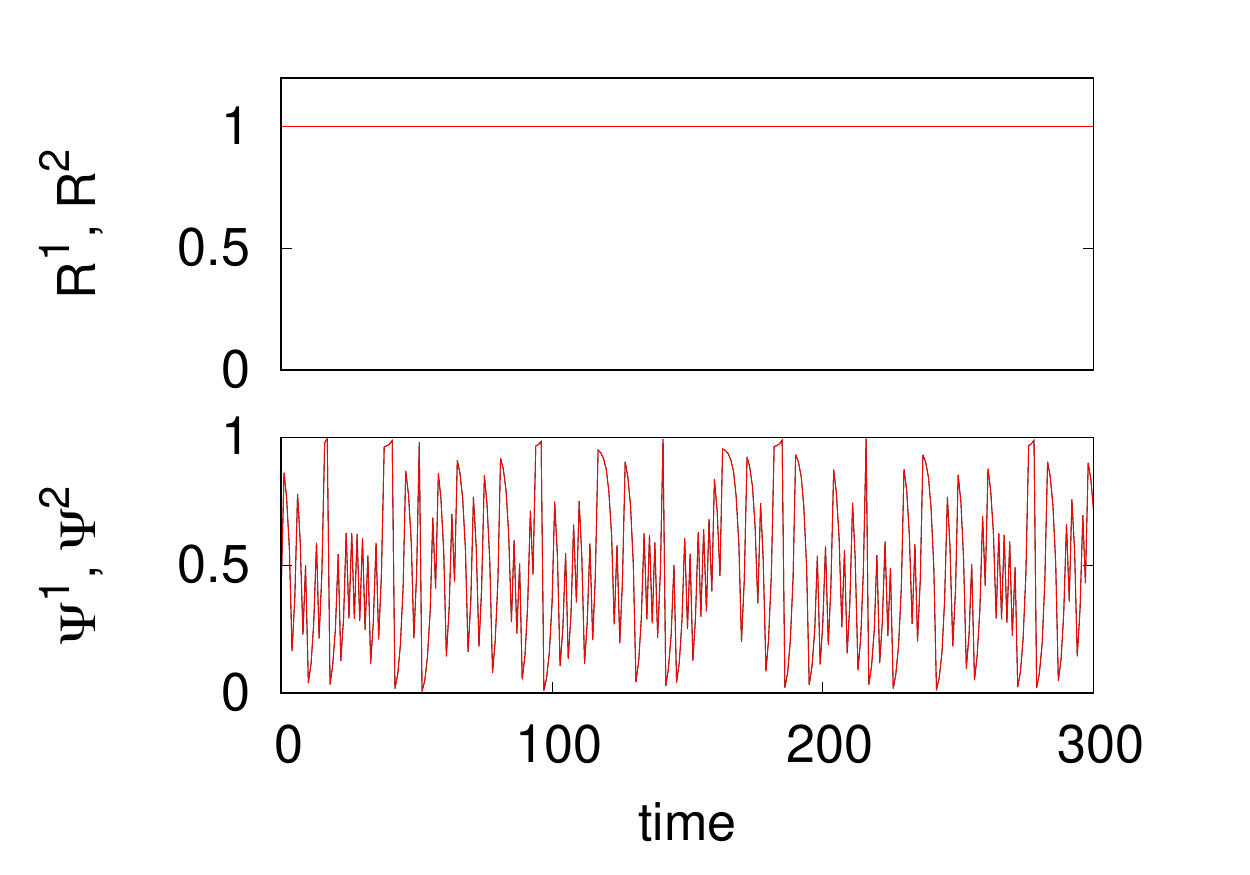}&
\hspace{-0.5cm}\includegraphics[scale = 0.44]{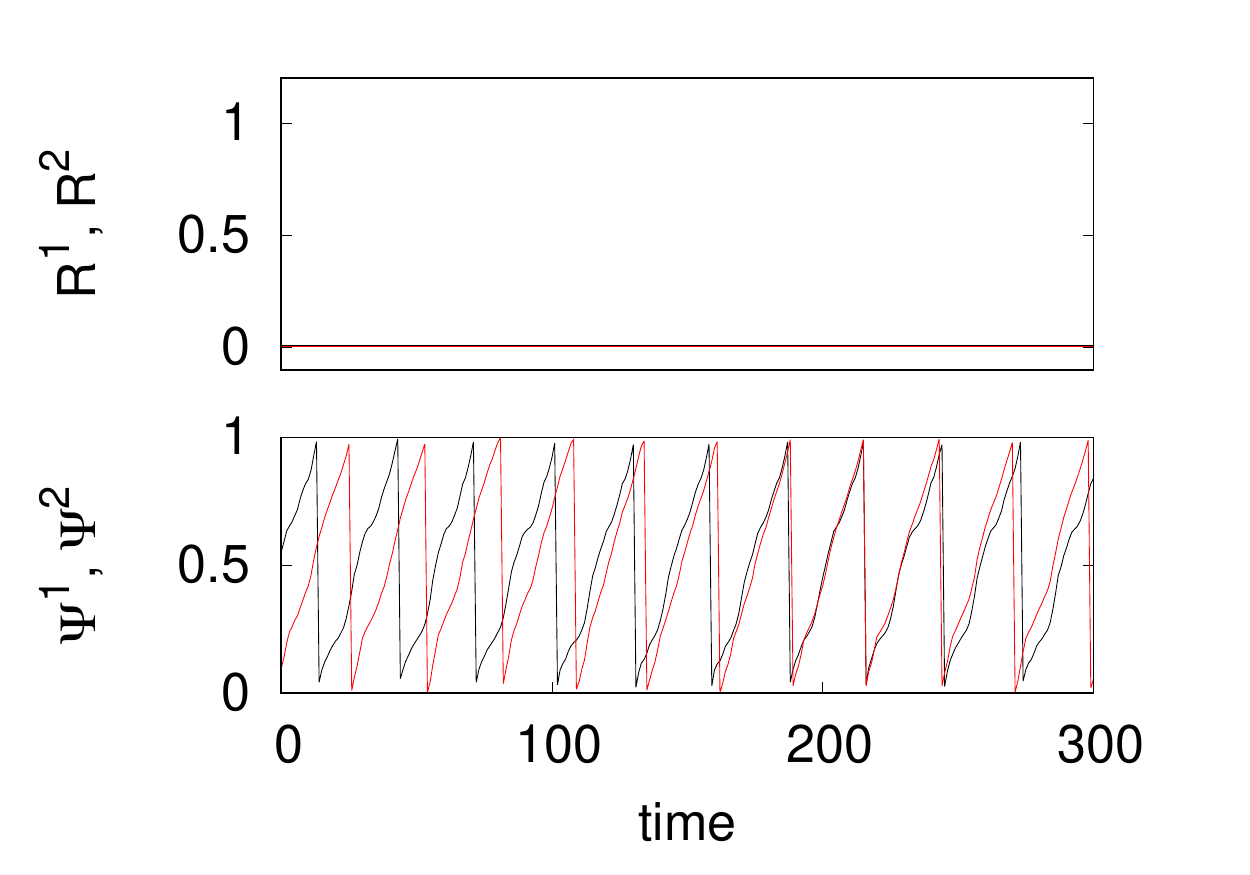}\\
 (d) & (e) & (f)\\
 \end{tabular}
\caption{\label{fig: states}\footnotesize (color online) Snapshots of the (a) two cluster phase configuration at $K = 10^{-2}, \epsilon_1 = 0.93$ (b) fully synchronised state at $K = 10^{-2}, \epsilon_1 = 0.45$ and (c) fully de-synchronised state at $K = 10^{-5}, \epsilon_1 = 0.75$ are shown. The variations of $R^{1}, R^{2}, \Psi^{1}, \Psi^{2}$ are shown for (d) two phase clustered state, (e) globally synchronised state and (f) fully desynchronised state at same parameters. Other parameters viz. $\Omega, N$ were kept fixed at $0.27$ and $150$ respectively. All of the above phase configurations were obtained for the same set of initial conditions.}
\end{figure*}

\begin{figure*}
\centering 
\begin{subfigure}{0.5\textwidth}
\centering\includegraphics[scale = 0.19]{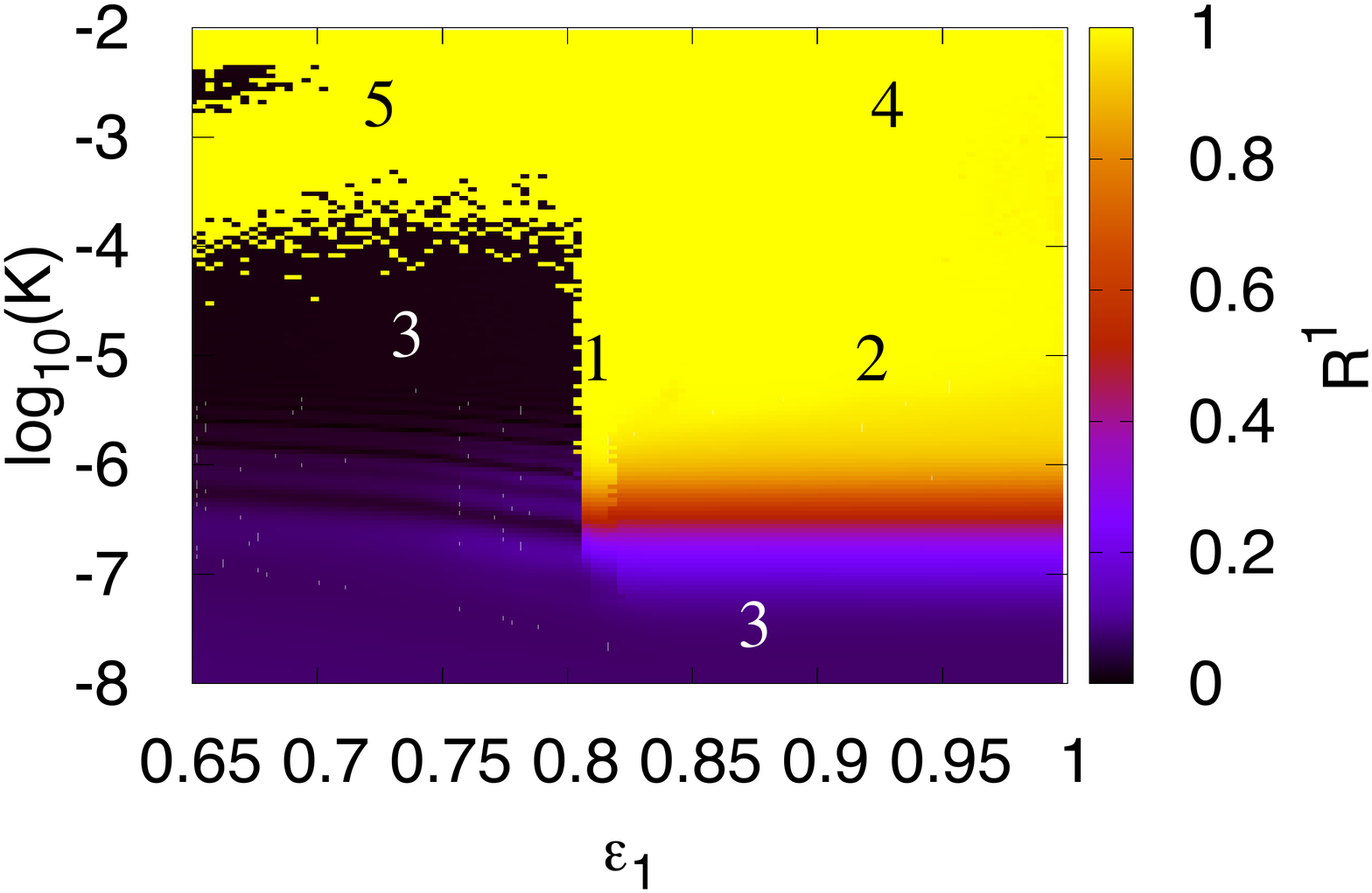}
\caption{}
\end{subfigure}%
\begin{subfigure}{0.5\textwidth}
\centering\includegraphics[scale = 0.19]{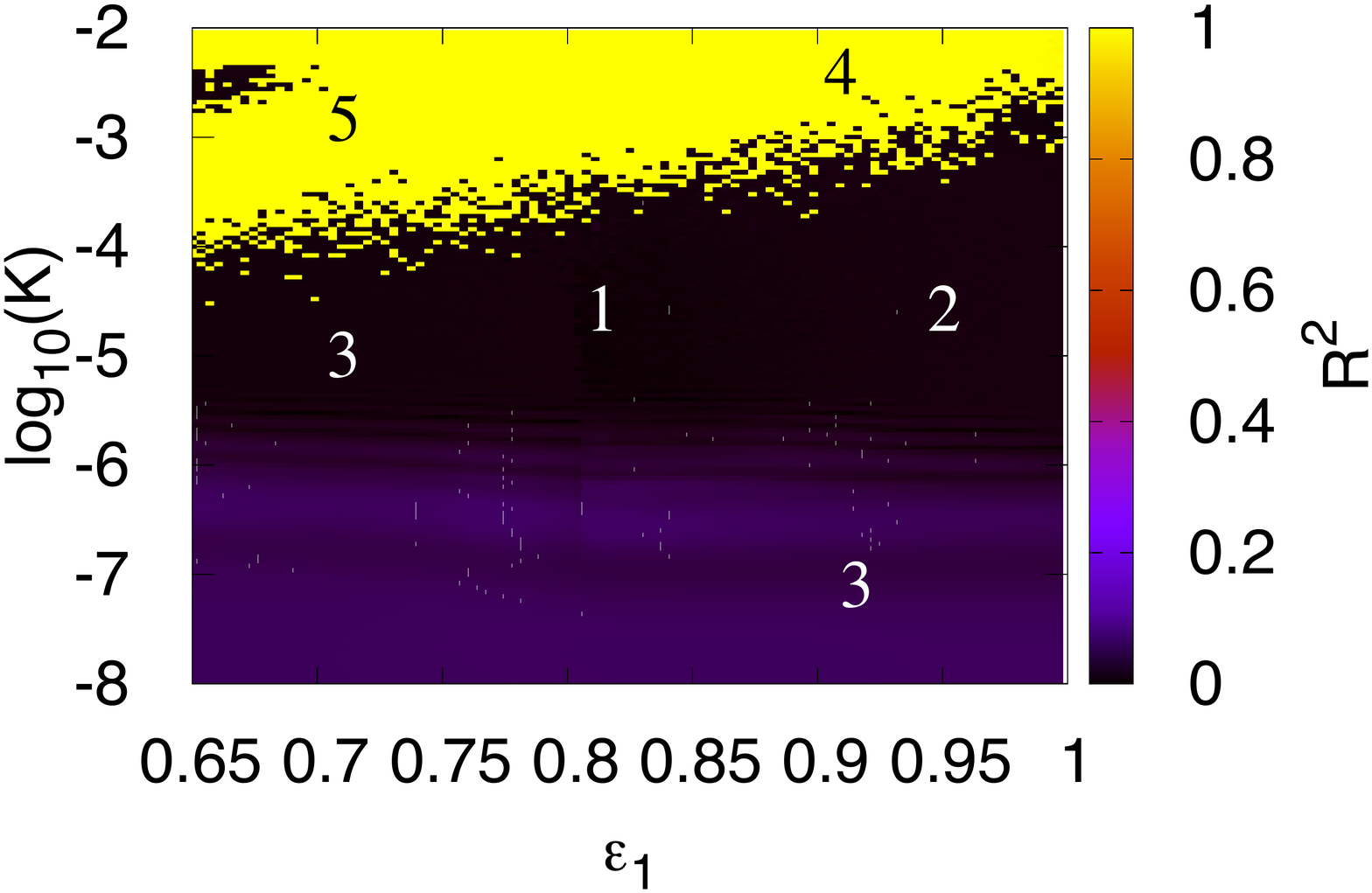}
\caption{}
\end{subfigure}
\caption{\label{fig: order_zoom}\footnotesize (color online) The order parameters (a) $R^1$ corresponding to group $1$, and (b) $R^2$ for group $2$ are calculated and plotted between the parameter region $0.65 < \epsilon_1 < 1.0$, $10^{-8} < K < 10^{-2}$ for $\Omega = 0.27$. At each of the parameter values we use a random initial condition and iterate the system for $4 \times 10^6$ time steps after which we calculate the order parameters $(R, R^1, R^2)$ and average them over $10^5$ time steps. The chimera states of case 1 and 2 are found in the region where the subgroup order parameters take values $R^1 \approx 1, R^2 \approx 0$. We have labelled the phase diagram by the final state seen in each region as defined in the text.} 
\end{figure*}

The chimera state is seen in the region where $R^1 \approx 1, R^2 \approx 0$. These show the existence of the chimera states (Fig. \ref{fig: initial}.(c)) in a region in $K, \epsilon_1$ space approximately given by $0.8 < \epsilon_1 < 1, 10^{-6} < K < 10^{-3}$ surrounded by other phase configurations around it whose snapshots are shown in Fig. \ref{fig: states}. A magnified version of this phase diagram around this region is shown in Fig. \ref{fig: order_zoom}. Five distinct types of phase configurations can be found in the phase diagram of Fig. \ref{fig: order_zoom}. These are chimera states, two clustered states, globally synchronised states and fully desynchronised states. The details of these dynamical states are as follows, 

\par
\begin{enumerate}
\item \textbf{Case 1 and case 2 : Chimera states (Fig. \ref{fig: initial})}: We obtain a chimera state when either $R^{1}$ or $R^{2}$ is one and the value of the other quantity is near zero. We get this condition at several of the parameter values for $\Omega = 0.27$. In particular when $-6 < \log_{10}K < -4$ and $0.8 < \epsilon_1 < 1.0$, at some parameters we find, case 1 : $R^{1} = 1$ and $R^{2} \approx 0$ (see Fig. \ref{fig: initial}.(c)) which indicates the chimera states with pure synchronisation in the synchronised group. Case 2 corresponds to chimera states with defects in the synchronised group for which we find $R^{1} \lesssim 1$ and $R^{2} \approx 0$ (Fig. \ref{fig: initial}.(f)). The temporal variation of $R^1, R^2$ also shows this behaviour. The variation of $\Psi_1$ and $\Psi_2$ with time shows that the variation of the average phases of the phase synchronised and desynchronised group are qualitatively different (see Figs. \ref{fig: initial}.c and f). The mirrored version of these chimera states where maps of group two phase synchronises while maps in group one become phase desynchronised are denoted as Case 1" and case 2". 

\item \textbf{Case 3 : Fully desynchronised states (Figs. \ref{fig: states}.(c), (f))}: These are found at those parameter values where $R^{1}$, $R^{2}$, $R$ are approximately zero. At these parameter values, all the maps in both the groups are temporally and spatially phase desynchronised. The temporal variation of $\Psi^{1}, \Psi^{2}$ suggest that the average phase of both the groups are approximately periodic. They are observed approximately for $\log_{10}K < -6$ and in the region $\log_{10}K < -4$ for $\epsilon_1 < 0.8$.

\item \textbf{Case 4 : Two clustered states (Fig. \ref{fig: states}.(a))}: We find that $R^{1} = 1$ and $R^{2} = 1$ in the parameter region approximately given by $-4 < \log_{10}K < 2$ and $0.65 < \epsilon_1 < 1$. The phases of the maps in each of the groups are such that they are spatially phase synchronised as suggested by the temporal variation of $R^1, R^2$ while the phases at which they synchronise are not equal as indicated by $\Psi^1, \Psi^2$ (see Fig. \ref{fig: states}.(d)). Figure \ref{fig: states}.(d) also suggests that each of these phase clusters do not synchronise to a temporally fixed phase value as can be seen from the variation of the average phases $\Psi^{1}, \Psi^{2}$ (see Fig. \ref{fig: states}.b). 

\item \textbf{Case 5 : Globally synchronised states (Fig. \ref{fig: states}.(b))}: These are characterised by the order parameter values when all three quantities, $R^{1}, R^{2}, R$ are approximately one. They can be seen mostly above $K \approx 10^{-3}$ for $\epsilon_1$ below $0.8$. The temporal variation of the average phases of each of the groups, $\Psi^{1}, \Psi^{2}$ in Fig. \ref{fig: states}.(e) suggests that all the maps are spatially phase synchronised at all time steps although the phase at which they synchronise is not a temporal fixed point similar to the temporal variation of the two clustered state. 

\end{enumerate}

\par
\begin{figure*}
\begin{subfigure}{0.5\textwidth}
\centering\includegraphics[scale = 0.5]{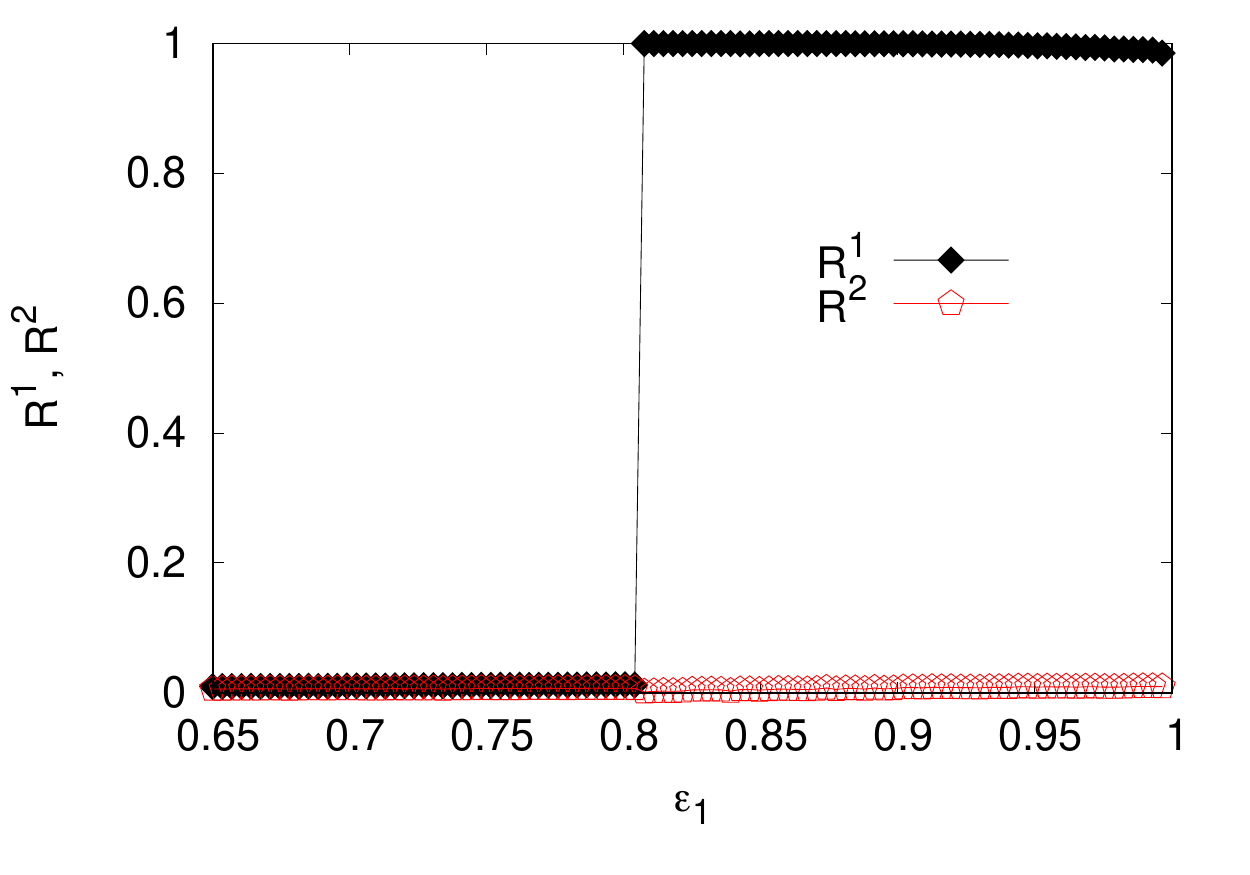}
\caption{}
\end{subfigure}%
\begin{subfigure}{0.5\textwidth}
\centering\includegraphics[scale = 0.5]{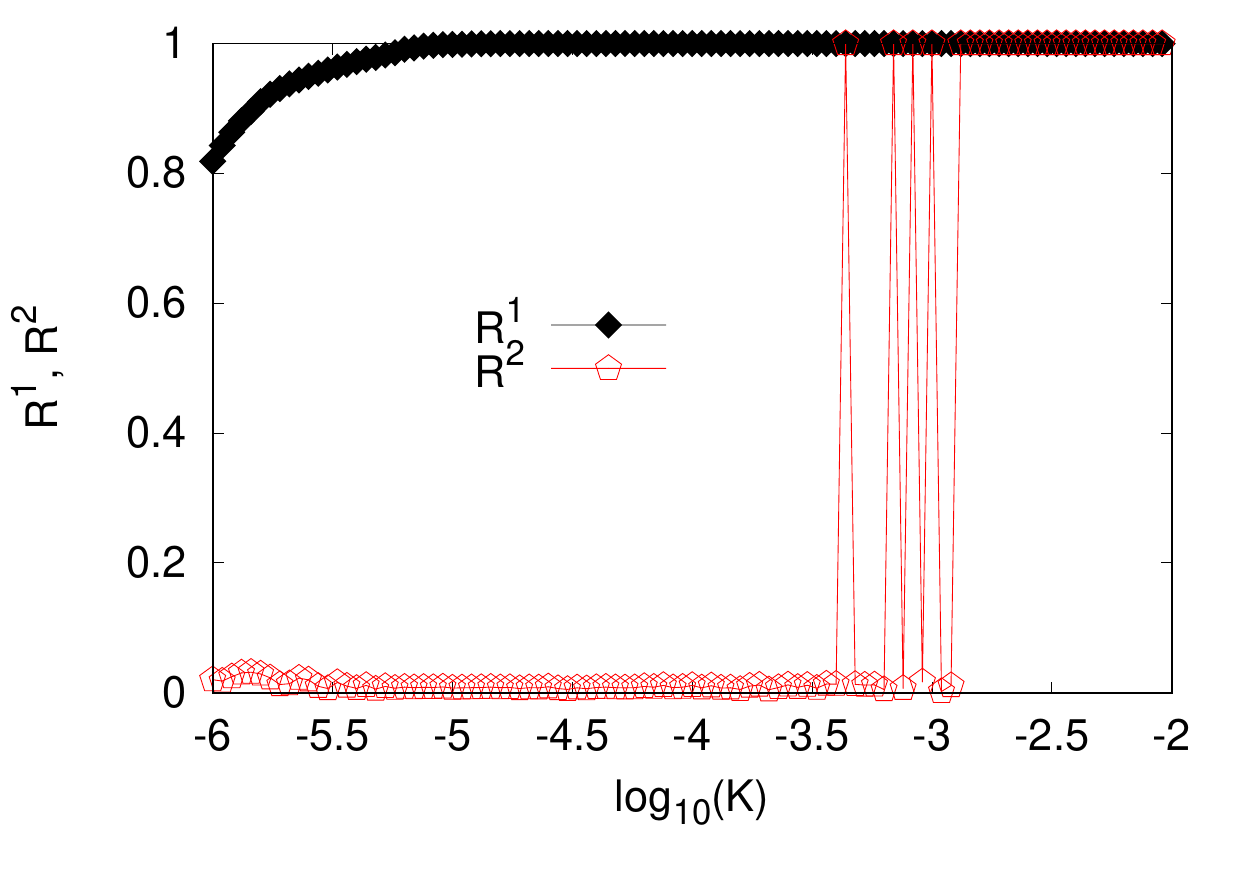}
\caption{}
\end{subfigure}
\caption{\label{fig: density_1}\footnotesize (color online) The variation of the group wise order parameters is plotted for (a) the parameter values  $K = 10^{-5}, \Omega = 0.27, N = 150$ as $\epsilon_1$ varies between $0.65$ and one. (b)The order parameters $R^1$ and $R^2$ are plotted as $K$ varies between $10^{-6}$ and $10^{-2}$ for the parameter $\epsilon_1 =0.93$ with $\Omega = 0.27$.}
\end{figure*}

 In this paper we are mainly interested in the region of the parameter space where the chimera states are seen and its transition to other phase configurations which are shown in Figs. \ref{fig: states}. Figure \ref{fig: order_zoom} shows that the fully desynchronised states seen in the region $-5.5 < \log_{10}K < -4$ and $0.65 < \epsilon_1 < 0.8$ transform to chimera states at $\epsilon_1 = 0.8$. The global phase desynchronised state seen between $-8 < \log_{10}K < -5.5$ and $0.8 < \epsilon_1 < 1$ transforms to chimera states as $\log_{10}K$ increases beyond $-5.5$. Between the parameter values $-4 < \log_{10}K < -3$ and $0.65 < \epsilon_1 < 1$ the chimera states transform to two clustered states. The transitions between these phase configurations due to the variation of the parameters $K, \epsilon_1$ is better understood from the variation of the order parameters $R^1, R^2$ at different cross sections of the phase diagram in the Fig. \ref{fig: order_zoom}. Figure \ref{fig: density_1}.(a) shows the variation of $R^{1}_{n}$ and $R^{2}_{n}$ with values of $\epsilon_1$ lying in the range between $0.65$ and one for $K = 10^{-5}$. It can be seen that both the subgroup order parameters  $R^1, R^2$ take values near zero when $\epsilon_1$ is less than 0.8. These values of the order parameters suggest that the system is in a fully phase desynchronised phase configuration for this range of $\epsilon_1$ and $K$ values. When the parameter $\epsilon_1> 0.8$ we see that $R^{1} = 1$ for group one and $R^{2} \approx 0$ zero for group two indicating a chimera phase configuration. When $\epsilon_1 \rightarrow 1$ we see from Fig. \ref{fig: density_1}.(a) that $R^{1} \lesssim 1$ while $R^2 \approx 0$. This indicates that some of the circle maps from the group one have phases that do not belong to the synchronised cluster at these values of $\epsilon_1$. As discussed earlier, this indicates the presence of a chimera phase state with defects in the synchronised group. We take another cross section of this phase diagram at the parameter $\epsilon_1 = 0.93$ in Fig. \ref{fig: density_1}.(b) that shows the variation $R^1$ and $R^2$ with $\log_{10}K$ as it increases from $-6$ to $-2$. We find that the subgroup order parameters take values $R^1 = 1$ and $R^2 \approx 0$ near $\log_{10}K \approx -5.7$ implying the existence of the chimera phase configuration till $\log_{10}K \approx -3.3$. When $\log_{10}K$ lies  between $-3.36$ and $-2.9$ we observe an interchange between these two states for a small variation of $K$. We find $R^1$ and $R^2$ both become one when $\log_{10}K > -2.7$. In the next section we discuss the properties of the chimera states shown in Fig. \ref{fig: initial}. 

\section{Basin stability}{\label{sec: BS}}
In the previous section, we have identified all the distinct spatiotemporal states found in different parameter regions of the phase diagram. We note that due to the large dimensional nature of the phase space, multiattractor solutions exist, and different initial conditions can go to different attractors at the same parameter values. The fraction of random initial conditions that go to a given attractor constitute a measure of the volume of its basin of attraction and also indicate the probability for a random initial condition to evolve to the attractor.  Recently, Menck et.al. \cite{menck2013} showed that the volume of the basin of attraction of an attractor can be interpreted as a measure of its global stability. In this section we discuss the basin stability of the states seen in the phase diagram of the previous section. The discussion of the basin stability of the chimera state is particularly interesting. 

\begin{figure*}
\centering
\begin{subfigure}{0.5\textwidth}
\centering\includegraphics[scale = 0.4]{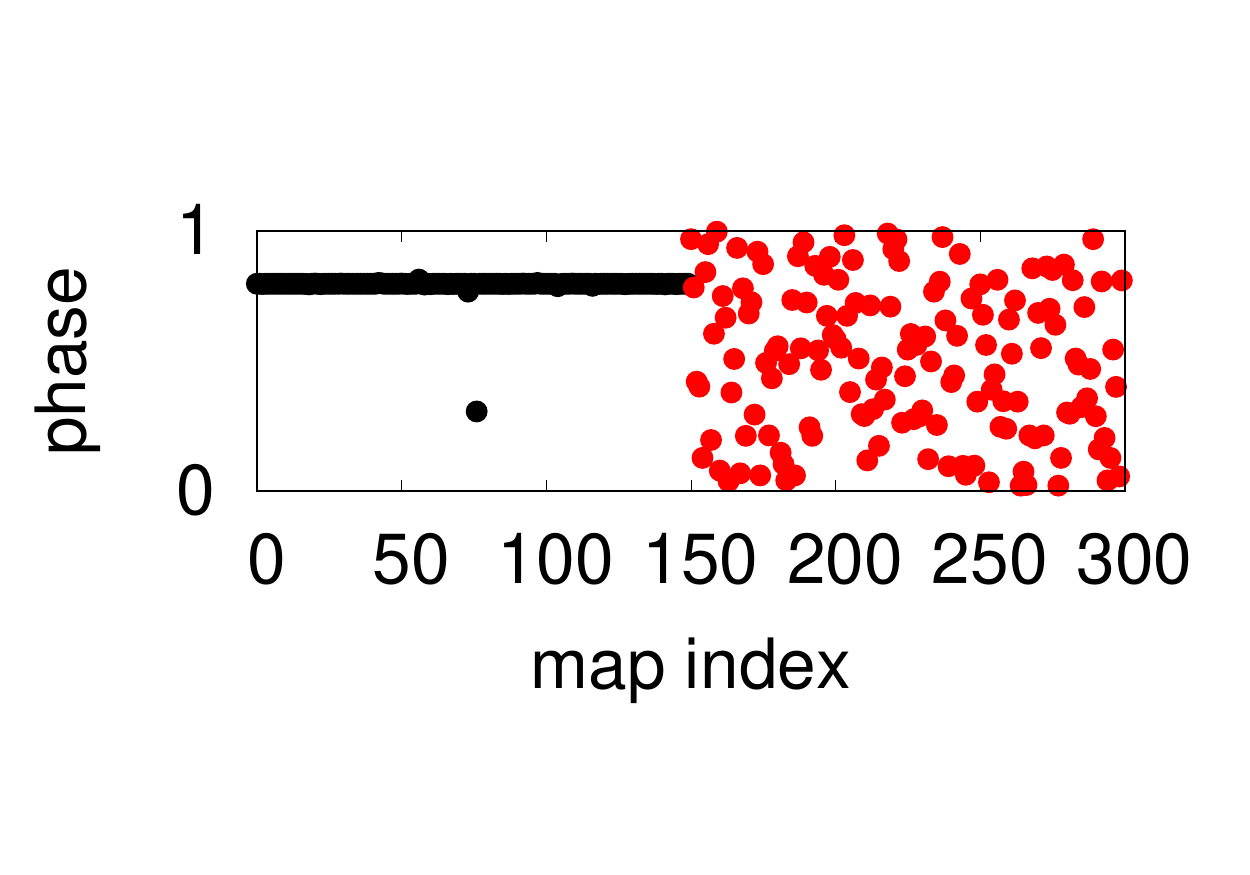}
\caption{}
\end{subfigure}%
\begin{subfigure}{0.5\textwidth}
\centering\includegraphics[scale = 0.4]{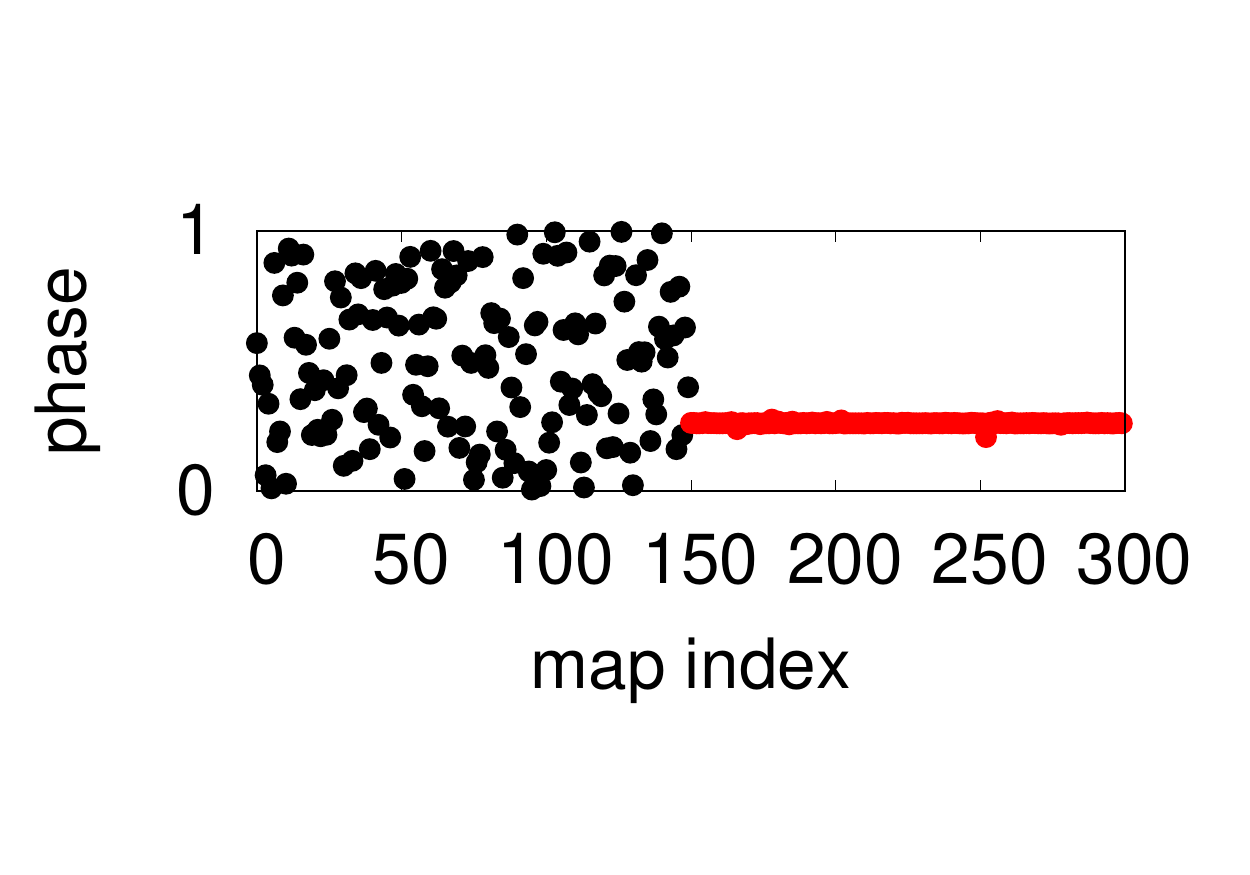}
\caption{}
\end{subfigure}
\caption{\label{fig: chim_symm} \footnotesize The snapshots of (a) the chimera state where group one are phase synchronised while group two is phase desynchronised and (b) chimera state where group one is phase desynchronised and maps in group two is phase synchronised. At parameters $\Omega = 0.27, K = 10^{-4}, \epsilon_1 = 0.94, N = 150$ two different initial conditions were used to obtain the above snapshots.}
\end{figure*}

\begin{figure*}
\centering\includegraphics[scale = 0.55]{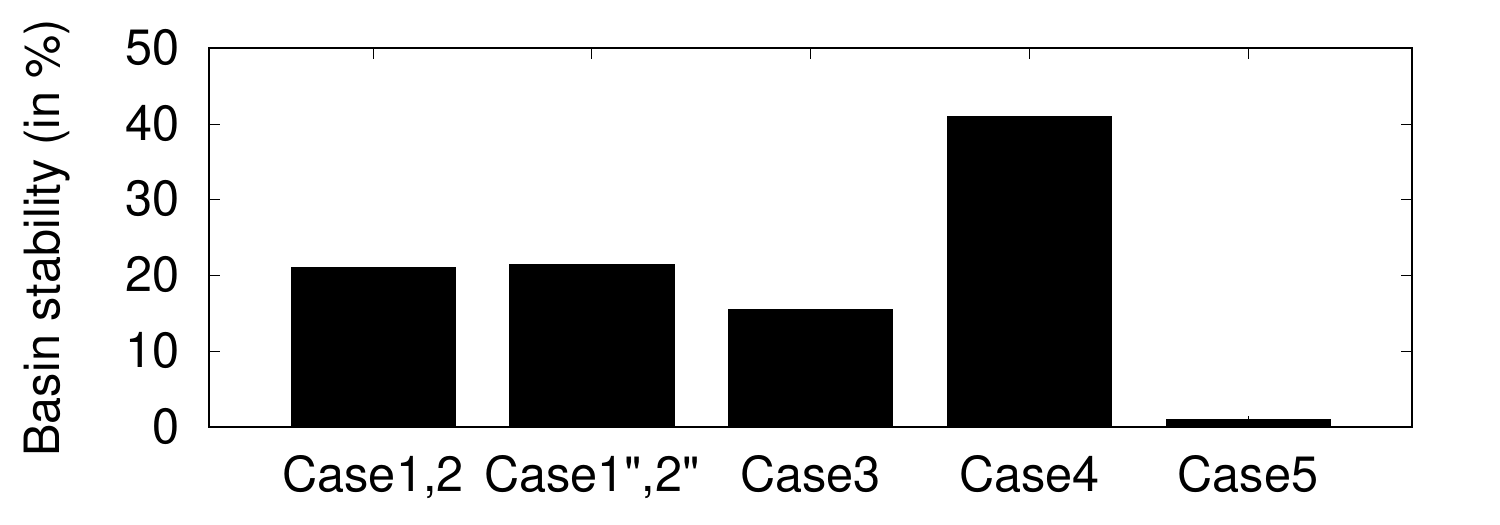}
\caption{\label{fig: histogram} \footnotesize Histogram of the fraction of the initial conditions that evolve to chimera states, two phase clustered state, fully phase desynchronised state and phase synchronised states. We use 400 initial conditions at parameter values $K = 10^{-4}, \Omega = 0.27, \epsilon_1 = 0.94, N = 150$. Here Case 1" and 2" are the mirrored versions of chimera states in Case 1 and 2. The volume of the basin of attraction of chimera state with synchronisation in group one and phase desynchronisation in group is $18.5\%$ of the total phase space while the same measure for the mirrored version is $19\%$. The basin volume of the fully phase desynchronised state is $15.5\%$ while for the two phase clustered state and fully phase synchronised state they are $46\%$ and $1\%$ respectively of the total phase space.}
\end{figure*}

Fig. \ref{fig: order_zoom} shows a large parameter region of the phase diagram containing the chimera state STI structures (case 1 and 2). We note that the chimera states shown in both Figs. \ref{fig: order_1} and \ref{fig: order_zoom} exhibit phase synchronisation in group one and STI structures in group two. Its clear that due to symmetry in the evolution Eq. \ref{sinecml} discussed earlier in section \ref{sec: model}, there exists a mirror version of this chimera where the nature of the dynamics of the groups are interchanged which can be accessed via a different set of random initial conditions (see figure \ref{fig: chim_symm}). The fraction of initial conditions which yields each of these symmetric configurations are also similar implying equal  basin volumes (see Fig. \ref{fig: histogram}). In addition to this, Fig. \ref{fig: histogram} plotted for   $K = 10^{-4}, \Omega = 0.27, \epsilon_1 = 0.94, N = 150$ indicates that due to the multistability of this system (Eq. \ref{sinecml}) a fraction  of the initial conditions evolve to the other states (case 3 - 5) , with basins of stability with volume proportional to the fraction in the histogram.

\par
\begin{figure*}
\centering
\begin{subfigure}{0.5\textwidth}
\centering\includegraphics[scale = 0.55]{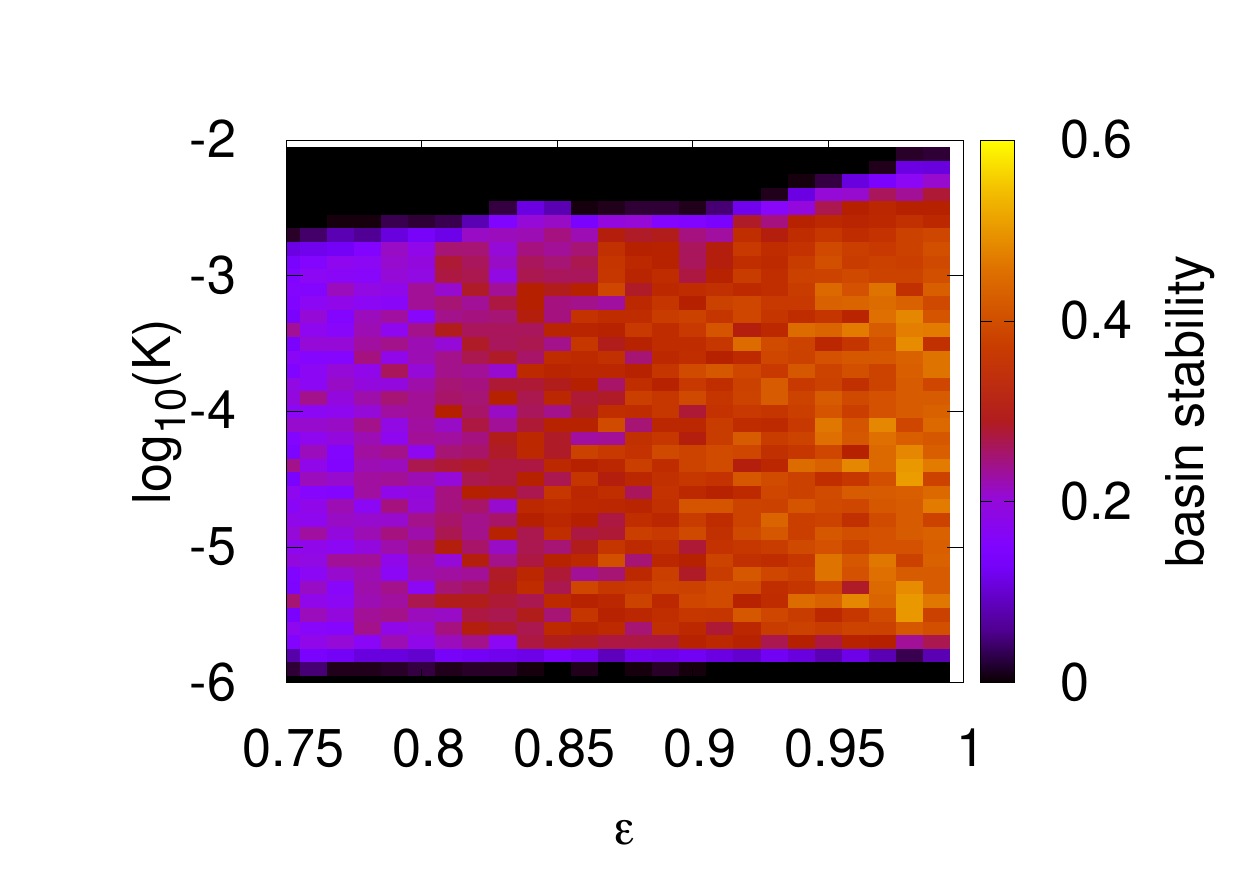}
\caption{\label{fig: chim_prob}}
\end{subfigure}%
\begin{subfigure}{0.5\textwidth}
\centering\includegraphics[scale = 0.55]{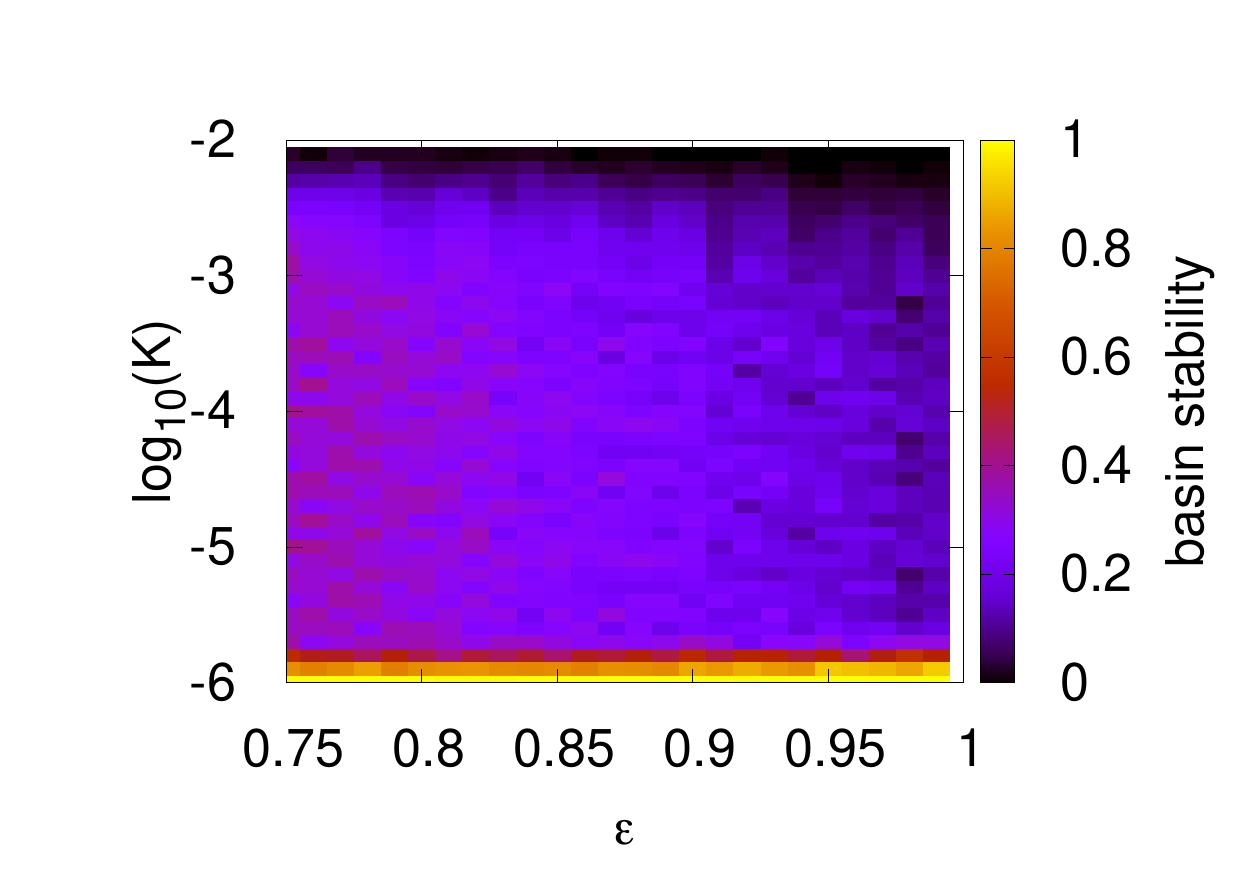}
\caption{\label{fig: desync_prob}}
\end{subfigure}
\caption{\label{fig: basin_stability_1} \footnotesize The basin stability of (a) chimera phase state (b) fully phase desynchronised state in $K - \epsilon_1$ space. The parameters $\Omega = 0.27, N = 150$ are kept fixed.}
 \end{figure*}

\begin{figure*}
\centering
\begin{subfigure}{0.5\textwidth}
\centering\includegraphics[scale = 0.55]{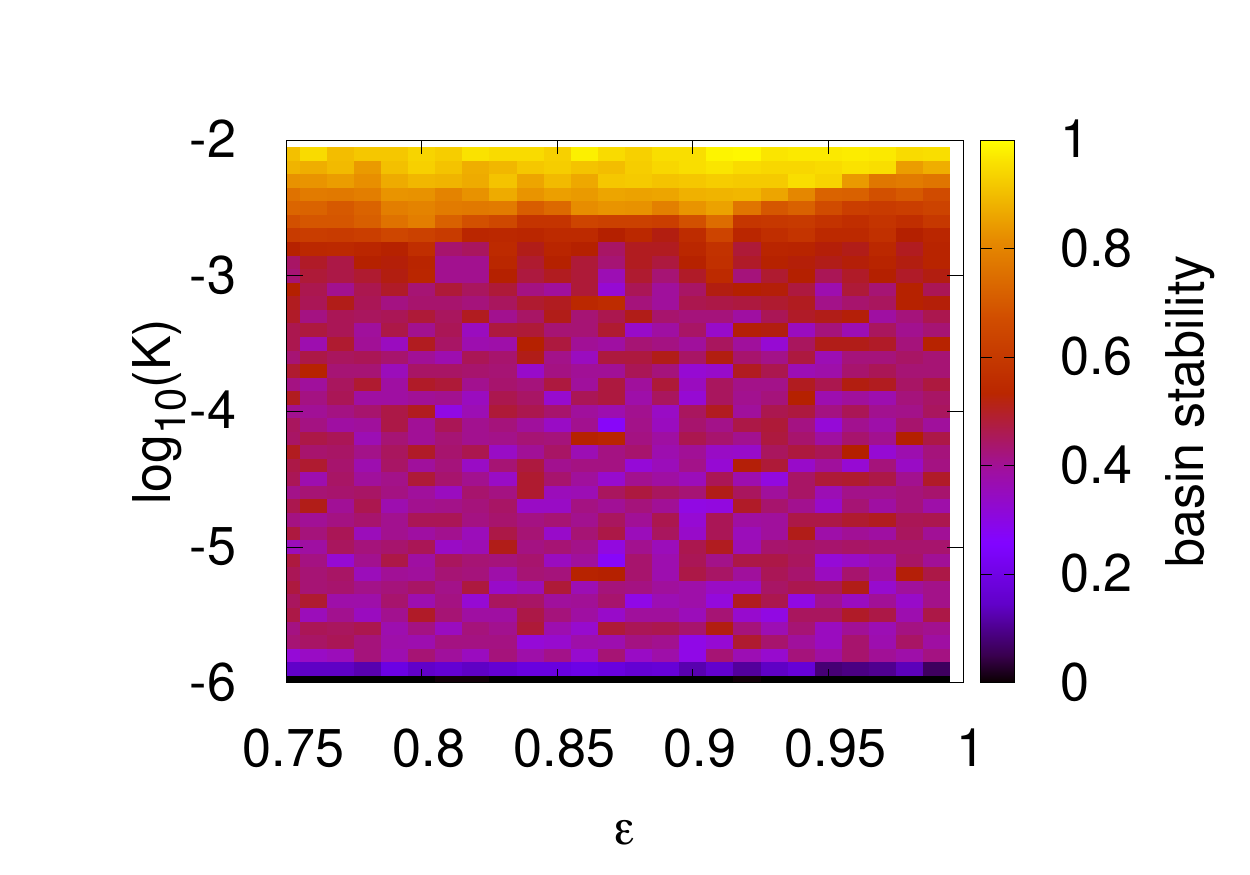}
\caption{\label{fig: two_phase_prob}}
\end{subfigure}%
\begin{subfigure}{0.5\textwidth}
\centering\includegraphics[scale = 0.55]{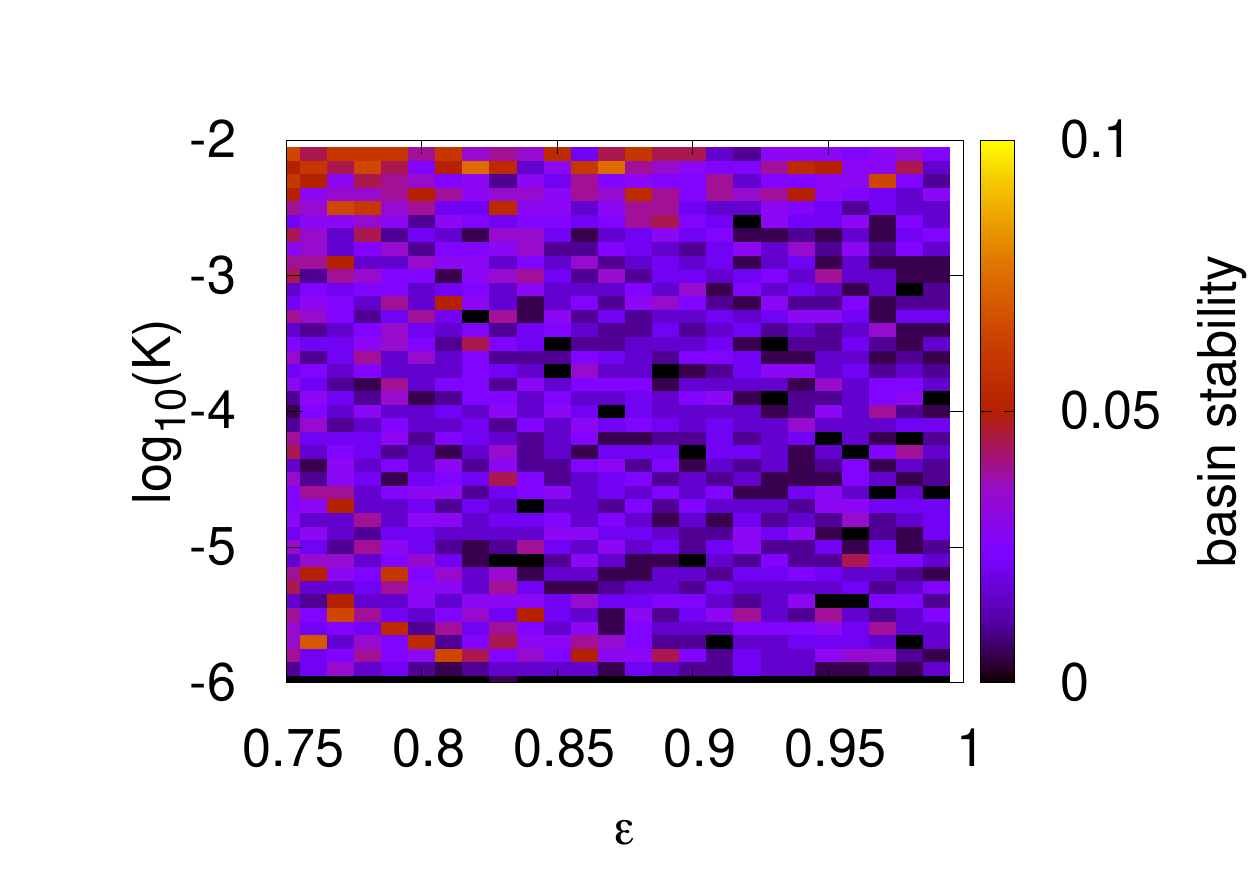}
\caption{\label{fig: sync_prob}}
\end{subfigure}
\caption{\label{fig: basin_stability_2} \footnotesize The basin stability of (a) two phase clustered state (b) fully phase synchronised state in the $K - \epsilon_1$ space. The parameters $\Omega = 0.27, N = 150$ are kept fixed.}
 \end{figure*}

 It is useful to examine  the basin stability of all the states observed  in the $K-\epsilon_1$ space. We examine a  $30\times30$ grid for the  range $-6 < \log_{10}K < -2$,$0.65 < \epsilon_1 < 1$, with  $\Omega$  fixed at $0.27$. At each grid point we choose 400 sets of initial conditions with $\theta$ values chosen  randomly between zero and one. The system in Eq. \ref{sinecml} is then evolved from each of these initial conditions for $4 \times 10^{6}$ time steps to the final state. The nature of the final state is then identified using the complex order parameters, $R, R^1, R^2$, which take the specific values for different final states as described in the previous section. Figs \ref{fig: basin_stability_1} and \ref{fig: basin_stability_2} show that  all attractors listed in cases 1 - 5 have a finite non-zero basin stability in the region bounded by $-6 < \log_{10}K < -2$ and $0.75 < \epsilon_1 < 1$ .

\begin{figure*}
\centering
\begin{subfigure}{0.5\textwidth}
\centering\includegraphics[scale = 0.55]{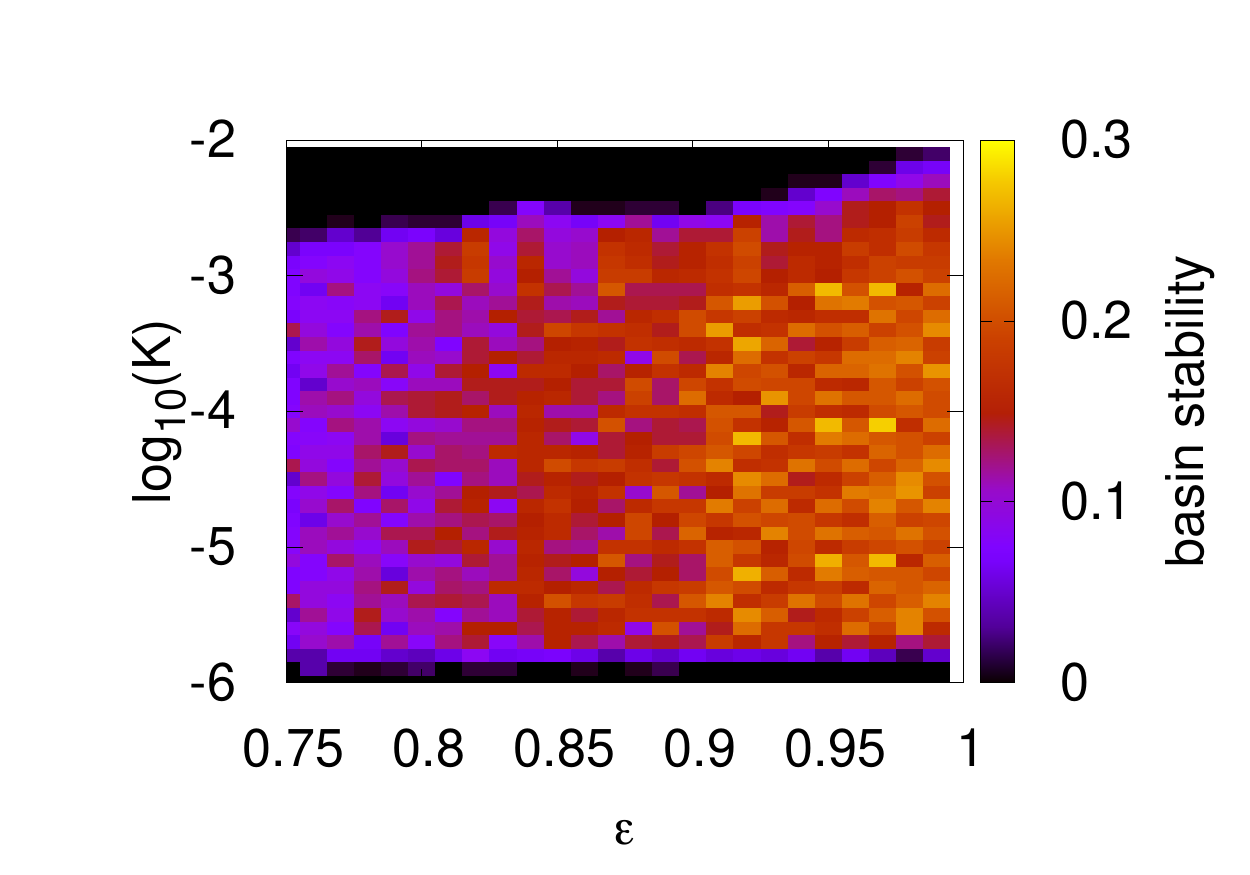}
\caption{}
\end{subfigure}%
\begin{subfigure}{0.5\textwidth}
\centering\includegraphics[scale = 0.55]{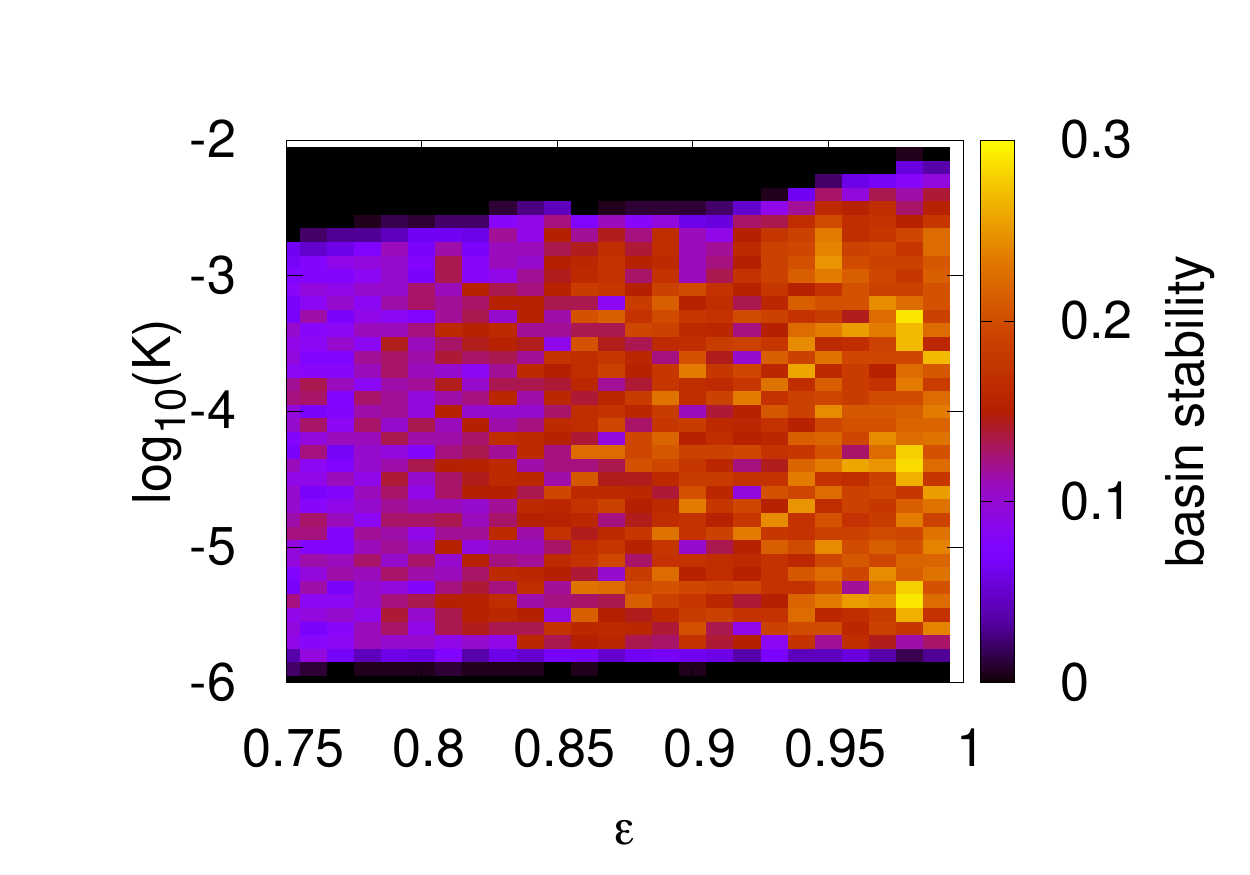}
\caption{}
\end{subfigure}
\caption{\label{fig: chim_symm_prob} \footnotesize The basin stability of (a) chimera state with phase synchronisation in group one and phase desynchronisation in group two and (b) chimera state where the distribution of phases are interchanged between the groups in the $K - \epsilon_1$ space. The parameters are $\Omega = 0.27, N = 150$.}
\end{figure*}

 Fig. \ref{fig: basin_stability_1}.a shows that the fraction of initial conditions that evolve to the  chimera state (case 1, 2)  varies between $0.2$ and $0.6$ in the region specified by $-5.5 < \log_{10}K < -3$ and $0.8 < \epsilon_1 < 1$. This fraction  is less than $0.2$ when $\epsilon_1 < 0.8$ and tends to zero when $\log_{10}K < -5.5$ and $\log_{10}K > -2$. In Fig.\ref{fig: basin_stability_1}.b we find that the globally phase desynchronised state has low basin stability in a significant portion of the parameter region of interest and the fraction of initial conditions that evolve to it less than 0.2 when $\log_{10}K > -6$. However it is near one when $\log_{10}K < -6$ for all values of $\epsilon_1$. The two phase clustered state is a favoured state when $K>10^{-3}$ (see Fig. \ref{fig: two_phase_prob}). The probability for completely random initial condition to evolve to a fully synchronised state is however low in the entire parameter space examined, as can be seen in  Fig. \ref{fig: basin_stability_2}.b. Figs. \ref{fig: chim_symm_prob} shows the basin stability of the two mirror chimera states. It is clear that the two states appear with approximately equal probability in the region of interest in the region $-5.5 < \log_{10}K < -3$ and $0.8 < \epsilon_1 < 1$. We note that the basin volume of the two phase cluster states is of similar magnitude in this region. These figures appear to indicate the existence of riddling in the basins of attraction of the final states. In future we hope to explore this in detail.  

\section{\label{sec: chim}Chimera states with STI like structures in the desynchronised group} 
The  region in the phase diagram where chimera states with spatiotemporally intermittent behaviour are seen in the $K, \epsilon_1$ parameter space for $\Omega = 0.27$ is clearly identified in the phase diagram of Fig. \ref{fig: order_zoom}. As mentioned earlier, this is the region where $R^1$ takes value $1$, and $R^2$ is zero. It can be seen from the space time plots and the temporal variation of the order parameter  $R^1$ (see Fig. \ref{fig: initial}) for this chimera state that the maps in the synchronised group are spatially phase synchronised but the phase at which they synchronise is not a temporal fixed point as shown by the variation of $\Psi^{1}$ (Figs. \ref{fig: initial}.c and f). The variation of $R^2, \Psi^2$ in Figs. \ref{fig: initial}.c and \ref{fig: initial}.(f) maps in the desynchronised group can be seen to be spatially and temporally desynchronised. Here we carry out the linear stability analysis of this chimera states for the parameters that where such solutions are seen and calculate the Lyapunov exponents.

\subsection{\label{sec: LS and LE}Linear stability analysis and Lyapunov spectrum}
We find the stability of the chimera states with spatiotemporal intermittent regions, by calculating the eigenvalues of the one step Jacobian matrix. 
\begin{equation}
J = \begin{bmatrix} 
	A & B \\
	C & D
      \end{bmatrix}
      \label{jacobian}	
\end{equation}
Here $A, B, C, D$ are $N \times N$ block matrices which have the form,
\begin{widetext}
\begin{equation}		
A = \begin{bmatrix}
	(1 + \frac{\epsilon_{1}}{N})f'(\theta_{n}^{1}(1)) & \frac{\epsilon_{1}}{N}f'(\theta_{n}^{1}(2)) & \cdots & \frac{\epsilon_1}{N}f'(\theta_{n}^{1}(N)) \\
	\frac{\epsilon_{1}}{N}f'(\theta_{n}^{1}(1))  &(1 + \frac{\epsilon_{1}}{N})f'(\theta_{n}^{1}(2)) & \cdots & \frac{\epsilon_1}{N}f'(\theta_{n}^{1}(N)) \\
	\cdots & \cdots & \cdots & \cdots \\
	\frac{\epsilon_{1}}{N}f'(\theta_{N}^{1}(1)) & \frac{\epsilon_{1}}{N}f'(\theta_{N}^{1}(2)) & \cdots & (1 + \frac{\epsilon_{1}}{N})f'(\theta_{N}^{1}(N))	
      \end{bmatrix}
\end{equation}
\begin{equation}		
D = \begin{bmatrix}
	(1 + \frac{\epsilon_{1}}{N})f'(\theta_{n}^{2}(1)) & \frac{\epsilon_{1}}{N}f'(\theta_{n}^{2}(2)) & \cdots & \frac{\epsilon_1}{N}f'(\theta_{n}^{2}(N)) \\
	\frac{\epsilon_{1}}{N}f'(\theta_{n}^{2}(1))  &(1 + \frac{\epsilon_{1}}{N})f'(\theta_{n}^{2}(2)) & \cdots & \frac{\epsilon_1}{N}f'(\theta_{n}^{2}(N)) \\
	\cdots & \cdots & \cdots & \cdots \\
	\frac{\epsilon_{1}}{N}f'(\theta_{N}^{2}(1)) & \frac{\epsilon_{1}}{N}f'(\theta_{N}^{2}(2)) & \cdots & (1 + \frac{\epsilon_{1}}{N})f'(\theta_{N}^{2}(N))	
      \end{bmatrix}
\end{equation}
\begin{equation}		
B = \begin{bmatrix}
	\frac{\epsilon_{2}}{N}f'(\theta_{n}^{2}(1)) & \frac{\epsilon_{2}}{N}f'(\theta_{n}^{2}(2)) & \cdots & \frac{\epsilon_2}{N}f'(\theta_{n}^{2}(N)) \\
	\frac{\epsilon_{2}}{N}f'(\theta_{n}^{2}(1)) & \frac{\epsilon_{2}}{N}f'(\theta_{n}^{2}(2)) & \cdots & \frac{\epsilon_2}{N}f'(\theta_{n}^{2}(N))  \\
	\cdots & \cdots & \cdots & \cdots \\
	\frac{\epsilon_{2}}{N}f'(\theta_{n}^{2}(1)) & \frac{\epsilon_{2}}{N}f'(\theta_{n}^{2}(2)) & \cdots & \frac{\epsilon_2}{N}f'(\theta_{n}^{2}(N)) 	
      \end{bmatrix}
\end{equation}
\begin{equation}		
C = \begin{bmatrix}
	\frac{\epsilon_{2}}{N}f'(\theta_{n}^{2}(1)) & \frac{\epsilon_{2}}{N}f'(\theta_{n}^{2}(2)) & \cdots & \frac{\epsilon_2}{N}f'(\theta_{n}^{2}(N)) \\
	\frac{\epsilon_{2}}{N}f'(\theta_{n}^{2}(1)) & \frac{\epsilon_{2}}{N}f'(\theta_{n}^{2}(2)) & \cdots & \frac{\epsilon_2}{N}f'(\theta_{n}^{2}(N))  \\
	\cdots & \cdots & \cdots & \cdots \\
	\frac{\epsilon_{2}}{N}f'(\theta_{n}^{2}(1)) & \frac{\epsilon_{2}}{N}f'(\theta_{n}^{2}(2)) & \cdots & \frac{\epsilon_2}{N}f'(\theta_{n}^{2}(N)) 	
      \end{bmatrix}
\end{equation} 
\end{widetext}

where $f'(\theta_{n}^{\sigma}(i)) =  1 - K\cos(2\pi\theta_{n}^{\sigma}(i))$ and $\sigma = 1, 2$ denote each group. The chimera states are seen in regimes where  the nonlinearity parameter $K$ takes low values $(< 10^{-3})$ (see the phase diagram in Fig. \ref{fig: order_zoom}).  Using this, and the fact that  $\left|cos2\pi\theta^{\sigma}_{n}(i)\right| \leq 1$, the quantity   $f'(\theta_{n}^{\sigma}(i)) = 1 - K\cos2\pi\theta_{n}^{\sigma}(i) \approx (1 - \alpha)$ where the upper bound on the value of $\alpha$ is $K$ in the chimera region. Using this approximation, the one step Jacobian matrix takes the form, 
\begin{equation}
J_{\text{c}} = \begin{bmatrix} 
	A & B \\
	B & A
      \end{bmatrix}
      \label{jacobian_block}	
\end{equation}
where,
 \begin{equation}
A = (1 - \alpha)\begin{bmatrix}
	(1 + \frac{\epsilon_1}{N}) & \frac{\epsilon_1}{N} & \cdots & \frac{\epsilon_1}{N} \\
	\frac{\epsilon_1}{N}  &(1 + \frac{\epsilon_1}{N}) & \cdots & \frac{\epsilon_1}{N} \\
	\cdots & \cdots & \cdots & \cdots \\
	\frac{\epsilon_1}{N} & \frac{\epsilon_1}{N} & \cdots & (1 + \frac{\epsilon_1}{N})
     \end{bmatrix}
\end{equation}
\begin{equation}		
B = (1 - \alpha)\begin{bmatrix}
	\frac{\epsilon_2}{N} & \frac{\epsilon_2}{N} & \cdots & \frac{\epsilon_2}{N} \\
	\frac{\epsilon_2}{N} & \frac{\epsilon_2}{N} & \cdots & \frac{\epsilon_2}{N}  \\
	\cdots & \cdots & \cdots & \cdots \\
	\frac{\epsilon_2}{N} & \frac{\epsilon_2}{N} & \cdots & \frac{\epsilon_2}{N} 	
      \end{bmatrix}
\end{equation}
Here $J_{c}$ (Eq. \ref{jacobian_block}) is a $2N \times 2N$ block circulant matrix\cite{davis1994} which can be block diagonalised using a matrix $\textbf{P}$  \cite{chatterjee2000, davis1994}, 
 \begin{equation}
 \textbf{P} = F_{2} \otimes I_{N}
 \label{P}
 \end{equation}

  where $I_{N}$ is a $N \times N$ identity matrix and $F_{2}$ is a $2 \times 2$ Fourier matrix\cite{davis1994} of the form,
\begin{equation}
 F_{2} = \frac{1}{\sqrt{2}}\begin{bmatrix}
 					1 & 1\\
					1 & \omega
				      \end{bmatrix}
				      \label{fourier}
\end{equation}
 with $\omega = \exp (\frac{2\pi i}{2}) = \cos \pi + i \sin \pi = -1$. So we have 
 \begin{equation}
\textbf{P}^{-1}J_{c}\textbf{P} = J_{c}^* = \begin{bmatrix}
								A + B & 0\\
								0 & A - B
					        			\end{bmatrix}
								\label{sim_trans_1}
 \end{equation}
where
\begin{equation}
\small A + B\\
 = (1 - \alpha)\begin{bmatrix}
             (1 + \frac{\epsilon_1 + \epsilon_2}{N}) & \frac{\epsilon_1 + \epsilon_2}{N} & \cdots &  \frac{\epsilon_1 + \epsilon_2}{N} \\
            \frac{\epsilon_1 + \epsilon_2}{N} &  (1 + \frac{\epsilon_1 + \epsilon_2}{N})  & \cdots & \frac{\epsilon_1 + \epsilon_2}{N} \\
             \cdots & \cdots &\cdots & \cdots \\
             \frac{\epsilon_1 + \epsilon_2}{N} &  \frac{\epsilon_1 + \epsilon_2}{N} & \cdots & (1 + \frac{\epsilon_1 + \epsilon_2}{N})
	    \end{bmatrix}
\end{equation} 
\begin{equation}
\small A - B = (1 - \alpha)\begin{bmatrix}
             (1 + \frac{\epsilon_1 - \epsilon_2}{N}) & \frac{\epsilon_1 - \epsilon_2}{N} & \cdots & \frac{\epsilon_1 - \epsilon_2}{N} \\
            \frac{\epsilon_1 - \epsilon_2}{N} &  (1 + \frac{\epsilon_1 - \epsilon_2}{N}) & \cdots & \frac{\epsilon_1 - \epsilon_2}{N} \\
             \cdots & \cdots &\cdots & \cdots \\
             \frac{\epsilon_1 - \epsilon_2}{N} & \frac{\epsilon_1 - \epsilon_2}{N} & \cdots & (1 + \frac{\epsilon_1 - \epsilon_2}{N})
	    \end{bmatrix}
\end{equation}
$A + B$ and $A - B$ are block circulant matrices\cite{davis1994}. The $j$th eigenvalue $\lambda_{j}$ of the matrix $A + B$ and $A - B$ is given by,

\begin{align*}
E_{j}^{A \pm B} = (1 - \alpha)\Bigg( 1 + \frac{\epsilon_1 \pm \epsilon_2}{N} &+ \frac{\omega^{j} (\epsilon_1 \pm \epsilon_2)}{N} + \frac{\omega^{2j} (\epsilon_1 \pm \epsilon_2)}{N}\\ &+ \cdots + \frac{\omega^{(N - 1)j} (\epsilon_1 \pm \epsilon_2)}{N} \Bigg)\label{eigenvalue}
\end{align*}

where $\omega$ is the $N^{\text th}$ root of unity i.e. $\omega = \exp(\frac{2 \pi i}{N})$. Setting  $j = 0$ we obtain the zeroth eigenvalues of the matrices $A + B$, $A - B$. So,
\begin{equation}
\begin{split}
E_{0}^{A + B} &= \bigg(1 + \frac{\epsilon_1 + \epsilon_2}{N} \times N \bigg)(1 - \alpha) \\
				&= (1 + \epsilon_1 + \epsilon_2)(1 - \alpha)\\
				&= 2(1 - \alpha)
\end{split}
\end{equation}

\begin{equation}
\begin{split}
E_{0}^{A - B} &= \bigg(1 + \frac{\epsilon_1 - \epsilon_2}{N} \times N\bigg)(1 - \alpha) \\
				&= (1 + \epsilon_1 - \epsilon_2)(1 - \alpha)\\
				&= 2 \epsilon_1(1 - \alpha)
\end{split}
\end{equation}				
For any $j > 0$ we have 
\begin{equation}
\begin{split}
E_{j}^{A \pm B} &= (1 - \alpha)\bigg(1 + \frac{\epsilon_1 \pm \epsilon_2}{N}(1 + \omega^{j} + \omega^{2j} + \cdots + \omega^{(N - 1)j})\bigg) \\
			&= 1 - \alpha				
\end{split}
\end{equation}
where we use $\epsilon_1 + \epsilon_2 = 1$ and $1 + \omega^{j} + \omega^{2j} + \cdots + \omega^{(N - 1)j} = 0$. So the eigenvalues of the matrix $J_{c}^*$ for $K \rightarrow 0$, are $2(1 - \alpha)$, $2\epsilon_1(1 - \alpha)$ and $2N - 2$ fold degenerate eigenvalues $1 - \alpha$. Therefore the Lyapunov exponents are, 
\begin{equation}
\begin{split}
\lambda_{1} &= \frac{1}{\tau}\lim_{\tau \rightarrow \infty}\sum\limits_{1}^{\tau}\ln2(1 - \alpha) \approx 0.693 + \ln(1 - \alpha)\\
\lambda_{2} &= \frac{1}{\tau}\lim_{\tau \rightarrow \infty}\sum\limits_{1}^{\tau}\ln(2\epsilon_1(1 - \alpha)) = \ln(2\epsilon_1) + \ln(1 - \alpha)\\
\lambda_{j} &= \frac{1}{\tau}\lim_{\tau \rightarrow \infty}\sum\limits_{1}^{\tau}\ln(1 - \alpha) = \ln(1 - \alpha)\,\,\,\,\,\, \text{for all $j = 3, \cdots, 2N$}
\end{split}
\label{LE}
\end{equation}

\begin{figure*}
\centering
\begin{subfigure}{0.5\textwidth}
\centering\includegraphics[scale = 0.55]{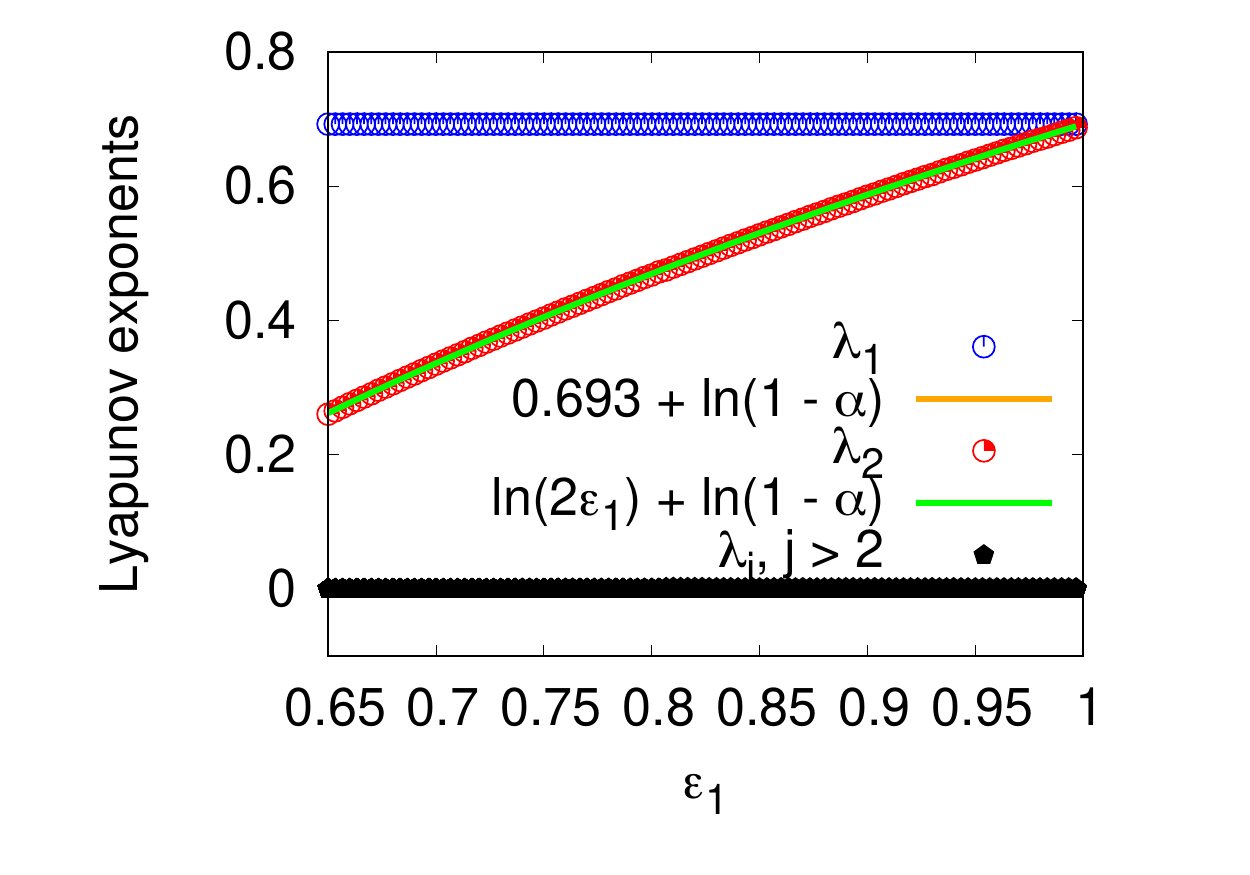}
\caption{\label{fig: LE_eps}}
\end{subfigure}%
\begin{subfigure}{0.5\textwidth}
\centering\includegraphics[scale = 0.55]{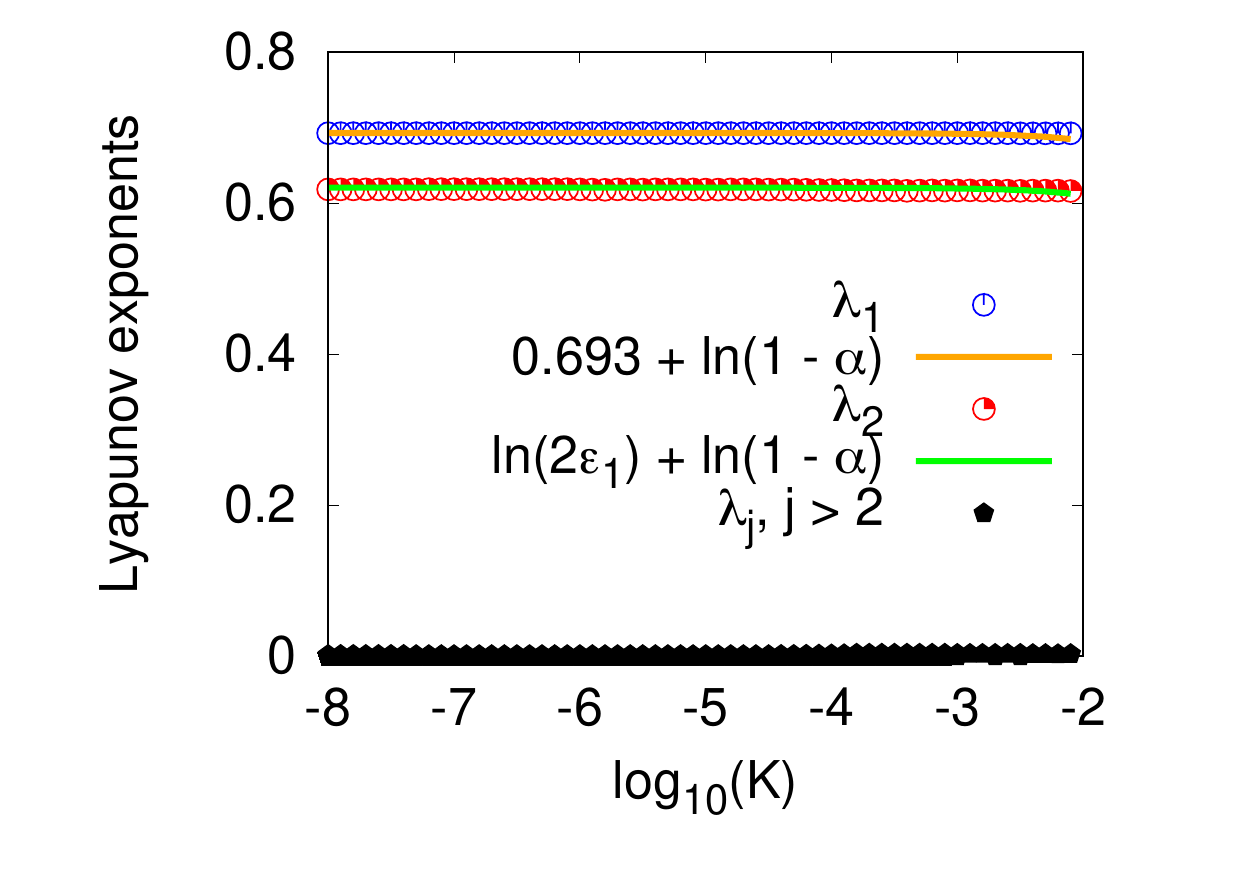}
\caption{\label{fig: LE_K}}
\end{subfigure}
\caption{\label{fig: LE_var} \footnotesize The variation of the largest Lyapunov exponent $\lambda_{1}$, the second largest Lyapunov exponent $\lambda_{2}$ and the rest of the  LE values are plotted between (a) $0.65 < \epsilon_1 < 1$ for $K = 10^{-5}$ and (b) $-8 < \log_{10}(K) < -2$ for $\epsilon_1 = 0.93$. In both the case the largest possible value of $\alpha = K$ is used. It can be clearly seen that the analytic value $\ln{2\epsilon_1} + \ln(1 - \alpha)$ matches with the numerically calculated  $\lambda_2$. All LE values are calculated via the Gram-Schmidt orthogonalisation method using the Jacobian for $10^{6}$ time steps after a transient of $3 \times 10^{6}$ steps. Other parameters are fixed at $\Omega = 0.27, N = 150$. }
\end{figure*}

Fig. \ref{fig: LE_eps}  plots the variation of Lyapunov exponents with $\epsilon_1$ for $K = 10^{-5}$. It is clear that the Lyapunov exponents (Eq. \ref{LE}) match the numerical values obtained from the numerical evolution  in the chimera regime, i.e.  $(-5.5 \lesssim \log_{10}K \lesssim -3)$. It is clear from Fig. \ref{fig: LE_K} that $\lambda_1, \lambda_2$ calculated in Eq. \ref{LE} start to  deviate from numerically calculated values when $\log_{10}{K} > -3$, as in this range the approximation to the Jacobian is not correct, i.e. $J \not\approx J_{\text{c}}$.  We note that two of the Lyapunov exponents obtained in range studied, viz $\lambda_1$ and $\lambda_2$ are positive.  Hence the maps in the chimera regime show hyperchaotic behaviour. We note that chimera state with hyperchaotic temporal dynamics has been observed earlier in coupled oscillator systems \cite{wolfrum2011}, where again a hyperchaotic STI chimera has been seen. We note that this is one of the few instances where the temporal dynamics of the chimera state with spatiotemporally intermittent structure is found to be hyperchaotic in nature. The stability analysis shown here applies to all the final states that appear in the region of the phase diagram when $K$ is negligibly small. The linear stability analysis of globally synchronised state (Case 4) and two clustered state (Case 5) is carried out for arbitrary value of $K$ in appendix \ref{app: ls}. 

\par
\begin{figure*}
\centering 
\begin{subfigure}{0.5\textwidth}
\centering\includegraphics[scale = 0.55]{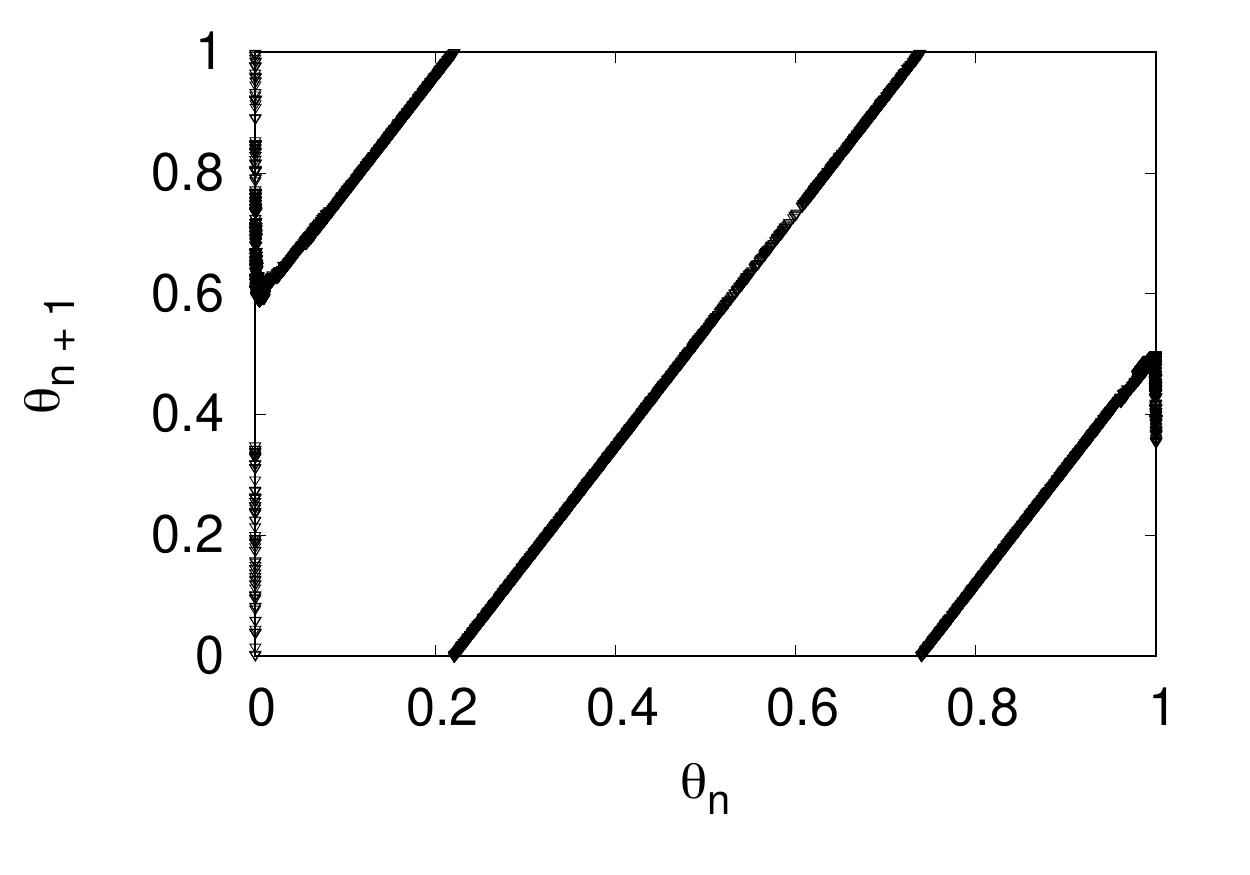}
\caption{\label{fig: return_1}}
\end{subfigure}%
\begin{subfigure}{0.5\textwidth}
\centering\includegraphics[scale = 0.55]{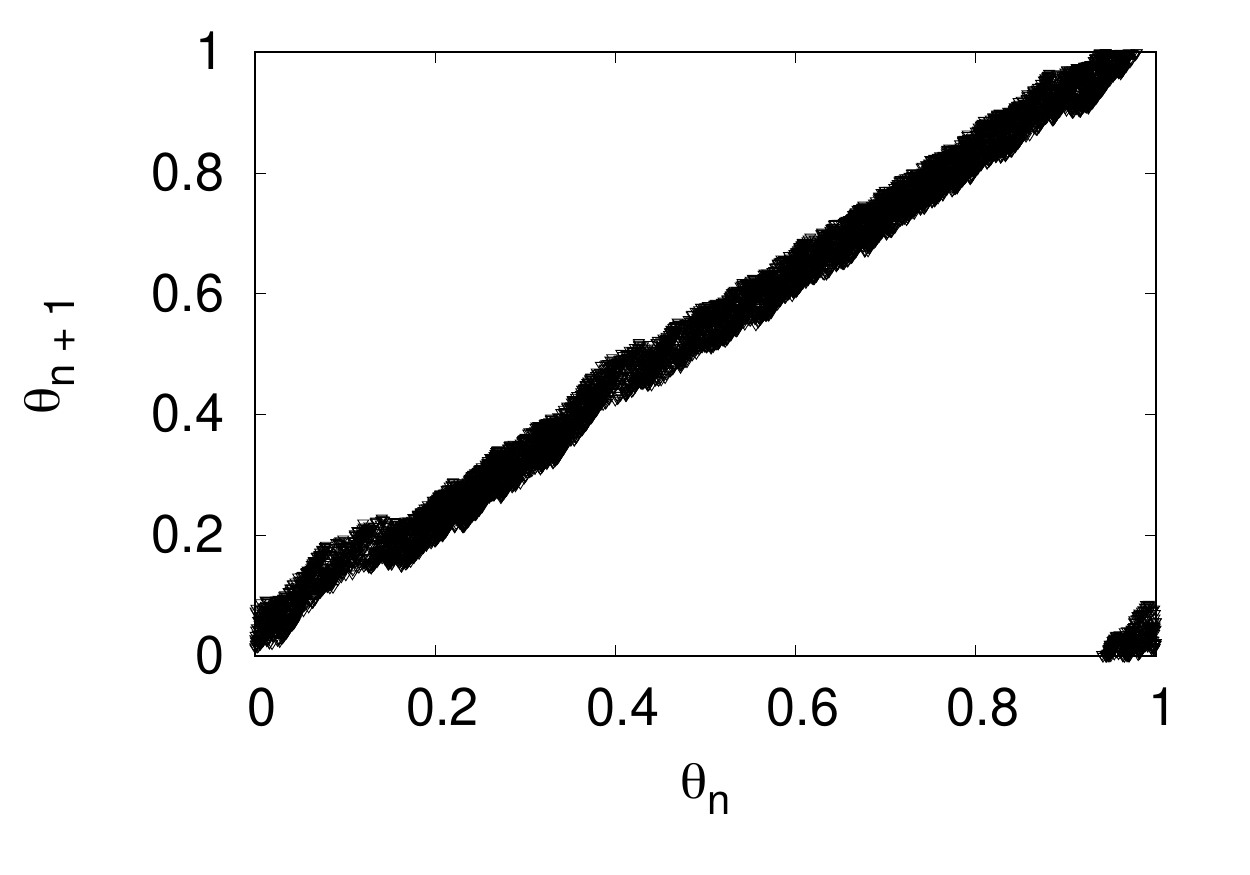}
\caption{\label{fig: return_2}}
\end{subfigure}
\caption{\label{fig: return} \footnotesize (color online) The return map of site 10 from (a) group one and (b) group two when the CML is already evolved in to the chimera state at parameters $K = 10^{-5}, \Omega = 0.27, \epsilon_1 = 0.93, N = 150$.}
\end{figure*}

 We examine the temporal dynamics of the chimera states via the site return maps by randomly choosing a typical site from each of the groups (see figure \ref{fig: return}). We observe that there is a distinct difference between the return map of a site from group one and group two. The return maps for groups one and two show non-banded and banded structures respectively. The noninvertible nature of the return map of a site belonging to the synchronised group can be clearly seen in Fig. \ref{fig: return_1}.  The space time behaviour of the phases of the circle maps in the desynchronised group suggest the existence of synchronised islands with  identical phases inside clusters of spatiotemporally phase desynchronised sites. We analyse this spatiotemporal intermittent structure of the incoherent group in the next section. 
 
\subsection{\label{sec: laminar_burst} Identifying the laminar and burst sites}
We begin the analysis of the spatiotemporally intermittent structure in the incoherent group of the chimera state by identifying the intermittent synchronised islands (laminar) within the desynchronised phases (burst) in the incoherent group. The existence of global coupling between the maps imply that the neighbourhood of each map is essentially the entire system. The coupling terms in the evolution Eq. \ref{sinecml} also show that phase of any map at a time step depends on phases of all maps in the system at previous time step. Therefore in order to locate the intermittent synchronised sites we must consider the all the maps at a given time step as well as the previous time step. 

We consider any two sites $(i, j)$ as laminar sites when the phases of the circle maps at these sites are such that the quantity $\Delta^{\sigma, \sigma'}_{ij} = \left|\frac{1}{2}\left| \exp (2\pi i \theta^{\sigma}(i)) + \exp(2\pi i \theta^{\sigma'}(j))\right| - 1\right|$ is less than an assigned cutoff value set by the parameter $\delta$. The quantity $\Delta^{\sigma, \sigma'}_{ij}$, which can also be considered as a two site order parameter (compare with the definition of $R^1, R^2, R$ of the group-wise and global order parameter given in Eq. \ref{global}), is used instead of directly computing the phase difference because $\Delta^{\sigma, \sigma'}_{ij}$ takes into account the fact that equation \ref{sinecml} has a modulo one operation. It is necessary to take account of the global coupling topology to identify the laminar and burst sites. By taking into account of the global coupling topology, we identify the laminar and burst sites in the spatiotemporal variation of the phases of the CML in two steps which we describe here :

\begin{figure*}
\centering
\begin{subfigure}{0.5\textwidth}
\centering\includegraphics[scale = 0.2]{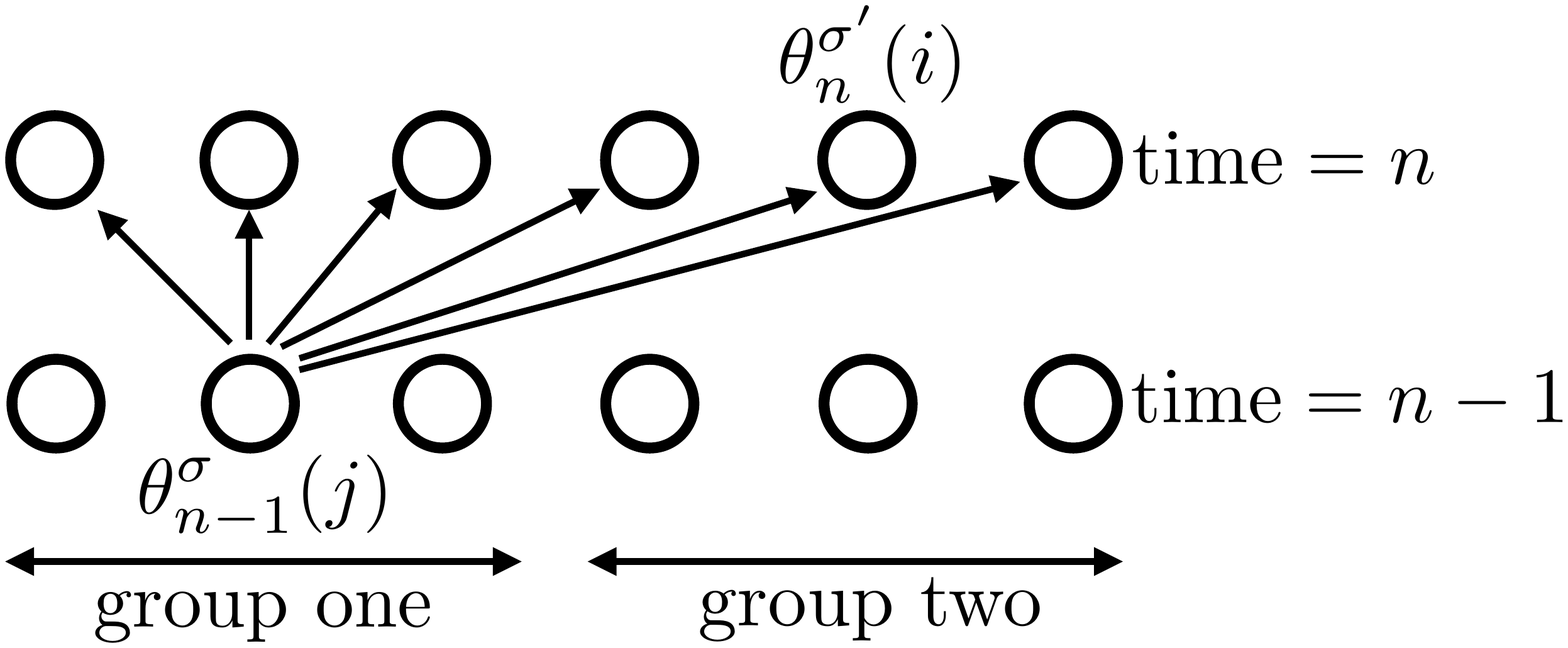}
\caption{}
\end{subfigure}%
\begin{subfigure}{0.5\textwidth}
\centering\includegraphics[scale = 0.2]{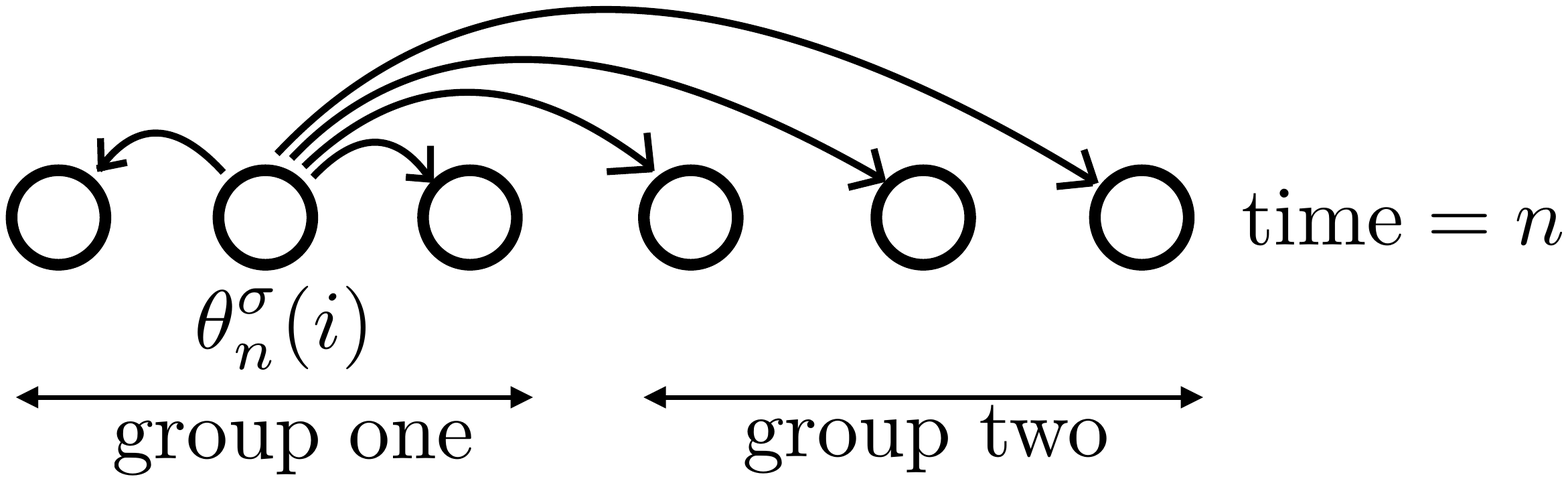}
\caption{}
\end{subfigure}
\caption{\label{fig: infec}\footnotesize (color online) A schematic of the method for identifying the laminar and burst sites. Each of these arrows in the above diagrams indicate the pair of sites between which the condition $\Delta^{\sigma, \sigma'}_{ij}$ is checked} 
\end{figure*}

\begin{enumerate}
\item We consider the phases of the CML at two consecutive time steps, $n$ and $n - 1$. The phase of the map at site $i$ in group $\sigma$ at time step $n$, is denoted as $\theta_{n}^{\sigma}(i)$. We choose two sites each from time steps $n - 1$ and $n$ that can belong to any of the groups and they are denoted by $\theta_{n - 1}^{\sigma}(j)$ and $\theta_{n}^{\sigma'}(i)$. We now check if $\Delta^{\sigma, \sigma'}_{ij} < \delta$ for all $i = 1, 2, \cdots, N$ for both $\sigma' = 1, 2$ and label those lattice sites as laminar, if the corresponding phase, $\theta_{n}^{\sigma'}(i)$ satisfies the condition, $\Delta^{\sigma, \sigma'}_{ij} < \delta$. We also label the lattice site at $\theta_{n - 1}^{\sigma}(j)$ as laminar if at least one such $i$ is found for which $\Delta^{\sigma, \sigma'}_{ij} < \delta$ (see Fig. \ref{fig: infec}.(a) for reference). We repeat this method for $j = 1, 2, \cdots, N$ for $\sigma = 1, 2$. We thus check if there is any temporal infection between the sites at time step $n - 1$ and time step $n$. Once the laminar sites at time step $n$ are identified by this method we check if there is any spatial infection between sites. We describe this in next step.

\item Now, we calculate $\Delta^{\sigma, \sigma'}_{ij} = \left|\frac{1}{2}\left| \exp(i2\pi \theta_{n}^{\sigma'}(j)) + \exp(i2\pi \theta_{n}^{\sigma}(i))\right| - 1\right|$ for all $j = 1, 2, \cdots, N$ when $\sigma'  \neq \sigma$ and for $j = 1, 2, \cdots, N$ except $j \neq i$ when $\sigma' = \sigma$ and we check the condition $\Delta^{\sigma, \sigma'}_{ij} < \delta$. A simple schematic is shown in Fig. \ref{fig: infec}.(b) for clarification. We label $\theta_{n}^{\sigma}(i)$ as a laminar site at time step $n$ if the condition is satisfied at least once. 
\end{enumerate}

After checking the phases of the maps at all sites at time step $n$ for temporal and spatial infections for laminarity in a similar fashion, we move on to the phases of the maps in the next time step. The intermittent synchronised and burst sites in a given spatiotemporal variation of the maps can all be identified in this way. In the next section we find the distribution of laminar and burst segments in the incoherent group of the chimera states.   

\subsection{Distribution of laminar and burst lengths}
\begin{figure*}
\centering
\begin{subfigure}{0.5\textwidth}
\centering\includegraphics[scale = 0.55]{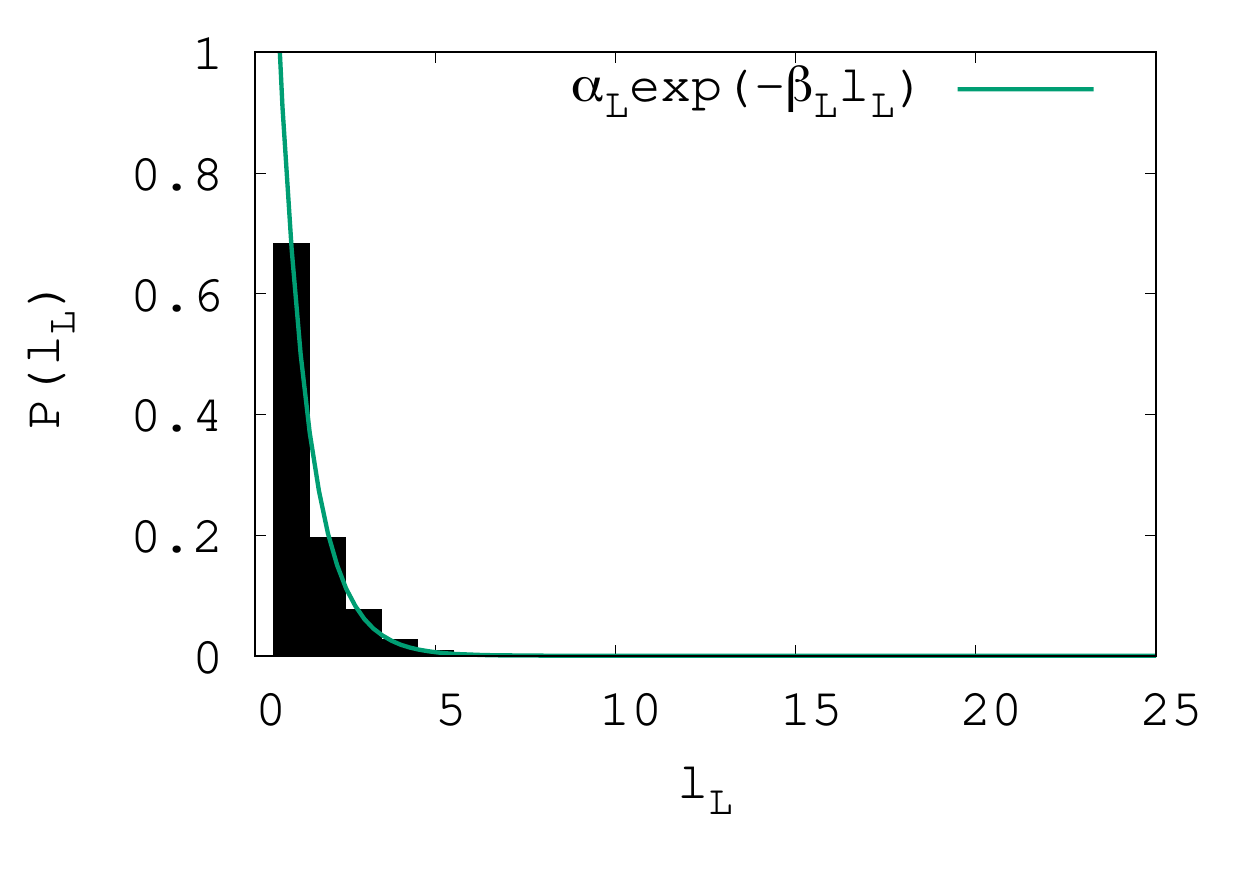}
\caption{}
\label{fig: dist_150_1}
\end{subfigure}%
\begin{subfigure}{0.5\textwidth}
\centering\includegraphics[scale = 0.55]{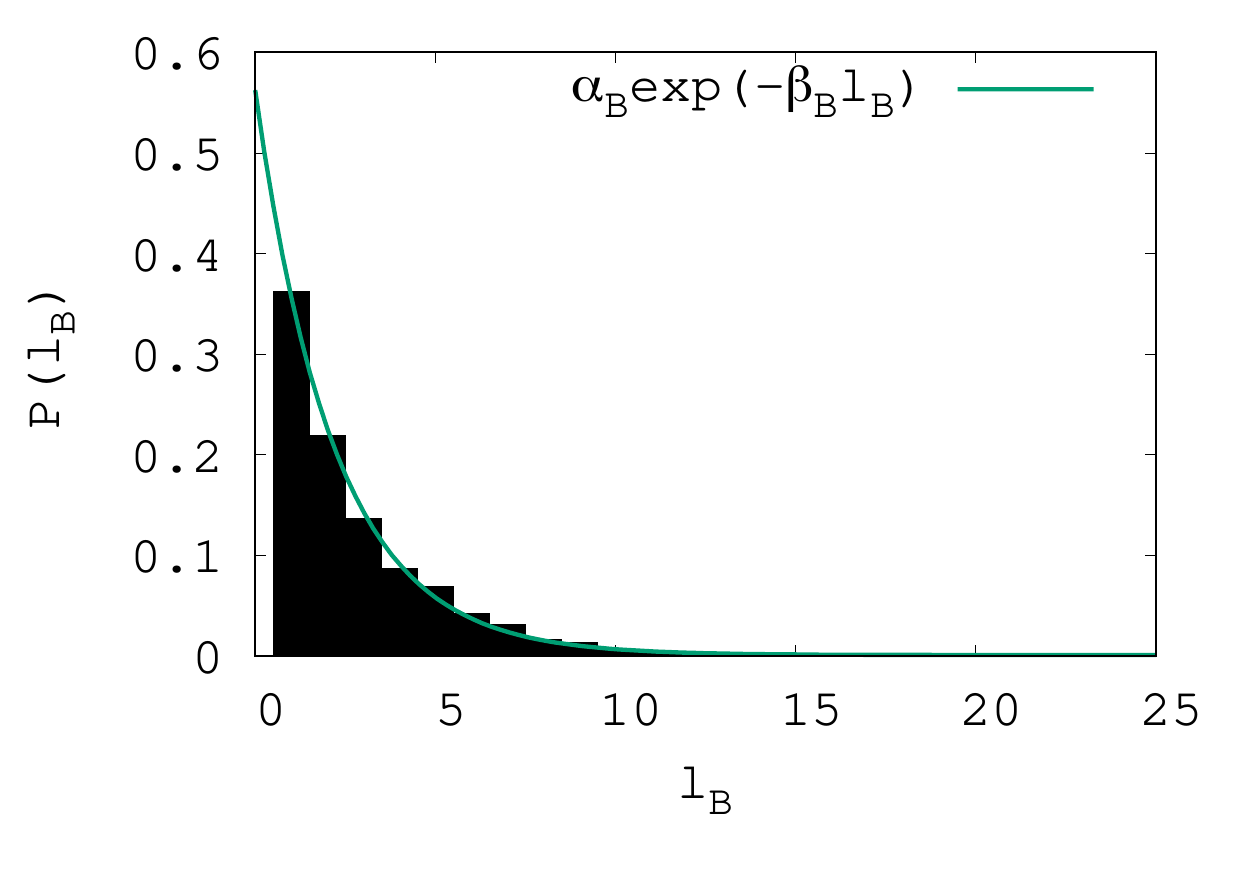}
\caption{}
\label{fig: dist_150_2}
\end{subfigure}
\caption{\label{fig: dist_150}\footnotesize The probability distribution of (a) laminar $(I_L)$ and (b) burst ($I_{B}$) segments in the spatiotemporal variation of the phases of incoherent maps in group two in the chimera state. The parameters are $K = 10^{-5}, \epsilon_1 = 0.93, \Omega = 0.27$ and $N = 150$. For the distribution of laminar segments, We find $P(I_L) \approx \alpha_L \exp(-\beta_L I_L)$ with $\alpha_L = 2.239 \pm 0.047, \beta_L = 1.19 \pm 0.018$. The distribution of the burst segments becomes, $P(I_B) \approx \alpha_B \exp(-\beta_B I_B)$ where $\alpha_B = 0.561 \pm 0.009, \beta_B = 0.456 \pm 0.009$.}
\end{figure*}

\begin{figure*}
\centering
\begin{subfigure}{0.5\textwidth}
\centering\includegraphics[scale = 0.55]{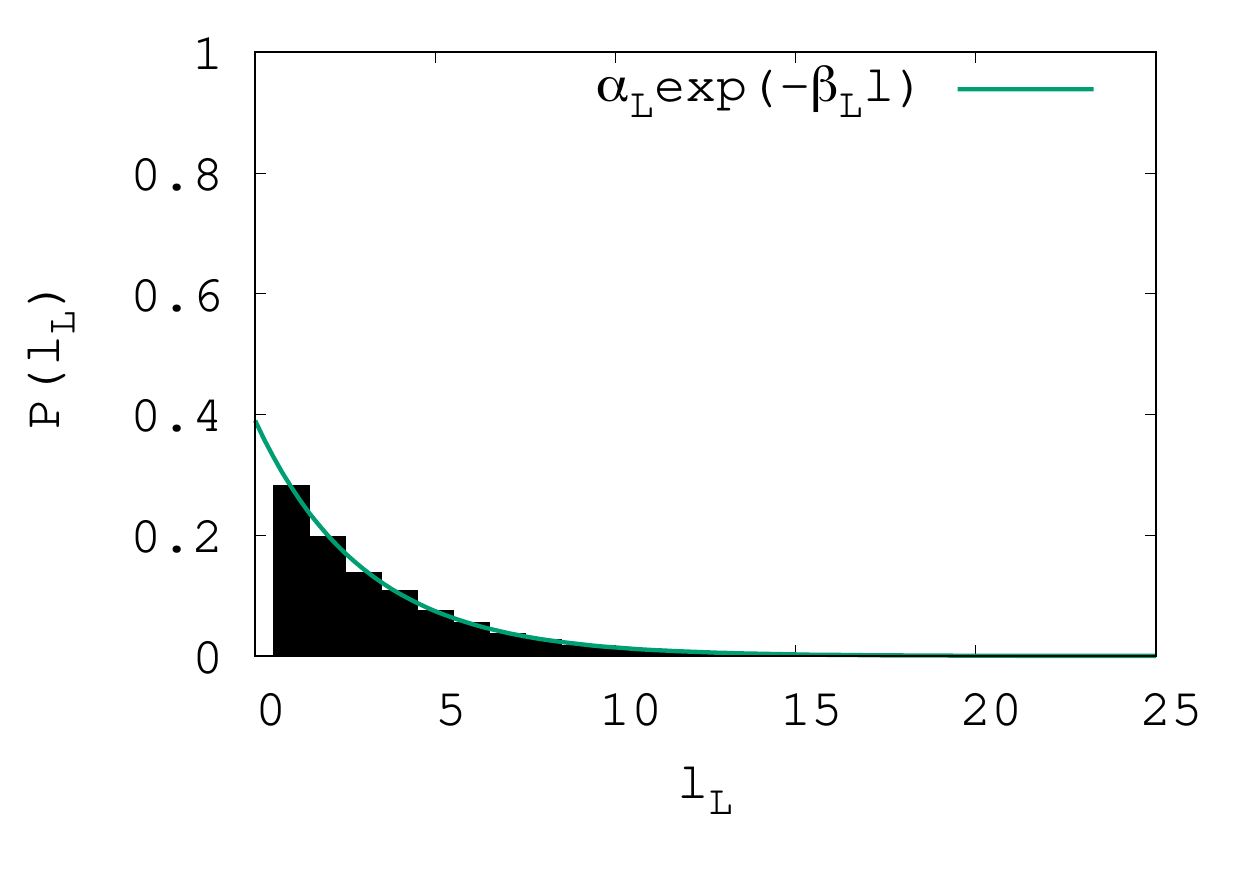}
\caption{}
\label{fig: dist_450_1}
\end{subfigure}%
\begin{subfigure}{0.5\textwidth}
\centering\includegraphics[scale = 0.55]{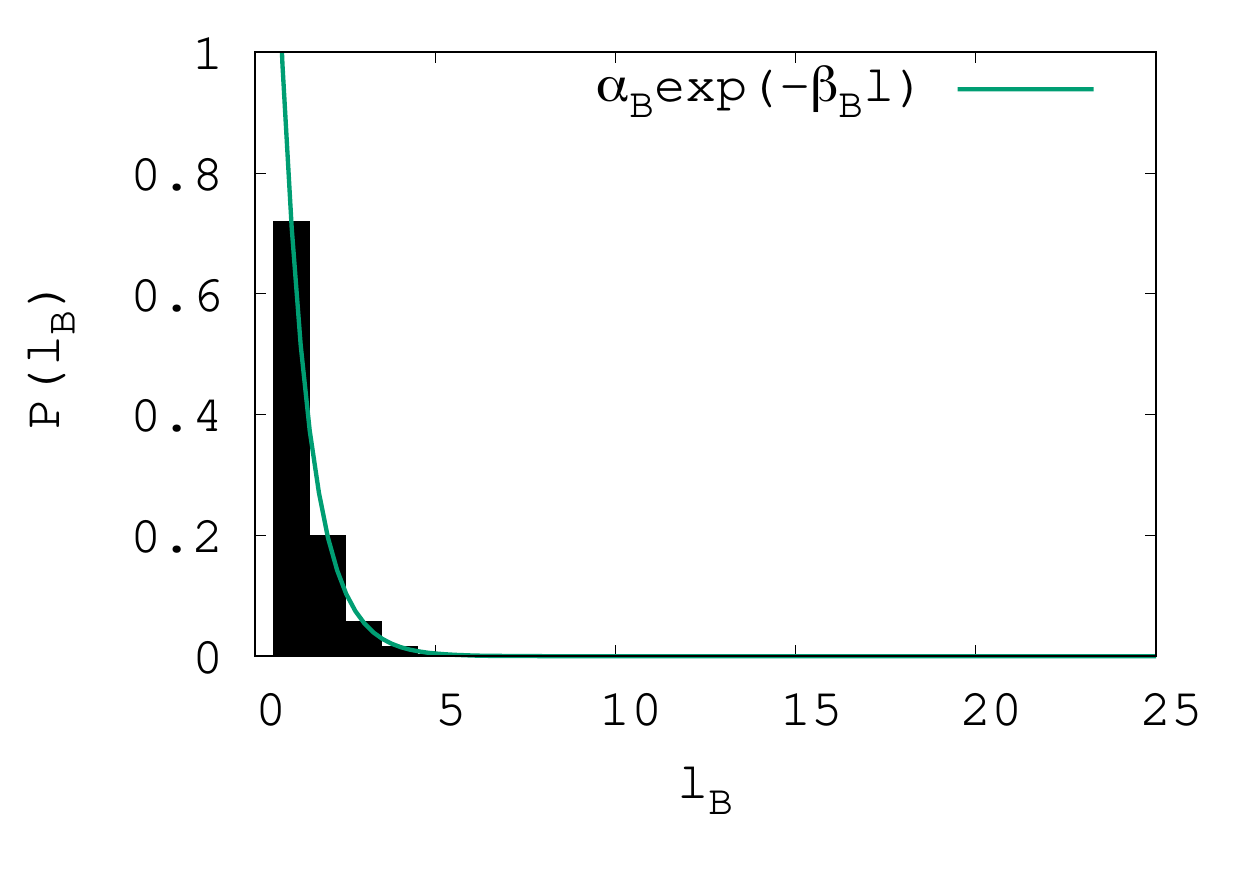}
\caption{}
\label{fig: dist_450_2}
\end{subfigure}
\caption{\label{fig: dist_450}\footnotesize The probability distribution of (a) laminar $(I_L)$ and (b) burst ($I_{B}$) segments in the spatiotemporal variation of the phases of incoherent maps in group two in the chimera state. The parameters are $K = 10^{-5}$, $\epsilon_1 = 0.93$, $\Omega = 0.27$ and $N = 450$. For the distribution of laminar segments, We find $P(I_L) \approx \alpha_L \exp(-\beta_L I_L)$ with $\alpha_L = 2.239 \pm 0.047$, $\beta_L = 1.19 \pm 0.018$. The distribution of the burst segments becomes, $P(I_B) \approx \alpha_B \exp(-\beta_B I_B)$ where $\alpha_B = 0.561 \pm 0.009$, $\beta_B = 0.456 \pm 0.009$.}
\end{figure*}

{\label{sec: LB}}
We note that the co-evolving maps are placed at consecutively numbered sites on a one dimensional lattice, with maps situated at sites $1$ to $N$ being identified as belonging to one group and maps from $N+1$ to $2N$ being identified as belonging to the other subgroup. The maps are coupled globally, with the intra group and intergroup couplings taking distinct values. Thus,  maps at consecutive sites, are influenced by the behaviour of all the other maps, with the crucial element being the number of maps in each subgroup whose phase angles are in the laminar or burst phase  as defined by the pairwise order parameter $\Delta^{\sigma, \sigma'}_{ij}$. It is interesting to see the distribution of laminar and burst lengths, viz. the distribution of the lengths of coherent and incoherent segments under these circumstances. Here, we find the number of consecutive sites which are laminar or burst sites at a given time in the phase desynchronised subgroup region during the evolution of chimera states. Thus the length of a laminar $(l_L)$/burst $(I_B)$ segment can vary from $0$ to $N$.
The probability $P(l_L)$ for a laminar segment of length $l_L$ to exist during a given time interval is the ratio of the total number of laminar segments of all lengths in this time period. The resulting distribution $P(l_L)$ of the laminar segments as well as the burst segments $P(l_B)$ is plotted in  Figs. \ref{fig: dist_150} and \ref{fig: dist_450}, and is seen to follow exponential behaviour $\alpha\exp(-\beta l)$ irrespective of system size. Thus, long laminar and burst segments are not very probable. 
This is unlike the case of spatiotemporal intermittency in systems with diffusive coupling where power law distributions are seen for laminar lengths \cite{chate1988, zahera2005, zahera2006}. Table \ref{table_two} lists the values of $\alpha_L, \beta_{L}$ for laminar segments and $\alpha_B, \beta_{B}$ for the burst segments for different system sizes.

\begin{table*}
\caption{\label{table_two}The values of the $\alpha_L$, $\beta_L$, $\alpha_B$ and $\beta_B$ for the distribution of laminar and burst segments in the incoherent group of the chimera states for different system sizes. }
\begin{ruledtabular}
\begin{tabular}{ccccc}
System size $(2N)$& $\alpha_L$ & $\beta_L$ & $\alpha_B$ & $\beta_B$\\
\hline
$200$& $4.3189 \pm 0.0021$& $1.6721 \pm 0.0005$& $0.2627 \pm 0.0009$&$ 0.2336 \pm 0.0011$\\
$400$& $1.3780 \pm 0.0018$& $0.8677 \pm 0.001$& $0.826 \pm 0.0032$& $0.6029 \pm 0.0024$\\
$600$& $0.7475 \pm 0.0075$& $0.5509 \pm 0.006$& $1.3291 \pm 0.0063$& $0.8441 \pm 0.0037$\\
$900$& $0.3896 \pm 0.0035$& $0.3315 \pm 0.0042$& $2.5695 \pm 0.0058$& $1.273 \pm 0.002$\\
\end{tabular}
\end{ruledtabular}
\end{table*}

\section{\label{sec: transition} Signatures of the transition from the chimera state and reproduction of the phase diagram}
\begin{figure*}
\centering
\begin{subfigure}{0.3\textwidth}
\centering\includegraphics[scale = 0.45]{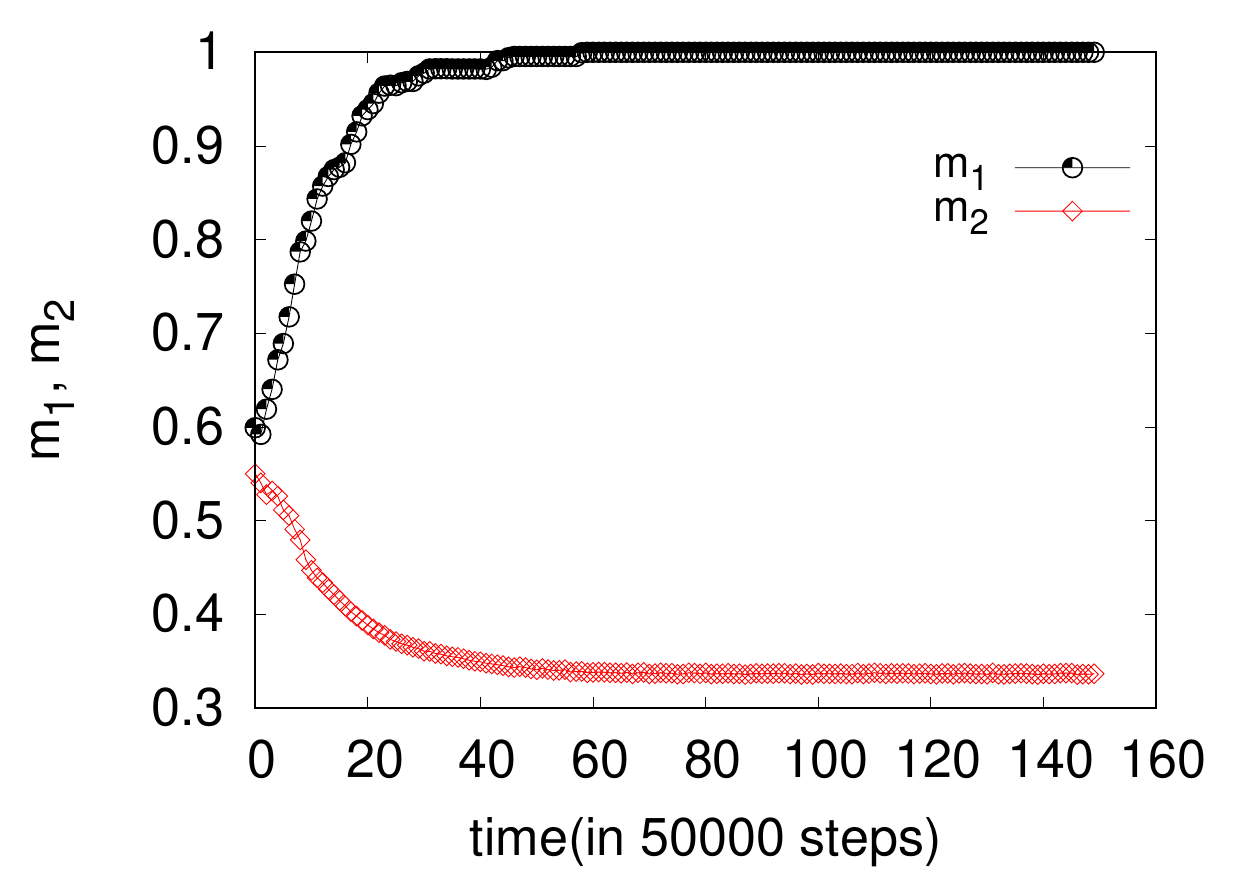}
\caption{}
\label{fig: m_chim_defect}
\end{subfigure}%
\begin{subfigure}{0.3\textwidth}
\centering\includegraphics[scale = 0.45]{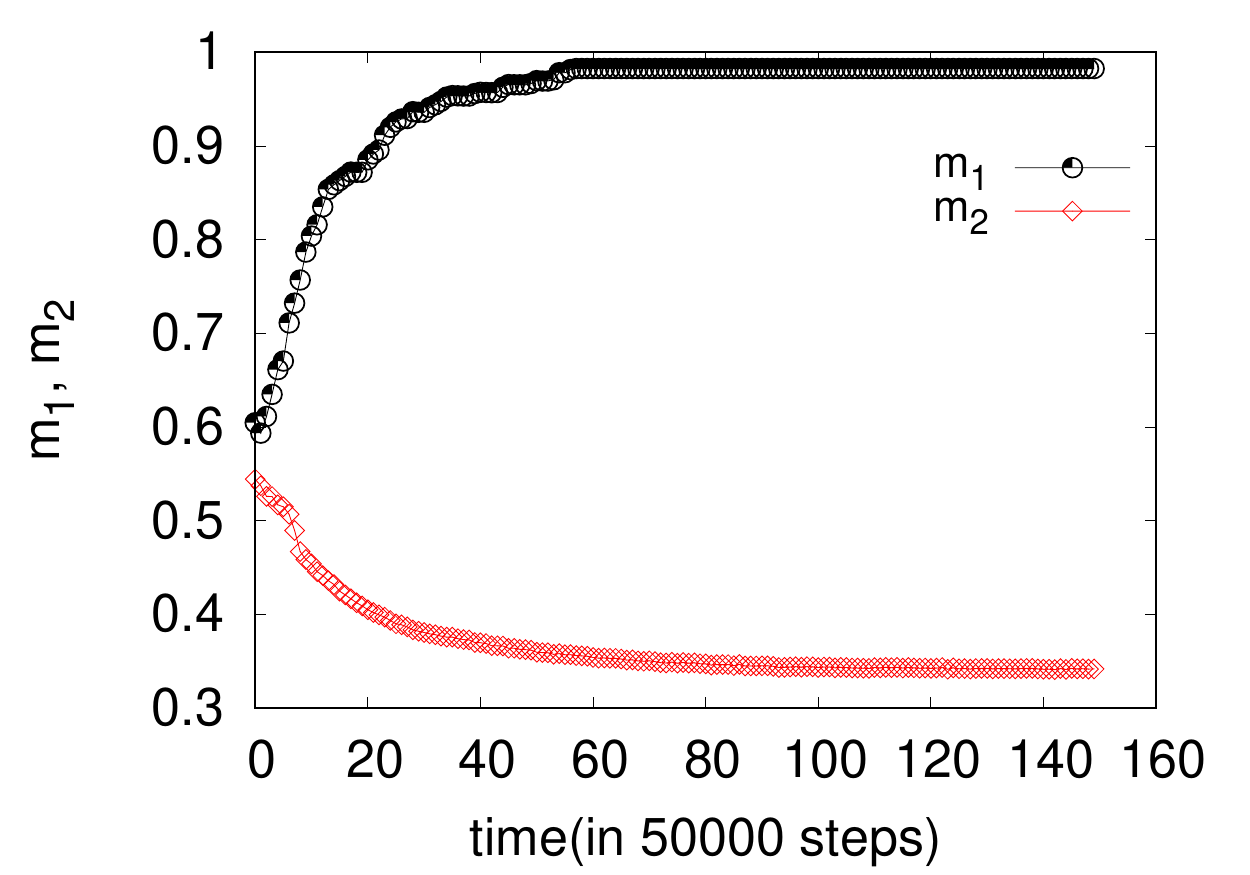}
\caption{}
\label{fig: m_chim_pure}
\end{subfigure}%
\begin{subfigure}{0.3\textwidth}
\centering\includegraphics[scale = 0.45]{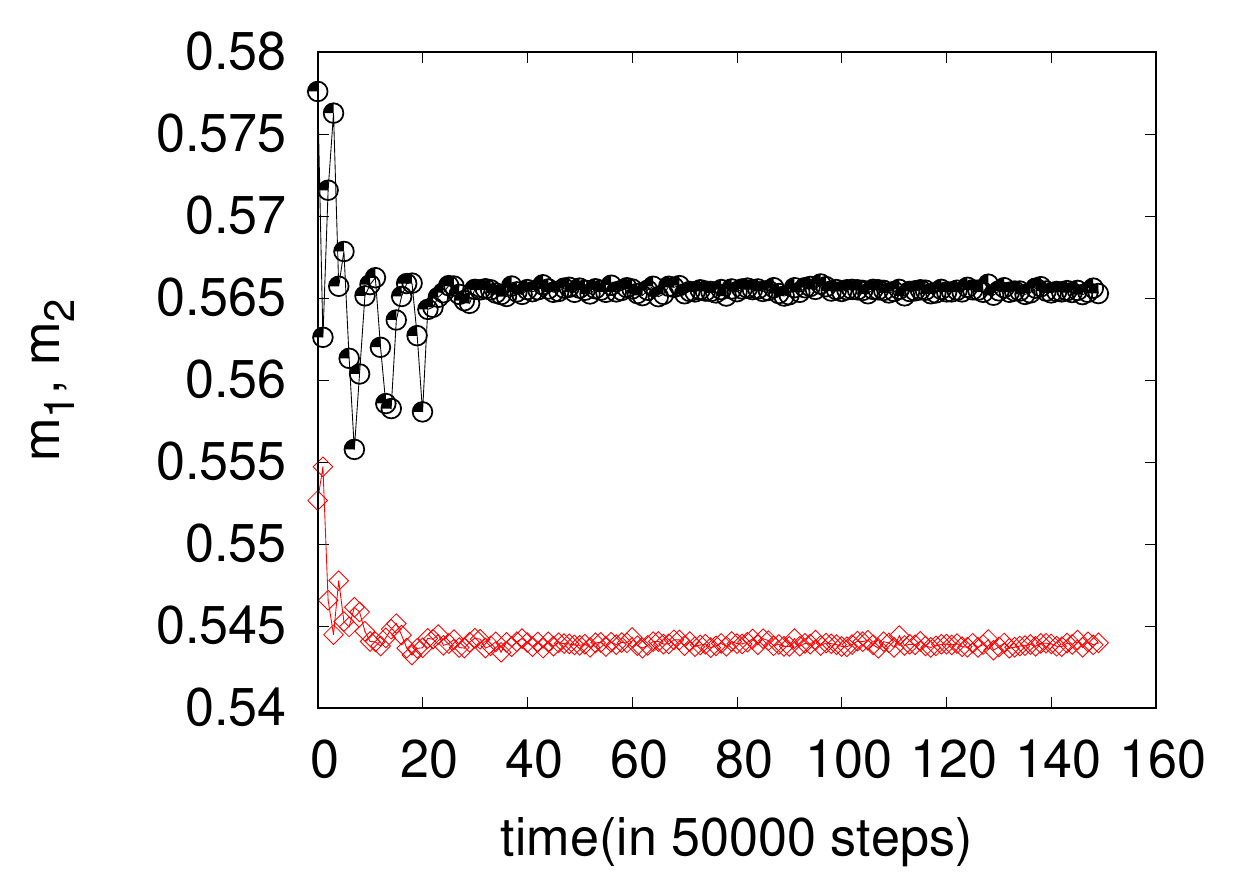}
\caption{}
\label{fig: m_full_desync}
\end{subfigure}
\caption{\label{fig: m_transient} The variation of $m_1$ and $m_2$ for during the evolution of (a) case 1 : chimera states with purely synchronised subgroup $(\epsilon_1 = 0.82)$, (b) case 2 : chimera states with defects in the synchronised subgroup $(\epsilon_1 = 0.93)$ and (c) case 3 : fully phase synchronised state $(\epsilon_1 = 0.75)$. Other parameters are kept fixed at $K = 10^{-5}$, $\Omega = 0.27$, $N = 150$. For each of the above cases the system is evolved from a completely random initial condition and laminar/burst sites are identified during the evolution. The average fraction of laminar sites is calculated after each $n' = 50000$ time steps. }
\end{figure*}

We have noted earlier that the crucial element which governs whether the map at a given site remains in the laminar or burst state, after time evolution from step $n$ to step $n+1$, is the number of maps in each subgroup whose phase angles are in the laminar or burst phase, at step $n$, as defined by the pairwise order parameter. This also turns out to be the key element in the existence of the spatiotemporally intermittent chimera. The phase diagram of Fig. \ref{fig: order_zoom} which focusses on the chimera region and its boundaries, is constructed using  the global order parameter  $(R_n$, and group-wise order parameters  $ R^{1}_{n}, R^{2}_{n})$ which differentiate between fully phase synchronised configurations, partially phase synchronised configurations (e.g. chimera states) and fully phase desynchronised configurations. This phase diagram, as well as the cross section taken at $\log_{10}K = -5.5$ (see Fig. \ref{fig: density_1}.a) show that at $\epsilon_1 \approx 0.8$ there is a transition from the fully desynchronised state to chimera states. Here we show that an identical phase diagram, and the signatures of these transitions can be reproduced using the average fraction of laminar sites $m_\sigma$  which is defined to be $m_\sigma = \frac{1}{n'}\sum\limits_{n}^{n + n'}\frac{x_\sigma(n)}{N}$ where $x_\sigma(n)$ is the number of laminar sites at any time step $n$ \footnote{We calculate the average because the fraction $\tfrac{x_\sigma}{N}$ fluctuates at consecutive time steps while its average, $m_{\sigma}$ tends to a constant value as the system reaches the final solution}. If this quantity is plotted as a function of time as in Fig. \ref{fig: m_transient} it can be seen  that after an initial transient, $m_\sigma$  settles to the  fixed values shown in table \ref{table_one}.  These are the $m_\sigma$ values in the chimera state and fully phase desynchronised state. 

\begin{table}
\caption{\label{table_one}The parameters of the system are kept fixed at $K = 10^{-5}, \Omega = 0.27, N = 150$ for all three cases below.}
\begin{ruledtabular}
\begin{tabular}{ccc}
Attractor& $m_{1}$ (numerical) & $m_{2}$ (numerical)\\
\hline
Case 1\footnote{$\epsilon_1 = 0.82$, chimera states with a purely synchronised subgroup}&1.0&0.337\\
Case 2\footnote{$\epsilon_1 = 0.93$, chimera states with defects in the synchronised subgroup}&0.982&0.344\\
Case 3\footnote{$\epsilon_1 = 0.75$, fully phase desynchronised state}&0.565&0.544\\
\end{tabular}
\end{ruledtabular}
\end{table}

\par
\begin{figure*}
\centering
\begin{subfigure}{0.5\textwidth}
\centering\includegraphics[scale = 0.55]{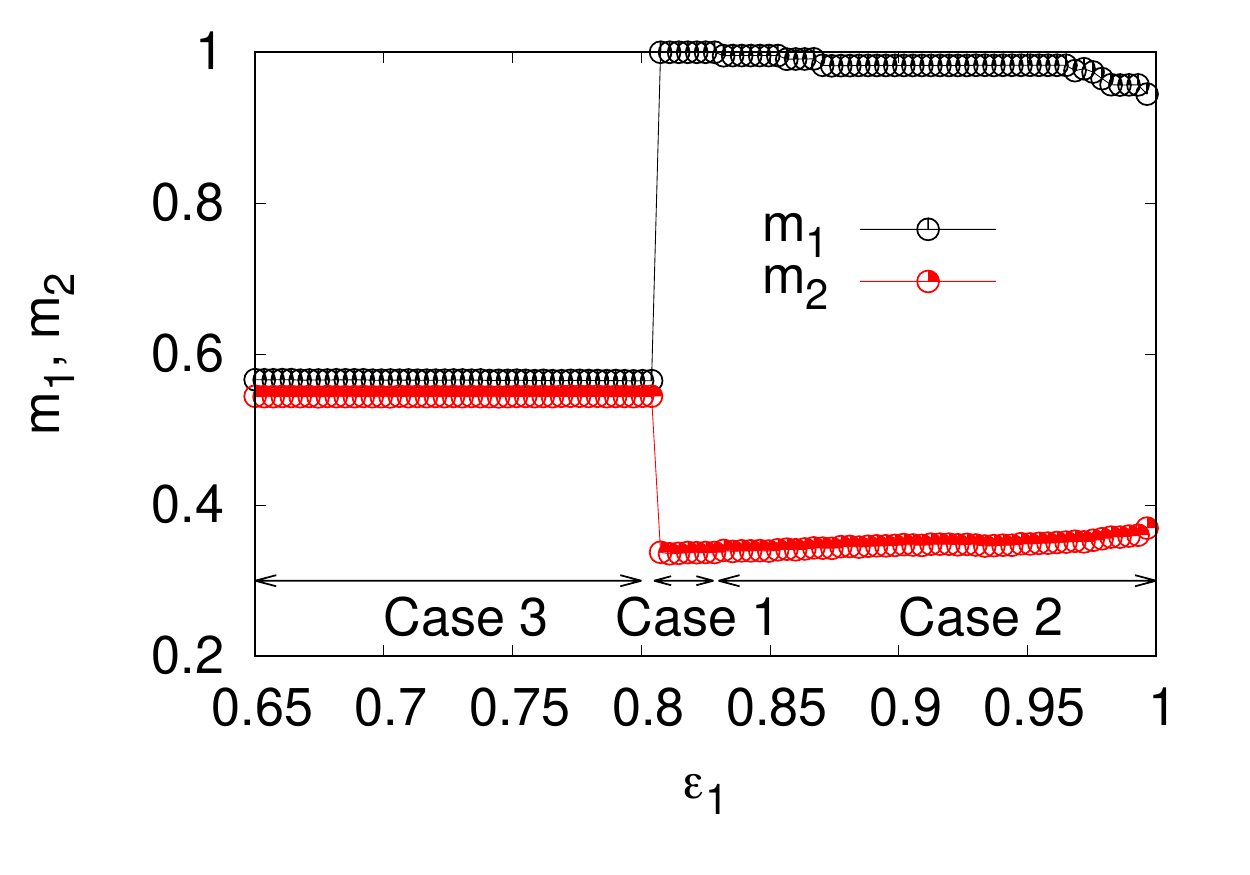}
\caption{\label{fig: density_2_1}}
\end{subfigure}%
\begin{subfigure}{0.5\textwidth}
\centering\includegraphics[scale = 0.55]{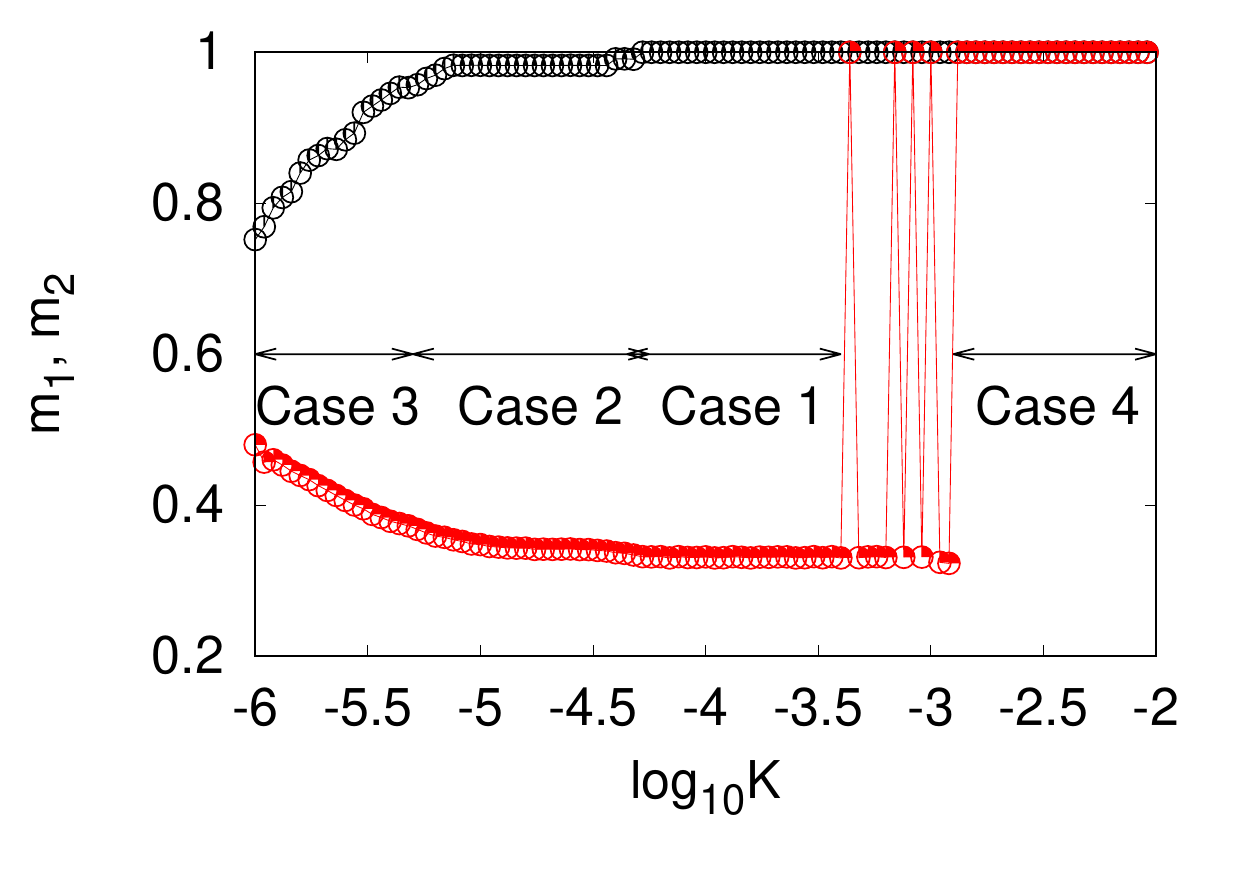}
\caption{\label{fig: density_2_2}}
\end{subfigure}
\caption{\label{fig: density_2} \footnotesize (color online) (a) The average fractions ${m}_1, {m}_2$ are calculated at parameters $N = 150, K = 10^{-5}, \Omega = 0.27$. The signature of the transition from the fully phase de-synchronisation to chimera phase state can be seen at $\epsilon_1 = 0.8$. Here the value of ${m}_{1}$ becomes one while ${m}_{2}$ remains less than one as expected for the chimera phase state in the CML. (b) The variation of the average fraction ${m}_{1}$ and ${m}_{2}$ are plotted as $\log_{10}K$ varies between $-6$ and $-2$ keeping $\epsilon_1$ fixed at $0.93$ with $\Omega = 0.27$. Chimera states with defects in the synchronised group appear when $\log_{10}K < -4.27$ while chimera states with purely phase synchronised group appear between $-4.3 < \log_{10}K < -3.4$. When $\log_{10}K$ is between $-3.36$ and $-2.9$ we observe that the system fluctuates between purely phase synchronised chimera state and two clustered state with small variations of $K$. The system settles to the two clustered state when $\log_{10}K > -2.9$. }
\end{figure*}

 Figs. \ref{fig: density_2_1} and \ref{fig: density_2_2} show  the variation of $m_\sigma$ with $\epsilon_1$ and $K$ respectively. The variation of ${m}_1, {m}_2$ clearly indicates the transition from the fully phase desynchronised state to the chimera phase state in the CML (see Fig. \ref{fig: density_2}.(a)) with increasing values of $\epsilon_1$. Here,  ${m}_{1}, {m}_{2} \approx 0.55$ (phase desynchronized values) when $\epsilon_1 < 0.81$ and ${m}_{1} \approx 1, {m}_{2} \approx 0$ (chimera state) when $\epsilon_1 > 0.81$. In fact between  the parameter values $\epsilon_1 = 0.808$ and $0.828$ we observe that ${m}_{1} = 1$ and ${m}_{2} \approx 0$ which identifies a chimera state with a purely phase synchronised subgroup. When $\epsilon_{1} > 0.828$ we find that ${m}_{1} \lesssim 1$ indicating that there are defects in the synchronised group. The number of defects slowly increases as $\epsilon_1$ increases to one for this fixed value of $K$. Comparing Figs. \ref{fig: density_1}.(a) and \ref{fig: density_2}(a) we can see that the quantity ${m}_{\sigma}$ can differentiate correctly between the chimera with a purely synchronised subgroup and the chimera state with defects in the synchronised subgroup. The variation of the order parameters in Fig. \ref{fig: density_1}.b shows another cross section taken at another parameter in the phase diagram, viz. $\epsilon_1 = 0.93$,  where similar behaviour is found in the variation of ${m}_1$ and ${m}_2$ (see Fig. \ref{fig: density_2}.(b)). To  compare this  through the quantity $m_\sigma$, we find that when $-5.2 < \log_{10}K < -4.3$ chimeras with defects in the phase synchronised cluster appear as ${m}_1 \lesssim 1, {m}_2 \approx 0$ in this range, while the chimera states with a purely synchronised subgroup appear when $-4.3 < \log_{10}K < -3.4$, since in this range of $K$, ${m}_1 = 1, {m}_2 \approx 0$. 

\begin{figure*}
\centering
\begin{subfigure}{0.5\textwidth}
\centering\includegraphics[scale = 0.7]{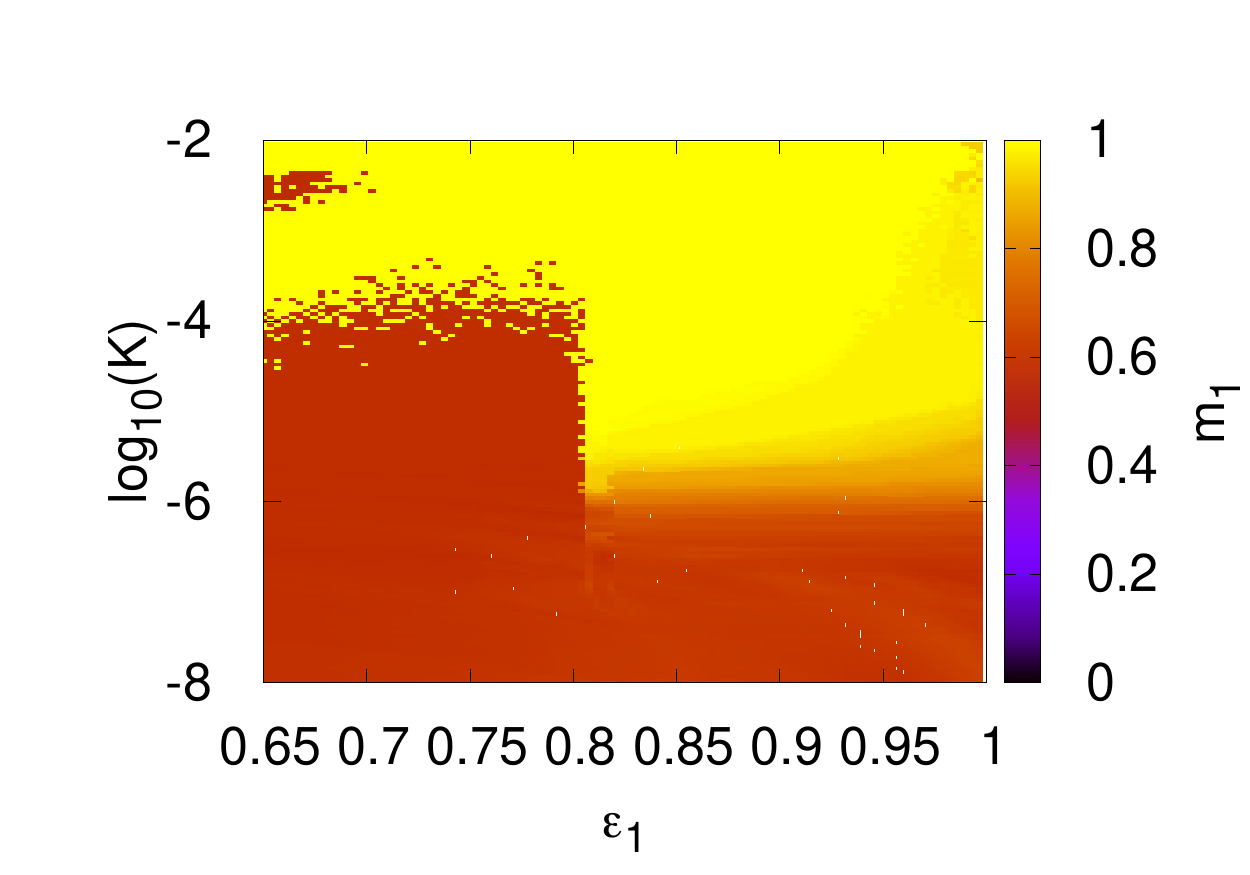}
\caption{}
\end{subfigure}%
\begin{subfigure}{0.5\textwidth}
\centering\includegraphics[scale = 0.7]{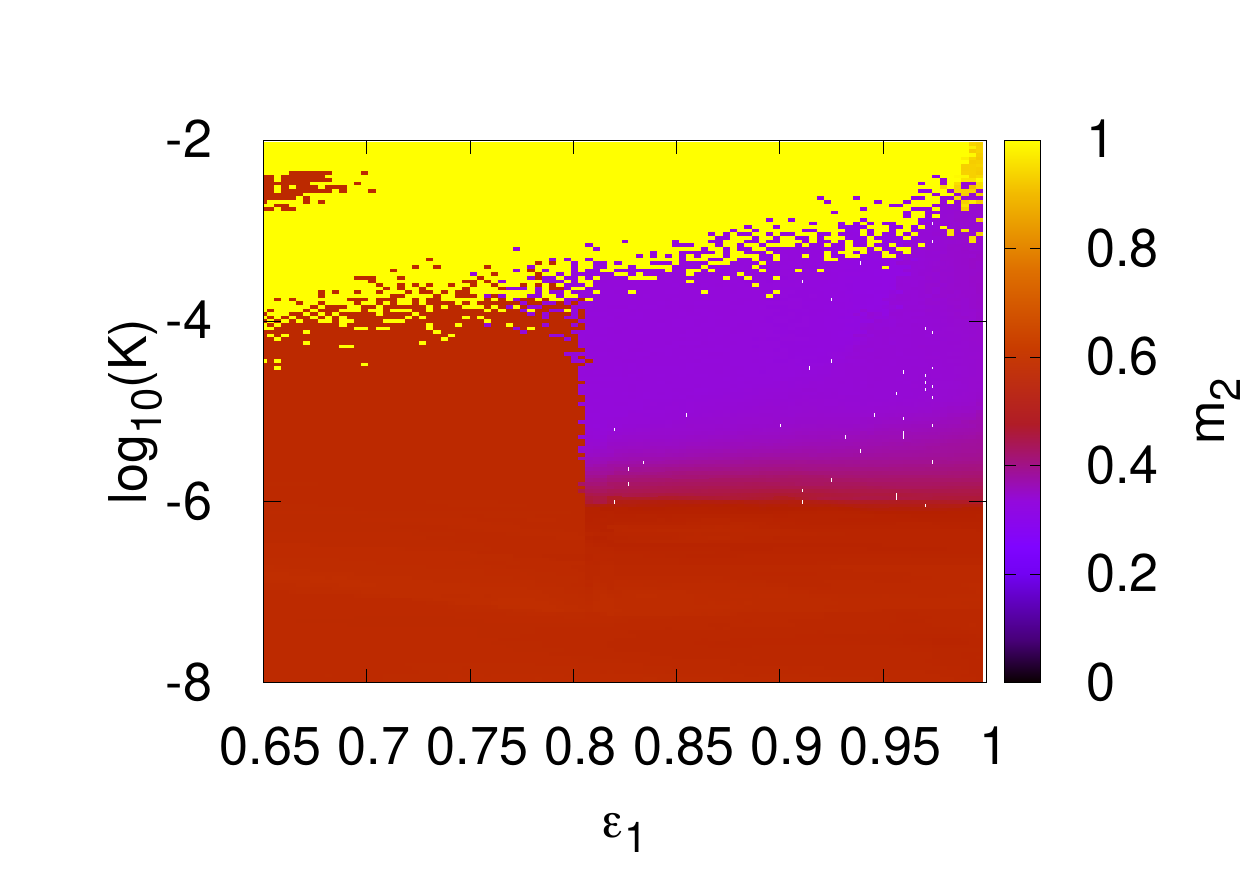}
\caption{}
\end{subfigure}
\caption{\label{fig: mean_phase_diag} \footnotesize (color online) The average fraction of laminar sites (a) ${m}_1$ and (b) ${m}_2$ are calculated over $10^{6}$ time after discarding initial $3\times10^{6}$ time steps at each parameter between $-8 < \log_{10}K < -2$, $0.65 < \epsilon_1 < 1$ with $\Omega = 0.27$ and $N = 150$. At each point a completely random initial condition between zero and one was used. These are used to identify the phases as explained in Sec VI.}
\end{figure*}

We now reproduce the phase diagram for the range of parameters given by $-8 < \log_{10}K < -2$ and $0.65 < \epsilon_1 < 1$ for $\Omega = 0.27$, using the quantities  ${m}_{1}$ and ${m}_2$, i.e. the average fraction of laminar sites in groups one and two. It can be seen from Figures \ref{fig: mean_phase_diag}.(a) and (b) that the average fraction of laminar sites correctly replicates the chimera configuration in the region approximately given by $0.8 < \epsilon_1 < 1$ and $-5.5 < \log_{10}K < -4$. We see that other types of configurations are also seen near the boundary of this region. Chimera configurations are seen for $\epsilon$ values, such that  $\epsilon_1 > 0.8$, whereas fully desynchronised configurations are seen for values of $\epsilon_1 < 0.8$.  Two clustered states are found  between $-4 < \log_{10}K < -3$ for $0.8 < \epsilon_1 < 1$ at the boundary of the parameter region which show the chimera states. We see that for this range of $\epsilon_1$ and for  $\log_{10}K > -3$, both $m_1$ and $m_2$ are one, indicating  the existence of two clustered states. Within the same range of $\epsilon_1$ if we decrease $K$ we see that defects start to appear in group one, as $m_1$ decreases from one. As $\log_{10}K$ approaches $-6$ the number of defects increases for this range of $\epsilon_1$. For  $\log_{10}K$ values  close to $-6$ the defects in group one cause the values  $m_1$ to be be comparable to $m_2$ implying that the chimera configuration is lost. Similarly, the fraction $m_1 \lesssim 1$ as $\epsilon_1$ increases from $0.8$ to one when $\log_{10}K$ is between the range $-5.5$ and $-4$ implying the appearance of defects in the synchronised group. The fully desynchronised phase configuration with $0.5 < m_1, m_2 < 0.6$ is seen at the parameters $-5.5 < \log_{10}K < -4$ and $\epsilon_1 < 0.8$ in Fig. \ref{fig: density_2}. Thus the average fraction of laminar sites calculated for the final state accurately reproduces the phase diagram of the CML in the region of interest and verifies the interaction between the sites during the spatiotemporal evolution of each attractor. 

\section{\label{sec: conclusion}Conclusion}
To summarise, we have analysed a system which shows novel chimera behaviour, viz. a mixed state with a synchronised part and a spatiotemporally intermittent part. This behaviour is seen in a coupled map lattice consisting of two groups of globally coupled sine circle maps with different values of intergroup coupling and intra-group coupling. The system, when evolved with random initial conditions, shows a variety of solutions in different regions of the parameter space. A phase diagram is obtained using the complex order parameter, and the basin stability of each type of solution in the context of multiattractor behaviour  is discussed. We note that the basin stability of the  chimera states, is large in the chimera region, with the chimera and its mirror version being equally probable at all parameter values. We note that the STI chimeras are seen at very small values of the nonlinearity parameter $K$ $(10^{-5.5} \lesssim K \lesssim 10^{-3})$, where the map behaviour is very close to the behaviour of coupled shift maps. Analytic techniques can be effectively applied in this regime, and confirm the results obtained numerically. The Lyapunov exponent spectrum in this regime is calculated by both methods, and turns out to have two positive exponents, confirming that the chimera seen here is a hyperchaotic chimera. We note that very few examples of hyperchaotic chimeras have been seen  earlier\cite{wolfrum2011}. The parameter values in this regime is similar to the regime where splay chimera states have been seen earlier, with splay initial conditions \cite{neelima2016}. However, none of the splay states observed show hyperchaotic behaviour. One application context where such low values of $K$ can be realised is that of coupled Josephson junction arrays with high values of capacitance \cite{josephson2017}. 

The spatiotemporally intermittent chimera seen here shows co-existing laminar and burst states, which are identified via a pairwise order parameter. The distribution of laminar and turbulent lengths drops off exponentially, due to global coupling, unlike the power law behaviour seen at some parameter values for locally diffusive coupling \cite{zahera2005, zahera2006}. The global nature of the coupling used here, and the distinct values of intergroup and intragroup coupling, implies  that the observed behaviour is dependent on the number of laminar and turbulent sites in each subgroup. The average fraction of laminar and burst sites saturates to steady state values after an initial transient. This average fraction can be used to construct the phase diagram in the vicinity of the chimera region. This phase diagram matches exactly the phase diagram constructed via the order parameter, confirming that the average number of laminar and turbulent sites is the crucial factor in the spatiotemporal dynamics of the chimera. A cellular automaton with global coupling which incorporates these features can be easily constructed. We hope to explore this approach in future work, and examine its consequences for the analysis of the chimera state. We also hope to explore the consequences of the hyperchaotic behaviour seen in the chimera state seen here, and its implications for experimental systems such as coupled laser models and Josephson junction arrays where such chimeras can be realised.            

\appendix
\section{\label{app: ls}Linear stability analysis of the globally synchronised state and the two phase clustered state}
\subsection{\label{app: sync} The globally synchronised state}
The analysis of the globally synchronised state has been carried out in Ref. \cite{nayak2011}. We summarise this over here. In order to carry out the linear stability analysis for the globally synchronised state, $\theta^{\sigma}_{n}(i) = \theta_{sync}$, $\forall \sigma = 1, 2$ and $i = 0, 1, 2, \cdots 2N$ at time step $n$, the one step Jacobian matrix (Eq. \ref{jacobian}) is converted to a block circulant form using a similarity transformation via a matrix given by a direct product of $2 \times 2$ Fourier matrix \cite{davis1994} and an $N \times N$ identity matrix. The transformed Jacobian is given as, 

\begin{widetext}
\begin{equation}
\small C + D\\
 = (1 - K\cos2\pi\theta_{\text{sync}})\begin{bmatrix}
             (1 + \frac{\epsilon_1 + \epsilon_2}{N}) & \frac{\epsilon_1 + \epsilon_2}{N} & \cdots &  \frac{\epsilon_1 + \epsilon_2}{N} \\
            \frac{\epsilon_1 + \epsilon_2}{N} &  (1 + \frac{\epsilon_1 + \epsilon_2}{N})  & \cdots & \frac{\epsilon_1 + \epsilon_2}{N} \\
             \cdots & \cdots &\cdots & \cdots \\
             \frac{\epsilon_1 + \epsilon_2}{N} &  \frac{\epsilon_1 + \epsilon_2}{N} & \cdots & (1 + \frac{\epsilon_1 + \epsilon_2}{N})
	    \end{bmatrix}
\end{equation} 
\begin{equation}
\small C - D = (1 - K\cos2\pi\theta_{\text{sync}})\begin{bmatrix}
             (1 + \frac{\epsilon_1 - \epsilon_2}{N}) & \frac{\epsilon_1 - \epsilon_2}{N} & \cdots & \frac{\epsilon_1 - \epsilon_2}{N} \\
            \frac{\epsilon_1 - \epsilon_2}{N} &  (1 + \frac{\epsilon_1 - \epsilon_2}{N}) & \cdots & \frac{\epsilon_1 - \epsilon_2}{N} \\
             \cdots & \cdots &\cdots & \cdots \\
             \frac{\epsilon_1 - \epsilon_2}{N} & \frac{\epsilon_1 - \epsilon_2}{N} & \cdots & (1 + \frac{\epsilon_1 - \epsilon_2}{N})
	    \end{bmatrix}
\end{equation}
\end{widetext}
The eigenvalues $\lambda_{j}$ of the matrix $C + D$ and $C - D$ is given by,
\newpage
\begin{widetext}
\begin{equation}
E_{j}^{C \pm D} = (1 - K\cos2\pi\theta_{\text{sync}})\Bigg( 1 + \frac{\epsilon_1 \pm \epsilon_2}{N} + \frac{\omega^{j} (\epsilon_1 \pm \epsilon_2)}{N} + \frac{\omega^{2j} (\epsilon_1 \pm \epsilon_2)}{N}+ \cdots + \frac{\omega^{(N - 1)j} (\epsilon_1 \pm \epsilon_2)}{N} \Bigg)
\label{eigenvalue}
\end{equation}
\end{widetext}
where $\omega$ is the $N^{\text th}$ root of unity i.e. $\omega = \exp(\frac{2 \pi i}{N})$. Setting  $j = 0$ we obtain the zeroth eigenvalues of the matrices $C + D$, $C - D$. So,
\begin{equation}
\begin{split}
E_{0}^{C + D} = 2(1 - K\cos2\pi\theta_{\text{sync}})
\end{split}
\end{equation}

\begin{equation}
\begin{split}
E_{0}^{C - D} = 2 \epsilon_1(1 - K\cos2\pi\theta_{\text{sync}})
\end{split}
\end{equation}				
For any $j > 0$ we have 
\begin{equation}
\begin{split}
E_{j}^{C \pm D} = (1 - K\cos2\pi\theta_{\text{sync}})				
\end{split}
\end{equation}
where we use $\epsilon_1 + \epsilon_2 = 1$ and $1 + \omega^{j} + \omega^{2j} + \cdots + \omega^{(N - 1)j} = 0$. So the eigenvalues of the matrix $J_{c}^*$ for $K \rightarrow 0$, are $2(1 - K\cos2\pi\theta_{\text{sync}})$, $2\epsilon_1(1 - K\cos2\pi\theta_{\text{sync}})$ and $2N - 2$ fold degenerate eigenvalues $1 - K\cos2\pi\theta_{\text{sync}}$. The eigenvalues of the Jacobian matrix for the shift map case $K = 0$ can be found from the above and they are, $2$, $2\epsilon_1$ and $2N - 2$ fold degenerate eigenvalues which are one. 

\subsection{\label{app: t_c} Two clustered state}
\par
\begin{figure}
\centering\includegraphics[scale = 0.6]{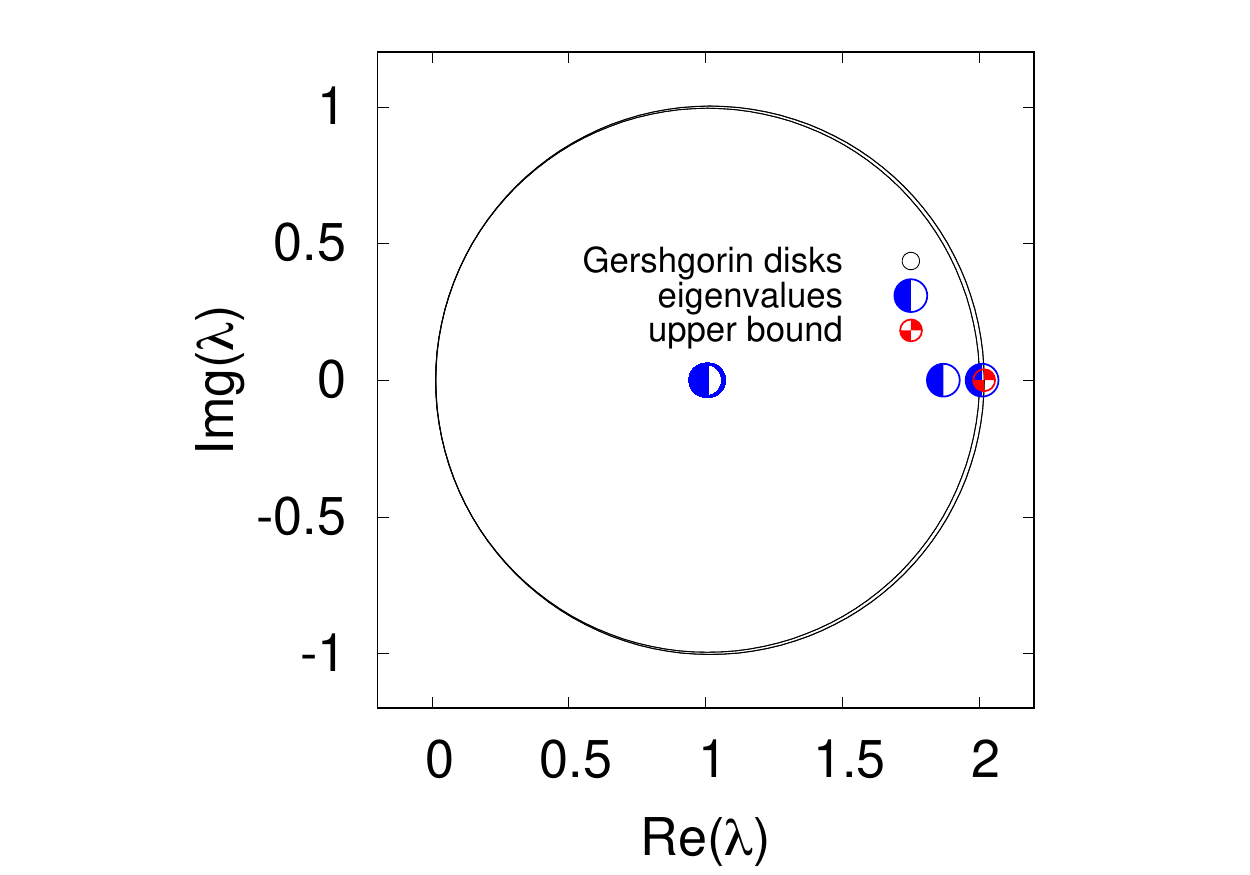}
\caption{\label{fig: gg} \footnotesize Eigenvalues and the Gershgorin disks of the Jacobian matrix of the system calculated for the two clustered state. The parameters are $\epsilon_1 = 0.93, \Omega = 0.27, K = 10^{-2}, N = 150$. }
\end{figure}

 Using the fact that the phases in group one take the values $\theta^{1}_{n}(i) = \theta_1$ and those in group two take the values $\theta^{2}_{n}(i) = \theta_{2}$ for all $i = 0, 1, 2, \cdots N - 1$, we find the eigenvalues of the Jacobian matrix in Eq. \ref{jacobian}. We verify the eigenvalue spectrum by calculating the upper bound on the largest eigenvalue using the Gershgorin theorem \cite{richard1994} analytically and checking if the entire eigenvalue spectra is less than the upper bound as discussed in Ref. \cite{neelima2016}. The Jacobian matrix in this case is given by, 
\begin{equation}
J_{\text{c}} = \begin{bmatrix} 
	E & F \\
	G & H
      \end{bmatrix}
      \label{jacobian_block}	
\end{equation}
Here $E, F, G, H$ are $N \times N$ block matrices which have the form,
\begin{equation}		
E = \begin{bmatrix}
	(1 + \frac{\epsilon_{1}}{N})f'(\theta_1) & \frac{\epsilon_{1}}{N}f'(\theta_1) & \cdots & \frac{\epsilon_1}{N}f'(\theta_1) \\
	\frac{\epsilon_{1}}{N}f'(\theta_1)  &(1 + \frac{\epsilon_{1}}{N})f'(\theta_1) & \cdots & \frac{\epsilon_1}{N}f'(\theta_1) \\
	\cdots & \cdots & \cdots & \cdots \\
	\frac{\epsilon_{1}}{N}f'(\theta_1) & \frac{\epsilon_{1}}{N}f'(\theta_1) & \cdots & (1 + \frac{\epsilon_{1}}{N})f'(\theta_1)	
      \end{bmatrix}
\end{equation}
\begin{equation}		
H = \begin{bmatrix}
	(1 + \frac{\epsilon_{1}}{N})f'(\theta_2) & \frac{\epsilon_{1}}{N}f'(\theta_2) & \cdots & \frac{\epsilon_1}{N}f'(\theta_2) \\
	\frac{\epsilon_{1}}{N}f'(\theta_2)  &(1 + \frac{\epsilon_{1}}{N})f'(\theta_2) & \cdots & \frac{\epsilon_1}{N}f'(\theta_2) \\
	\cdots & \cdots & \cdots & \cdots \\
	\frac{\epsilon_{1}}{N}f'(\theta_2) & \frac{\epsilon_{1}}{N}f'(\theta_2) & \cdots & (1 + \frac{\epsilon_{1}}{N})f'(\theta_2)	
      \end{bmatrix}
\end{equation}
\begin{equation}		
F = \begin{bmatrix}
	\frac{\epsilon_{2}}{N}f'(\theta_2) & \frac{\epsilon_{2}}{N}f'(\theta_2) & \cdots & \frac{\epsilon_2}{N}f'(\theta_2) \\
	\frac{\epsilon_{2}}{N}f'(\theta_2) & \frac{\epsilon_{2}}{N}f'(\theta_2) & \cdots & \frac{\epsilon_2}{N}f'(\theta_2)  \\
	\cdots & \cdots & \cdots & \cdots \\
	\frac{\epsilon_{2}}{N}f'(\theta_2) & \frac{\epsilon_{2}}{N}f'(\theta_2) & \cdots & \frac{\epsilon_2}{N}f'(\theta_2)	
      \end{bmatrix}
\end{equation}
\begin{equation}		
G = \begin{bmatrix}
	\frac{\epsilon_{2}}{N}f'(\theta_1) & \frac{\epsilon_{2}}{N}f'(\theta_1) & \cdots & \frac{\epsilon_2}{N}f'(\theta_1) \\
	\frac{\epsilon_{2}}{N}f'(\theta_1) & \frac{\epsilon_{2}}{N}f'(\theta_1) & \cdots & \frac{\epsilon_2}{N}f'(\theta_1)  \\
	\cdots & \cdots & \cdots & \cdots \\
	\frac{\epsilon_{2}}{N}f'(\theta_1) & \frac{\epsilon_{2}}{N}f'(\theta_2) & \cdots & \frac{\epsilon_2}{N}f'(\theta_1) 	
      \end{bmatrix}
\end{equation} 
where $f'(\theta_{\sigma}) =  1 - K\cos2\pi\theta_{\sigma}$ and $\sigma = 1, 2$ denote each group. The bounds on the eigenvalues are obtained by constructing the Gershgorin disks, whose centres have values given by the diagonal elements of the matrix of interest and whose radii are given by the sum of the off-diagonal elements in the row or column. The diagonal elements of the matrix $J_c$ are real and nonnegative which implies that the Gershgorin row region and the column region will consist of disks whose centers lie on the real axis. For the two phase clustered state the centre, $(c_{j})$ of the $j$th Gershgorin disk is given by, $c^{\sigma}_{j} = \left(1 + \frac{\epsilon_1}{N}\right)(1 - K\cos2\pi\theta_{\sigma})$, $\sigma = 1,$ for $j = \{0, 1, \cdots, N - 1 \}$ and $\sigma = 2$ \text{when} $j = \{ N, N + 1, \cdots, 2N \}$. The radius of the $j$th disc in the Gershgorin row region $r_{j}$ is,
\begin{equation}
\begin{split}
r^{1}_{j} = \frac{\epsilon_1(N - 1)}{N}(1 - K\cos\theta_1) &+ \epsilon_2(1 - K \cos\theta_2)\\ &\text{for}j = 0, 1, 2, \cdots, N - 1\\
r^{2}_{j} = \frac{\epsilon_1(N - 1)}{N}(1 - K\cos\theta_2) &+ \epsilon_2(1 - K \cos\theta_1)\\ &\text{for}j = N, N + 1, \cdots, 2N\\
\end{split}
\end{equation}
The radius of the $i$th disc in the column region is
\begin{equation}
\begin{split}
s^{1}_{i} = \left(\frac{\epsilon_1(N - 1)}{N} + \epsilon_2\right)&(1 - K\cos\theta_1)  \,\,\,\,\,\,\\ &\text{for}j = \{0, 1, 2, \cdots, N - 1\}\\
s^{2}_{i} = \left(\frac{\epsilon_1(N - 1)}{N} + \epsilon_2\right)&(1 - K\cos\theta_2)  \,\,\,\,\,\,\\ &\text{for}j = \{N, N + 1, \cdots, 2N\}\\
\end{split}
\end{equation}
Since the centres of every disc in the Gershgorin row and column region lie on the real axis, the two bounds set by the Gershgorin row and column regions are given by the two largest numbers at which the discs from each of these sets intersect the real axis i.e. 
$\text{max}(c^{\sigma}_{j} + r^{\sigma}_{j})$ and $\text{max}(c^{\sigma}_{i} + s^{\sigma}_{i})$ for $\sigma = 1, 2$ and $i, j = 0, 1, \cdots, 2N - 1$. The required bound on the eigenvalues is the minimum of these two values.
So the upper bound on the eigenvalues of Jacobian for the two clustered state is,
\begin{eqnarray*}
\text{min} \left( \text{max}\left( c^{\sigma}_{j} + r^{\sigma}_{j} \right), \text{max}\left( c^{\sigma}_{i} + s^{\sigma}_{i} \right)\right)
\end{eqnarray*}
for $\sigma = 1, 2$ and $i, j = 0, 1, 2, \cdots, 2N - 1$. Fig. \ref{fig: gg} shows that the largest eigenvalue as calculated numerically, almost saturates the upper bound of the system. 

\nocite{*}
\bibliography{CA_chim_final}

\end{document}